\documentclass[12pt,a4paper,psamsfonts]{article}
\usepackage{amsmath,amssymb,amsthm}
\usepackage{eucal}
\usepackage{breqn}

\numberwithin{equation}{section}

\usepackage{geometry}

\newcommand{\barz}{{\bar{z}}}

\newcommand{\be}{\begin{equation}}
\newcommand{\ee}{\end{equation}}
\newcommand{\bea}{\begin{eqnarray}}
\newcommand{\eea}{\end{eqnarray}}

\renewcommand{\d}{\delta}
\newcommand{\e}{\epsilon}
\newcommand{\G}{\Gamma}

\newcommand{\m}{\mu}
\newcommand{\n}{\nu}

\newcommand{\s}{\sigma}

\newcommand{\hlf}{\frac{1}{2}}

\newcommand{\non}{\nonumber}

\newcommand{\p}{\partial}

\newcommand{\w}{\wedge}
\newcommand{\Z}{\mathbb{Z}}

\newcommand{\tr}{\operatorname{tr}}

\newcommand{\U}{\operatorname{U}}

\newcommand{\lp}{\left(}
\newcommand{\rp}{\right)}

\newcommand{\hph}[1]{{\hphantom{#1}}}

\makeatletter
\renewcommand\section{\@startsection {section}{1}{\z@}%
                                   {-3.5ex \@plus -1ex \@minus -.2ex}
                                   {2.3ex \@plus.2ex}%
                                   {\normalfont\large\bfseries}}
\renewcommand\subsection{\@startsection{subsection}{2}{\z@}%
                                     {-3.25ex\@plus -1ex \@minus -.2ex}%
                                     {1.5ex \@plus .2ex}%
                                     {\normalfont\bfseries}}
\makeatother

\begin{document}

\begin{center}
\addtolength{\baselineskip}{.5mm}
\thispagestyle{empty}
\begin{flushright}
January 11, 2016

\hfill         MI-TH-1604
\end{flushright}

\vspace{20mm}

{\Large  \bf Three-Point Disc Amplitudes in the RNS Formalism}

\vspace{15mm}

\vskip 1.25 cm {
Katrin Becker\footnote{email address: kbecker@physics.tamu.edu}$^{}$,
Melanie Becker\footnote{email address: mbecker@physics.tamu.edu}$^{}$,
Daniel Robbins\footnote{email address: drobbins@physics.tamu.edu}}$^{}$ and
Ning Su\footnote{email address: suning@gmail.com}$^{}$\\
{\vskip 0.5cm $^{}$ \it George and Cynthia Mitchell Institute for Fundamental Physics and Astronomy, \\ Department of Physics, Texas A{\&}M University, \\ College Station, TX 77843, USA \\}

\vspace{20mm}

{\bf  Abstract}
\end{center}
We calculate all tree level string theory vacuum to Dp-brane disc amplitudes involving an arbitrary RR-state and two NS-NS vertex operators. This computation was earlier performed by K.~Becker, Guo, and Robbins for the simplest case of a RR-state of type $C^{(p-3)}$. Here we use the aid of a computer to calculate all possible three-point amplitudes involving a RR-vertex operator of type $C^{(p+1+2k)}$.

\vfill
\newpage








\section{Introduction}

D-branes have played many roles in string theory.  From the point of view of the string world-sheet they are simply boundary conditions, i.e.\ strings can end on the D-branes.  In practice, this means that if we compute string scattering amplitudes in a background with D-branes (including the type I string, which in this language is interpreted to have space-time-filling D9-branes), then we must include contributions from world-sheets with boundaries, in addition to the usual closed world-sheets.

Alternatively, from the point of view of the low-energy effective theory, D-branes host some degrees of freedom that are localized on the D-brane world-volume.  In this paper, we will only be considering separated D-branes, in which case the content of the world-volume theory is simply that of maximally supersymmetric super-Yang-Mills with gauge group $\U(1)$ for each D-brane. The full effective action is then a sum of a bulk action plus localized actions at each D-brane.  These localized actions involve both the world-volume fields and the bulk fields, and can be expanded in derivatives.  The details of that expansion are interesting in their own right as an example of an effective theory that admits many different dual perspectives.  But even more compellingly, there are examples in which the higher derivative couplings localized on D-branes play an essential role in determining the vacuum structure of string theory, such as in the F-theory duals of M-theory backgrounds on Calabi-Yau four-folds~\cite{Becker:1996gj,Dasgupta:1999ss}.  In the IIB description of these constructions, there are D7-branes which wrap four-cycles of the internal space.  These D7-branes host four-derivative bulk-field couplings of the schematic form
\be
\label{eq:D7Anomaly}
\int_{D7} C^{(4)}\w\tr(R\w R),
\ee
which lead to the $C^{(4)}$ tadpole equation.  This condition is crucial to get consistent solutions.  Similarly, there should be more four-derivative couplings which can contribute to charge cancellation in certain other flux backgrounds~\cite{Becker:2010ij,McOrist:2012yc,Maxfield:2013wka}.  For these reasons it is important to systematically compute the entire four-derivative effective action localized on a D-brane.

Of course, the world-sheet and effective theory perspectives are related.  The terms in the effective action can be computed by the relevant perturbative string scattering amplitudes.  For example, the coupling (\ref{eq:D7Anomaly}) can be obtained by computing a three-point disc amplitude with one R-R vertex operator and two graviton vertex operators~\cite{Craps:1998fn,Craps:1998tw,Stefanski:1998he,Bachas:1999um} (though there are other methods for deducing these particular couplings~\cite{Bershadsky:1995qy,Green:1996dd,Minasian:1997mm,Morales:1998ux,Scrucca:1999uz}).  As a preliminary step towards computing the full effective action, we need to compute all of the relevant string scattering amplitudes, as we do herein.

We calculate type II superstring scattering amplitudes on world-sheets with the topology of a disk, with closed or open string insertions. We are following references \cite{Becker:2011bw}, \cite{Becker:2011ar}, and \cite{Becker:2010ij}, where the formalism was developed and some simple amplitudes were computed. Similarly as done in these references,
the final goal is to extract information about the corresponding Dp-brane effective actions.
Some new aspects of these actions are discussed in a forthcoming paper~\cite{paper2}, where an interesting non-renormalization result for three-point functions involving a R-R field of type  \(C^{(p+5)}\) is presented.

\begin{figure}[!htb]
\label{fig:sugracab}
\begin{center}
\scalebox{.45}{\includegraphics{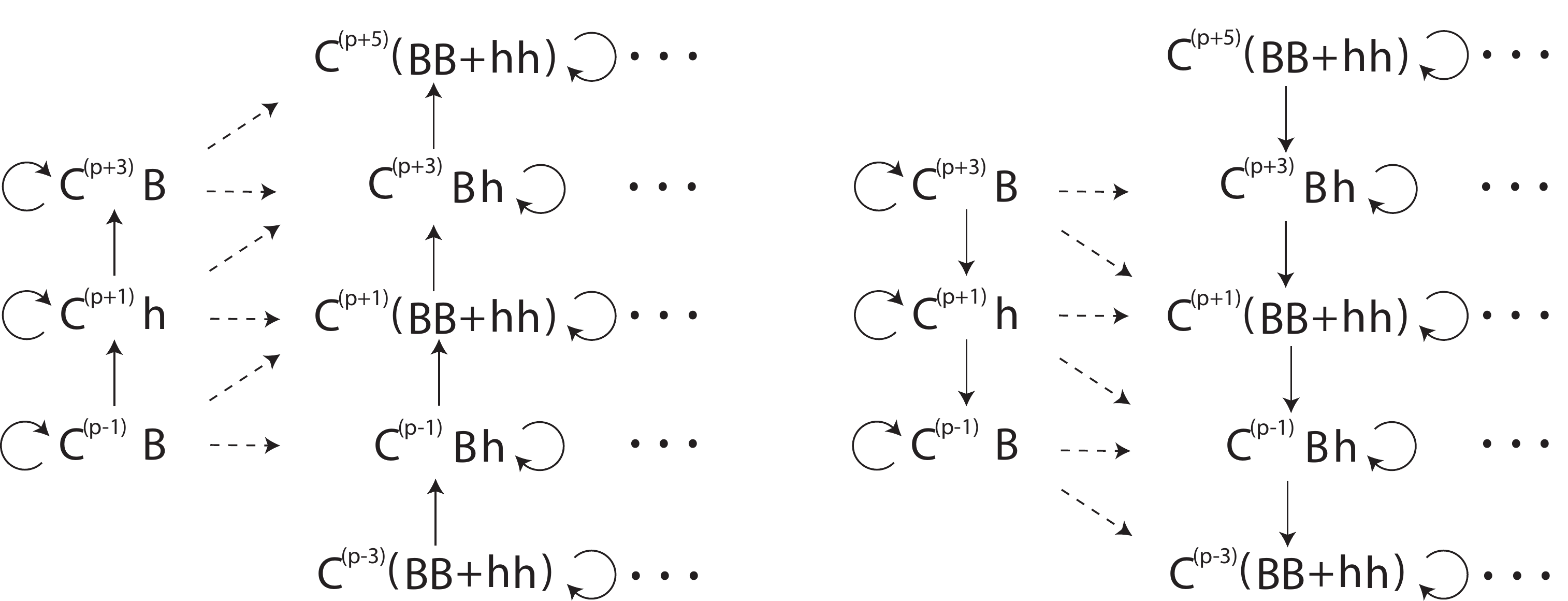}}
\caption{This is a reproduction of figure 7 from~\cite{Becker:2011ar}.  The arrows indicate amplitudes that are related by T-duality, either linearly (solid line) or non-linearly (dashed line).}
\end{center}
\end{figure}

The calculation of the two-point function involving one R-R state and one NS-NS state appeared in earlier papers \cite{Gubser:1996wt,Garousi:1996ad,Hashimoto:1996kf,Hashimoto:1996bf,Garousi:2010ki} or in the notation and conventions used herein in \cite{Becker:2011bw}.
The three-point amplitude involving one R-R field of type  \(C^{(p
-3)}\) was calculated in \cite{Becker:2011ar}.
Our goal here is to compute the most general tree level string theory vacuum to Dp-brane amplitude with insertion of an arbitrary R-R state and
various NS-NS vertex operators. We then restrict to the case of one R-R field and two
NS-NS fields. This amplitude is expressed in terms of the R-R potential \(C^{(p+1+2k)}\) and two
NS-NS fields.
The collection of these amplitudes is shown in Figure 7 of \cite{Becker:2011ar} (reproduced here in figure \ref{fig:sugracab}).
%

Because the amplitudes are invariant under a certain $\Z_2$ symmetry (combining reflection in the space-time directions that are normal to the brane with worldsheet parity), they are non-vanishing only if
\noindent
\begin{enumerate}
\item k is even and both NS-NS fields are antisymmetric or both are symmetric.
\item  k is odd and one of the NS-NS fields is symmetric and the other one is antisymmetric.
\end{enumerate}

In general, the coefficients of the amplitudes cannot be evaluated analytically, so we write them in a complex integral form. Due to conservation
of momentum and integration by parts, the integrals can be written in many different ways. We determine a minimum set of variables and express the final result in terms of them. Our choice makes the exchange symmetry of the two NS-NS fields apparent. This is a nontrivial check of our result,
since at the level of the vertex operators this symmetry is not manifest.

\section{Calculational Tools and Strategy}

To calculate the expression of the three-point function we are interested in we follow the notation and conventions presented in \cite{Becker:2011bw}.

\subsection{Manipulating the region of integration}

We construct the n-point correlator as in eqn.\ 3.1 of \cite{Becker:2011ar},
\begin{multline}\label{startingvertexexpr}
\left\langle 0\left|V_1(z_1,\bar{z}_1)V_2(z_2,\bar{z}_2)\left(\prod _{k=3}^n\int _{\mathbb{C}}d^2z_kU_k(z_k,\bar{z}_k)\right)\left(b_0+\tilde{b}_0\right)\right.\right.\\
\times\left.\left.\vphantom{\lp\prod_{k=3}^n\rp}\int_{|w|>\max \left(1\left/\left|z_i\right|\right.\right)}\frac{d^2w}{|w|^2}w^{-L_0}\bar{w}^{-\tilde{L}_0}\right|B\right\rangle,
\end{multline}
where \(U(z,\bar{z})\) is the integrated vertex operator defined as
\begin{equation}
U(z,\bar{z})=\left\{\tilde{b}_{-1},\left[b_{-1},V(z,\bar{z})\right]\right\},
\end{equation}
and \(V(z,\bar{z})\) are closed string vertex operators, whose explicit form is presented later in this section. Also, $|B\rangle$ is a boundary state which encodes the D-brane boundary conditions.  We would like to manipulate this expression to eliminate the explicit factor of $w^{-L_0}\bar{w}^{-\tilde{L}_0}$ and convert the integration region to something easier to work with.

In practice, all vertex operators have the form \(V(z,\bar{z})=V(z) \tilde{V}(\bar{z})\) (or a linear combination of such terms) with \(V=c U+\eta  W\), leading to  \(U(z,\bar{z})=U\tilde{U}\). Now we show the general correlator defined in (\ref{startingvertexexpr}) can be further simplified. We first use
\begin{equation}
w^{L_0}\bar{w}^{\tilde{L}_0}\mathcal{O}(z,\bar{z})w^{-L_0}\bar{w}^{-\tilde{L}_0}=w^h\bar{w}^{\tilde{h}}\mathcal{O}(z w,\bar{z}\bar{w}),
\end{equation}
for any conformal primary operator of weight $(h,\tilde{h})$,
to pull the propagator to the left. Then the correlator can be written as
\begin{multline}
\int _{|w|>\max (1/|z_i|)} d^2w|w|^{2n-6} \left\langle 0\left| V_1(w z_1,\bar{w} \bar{z}_1)V_2(w z_2,\bar{w}
\bar{z}_2)\vphantom{\lp\prod_{k=3}^n\rp}\right.\right.\\
\times\left.\left.\left(\prod _{k=3}^n\int _{\mathbb{C}}d^2z_kU_k(w z_k,\bar{w} \bar{z}_k)\right)\left(b_0+\tilde{b}_0\right)\right|B\right\rangle.
\end{multline}

We can use conformal symmetry to set \(z_1=\infty\), so it will not affect the \(|w|>\max(1/|z_i|)\) condition.  Then we can write \(\prod _{k=3}^n\int _{\mathbb{C}}d^2z_k=\sum_{\alpha=2}^n\int _{V_{\alpha }}\prod _{k=3}^nd^2z_k\), where \(\int _{V_{\alpha }}\) denotes the
region where \(z_{\alpha }=\min(\{z_i\})\). Also use \(\theta _{|w|>\max(1/|z_i|)}\) to denote the Heaviside
function \(\theta(|w|-\max(1/|z_i|))\). Then we can rewrite the integration \(\int _{|w|>\max(1/|z_i|)}
d^2w\) as \(\int _{\mathbb{C}}d^2w \theta _{|w|>\max(1/|z_i|)}\). The correlator becomes
\begin{multline}
\sum_\alpha\int _{\mathbb{C}}d^2w\, \theta _{|w|>1/|z_{\alpha }|} \,\theta _{|z_{\alpha }|=\min(|z_i|)}\,|w|^{2n-6}\left\langle 0\left|\vphantom{\lp\prod_{k=3}^n\rp} V_1(\infty ,\infty )V_2(w z_2,\bar{w} \bar{z}_2)\right.\right.\\
\times\left.\left.
\left(\int _{\mathbb{C}}\prod _{k=3}^nd^2z_kU_k(w z_k,w \bar{z}_k)\right)\left(b_0+\tilde{b}_0\right)\right|B\right\rangle.
\end{multline}
Next we can rescale the coordinates as \(z_2'=w z_2\), \(z'_k=w z_k\),
\begin{multline}
\sum_\alpha\int _{\mathbb{C}}\frac{d^2z_2' }{|z_2'|^2}\,\theta _{|z_{\alpha }'|>1}\,\theta _{|z_{\alpha }'|=\min(|z_i'|)}\,\left\langle 0\left|\vphantom{\lp\prod_{k=3}^n\rp}V_1(\infty ,\infty )V_2(w',\bar{w}')\right.\right.\\
\times\left.\left.\left(\int _{\mathbb{C}}\prod _{k=3}^nd^2z'_kU_k(z'_k,\bar{z}'_k)\right)\left(b_0+\tilde{b}_0\right)\right|B\right\rangle.
\end{multline}

On the other hand, we can conveniently rewrite the Heaviside function as \(\sum _{\alpha }\theta _{|z_{\alpha }'|>1}\theta _{|z_{\alpha }'|=\min(|z_i'|,w')}
=\prod _i\theta _{\left|z_i'\right|>1}\), leading to the expression
\begin{equation}\label{vertexexpr}
\int _{|z_2|>1}\frac{d^2z_2 }{|z_2|^2}\left\langle 0\left| V_1(\infty ,\infty )V_2(z_2,\bar{z}_2)\left(\prod _{k=3}^n\int
_{|z_k|>1}d^2z_kU_k(z_k,\bar{z}_k)\right)\left(b_0+\tilde{b}_0\right)\right|B\right\rangle,
\end{equation}
where we renamed the dummy variable.
This is the formula we use to calculate the three point amplitudes.

We can choose the picture charge for each vertex operator to an arbitrary value, as long as the total picture charge equals
$-2$. The amplitude
is independent on how these charges are precisely distributed. See section 4 of \cite{Becker:2011bw}
for a detailed discussion on these issues. We choose  \((-1/2,-1/2),(-1,0)\) for first two vertex
operators respectively, then \((0,0)\) for the last.

Let us now evaluate the amplitude with one R-R and two NS-NS vertex operators
\begin{equation}
\int _{|z_2|>1}\int _{|z_3|>1}\frac{d^2z_2d^2z_3}{|z_2|^2}\left\langle 0\left| V_{-\hlf,-\hlf}(\infty ,\infty )V_{-1,0}(z_2,\bar{z}_2)U_{0,0}(z_3,\bar{z}_3)\left(b_0+\tilde{b}_0\right)\right|B\right\rangle.
\end{equation}
Vertex operators for different picture charges appear in \cite{Becker:2011bw}. In particular for this case we use
\bea
V_{-\hlf,-\hlf} &=& f_{AB}:c\widetilde{c}e^{-\hlf\phi}S^Ae^{-\hlf\widetilde{\phi}}\widetilde{S}^Be^{ip_1X}:,\non\\
V_{-1,0} &=& \e_{2\,\m\n}:c\widetilde{c}e^{-\phi}\psi^\m\lp\bar{\p}X^\n-ip_2^\rho\widetilde{\psi}_\rho\widetilde{\psi}^\n\rp e^{ip_2X}:\non\\
&& \qquad -\hlf :ce^{-\phi}e^{\widetilde{\phi}}\widetilde{\eta}\psi^\m\widetilde{\psi}^\n e^{ip_2X}:,\\
U_{0,0} &=& \e_{3\,\m\n}:\lp\p X^\m-ip_3^\rho\psi_\rho\psi^\m\rp\lp\bar{\p}X^\n-ip_3^\s\widetilde{\psi}_\s\widetilde{\psi}^\n\rp e^{ip_3X}:.\non
\eea
Here $\e_2$ and $\e_3$ are the polarizations for the two NS-NS states, while
\be
f_{AB}=\lp\mathcal{C}\sum_n\frac{1}{n!}F^{(n)}_{\m_1\cdots\m_n}\G^{\m_1\cdots\m_n}\rp_{AB},
\ee
with $\mathcal{C}_{AB}$ being the antisymmetric charge conjugation matrix for the Clifford algebra of $(\G^\m)^A_{\hph{A}B}$.  In our amplitude we can also drop the second line in $V_{-1,0}$; it is only included in order to ensure that the vertex operator is BRST-invariant, but it doesn't have the right charges to contribute to this amplitude.

We will also use that
\be
\label{eq:GhostSector}
\lim_{z_1\rightarrow\infty}\left\langle 0\left|:c\widetilde{c}e^{-\hlf\phi}S^Ae^{-\hlf\widetilde{\phi}}\widetilde{S}^B(z_1,\bar{z}_1)::c\widetilde{c}e^{-\phi}(z_2,\bar{z}_2):\right.\right.=\left\langle A,B\left|z_2^\hlf\bar{z}_2\right.\right.,
\ee
where $A$ and $B$ are spinor indices and $\langle A,B|$ is the corresponding R-R vacuum.

After plugging in the vertex operators, we can separate the correlator into each sector (boson, fermion,
$bc$, and $\phi$ sectors) and do all possible Wick contractions. The evaluation of each sector appears in appendix \ref{app:Sectors}.

\subsection{The integrand of the amplitudes}
\label{subsec:Integrands}

After evaluating each sector, we see all the integrands can be spanned by the following set of integrals
\begin{equation}
I_{a,b,c,d,e,f}=\int _{|z_i|\leq 1}d^2z_2d^2z_3\widetilde{\mathcal{K}}\mathcal{K},
\end{equation}
where
\begin{equation}
\widetilde{\mathcal{K}}=\left|z_2\right|^{2a}\left|z_3\right|^{2b}\left(1-\left|z_2\right|^2\right)^c\left(1-\left|z_3\right|^2\right)^d\left|z_2-z_3\right|^{2e}\left|1-z_2\bar{z}_3\right|^{2f},
\end{equation}
\begin{multline}
\mathcal{K}=\left|z_2\right|^{2p_1p_2}\left|z_3\right|^{2p_1p_3}\left(1-\left|z_2\right|^2\right)^{p_2Dp_2}\left(1-\left|z_3\right|^2\right)^{p_3Dp_3}\\
\times\left|z_2-z_3\right|^{2p_2p_3}\left|1-z_2\bar{z}_3\right|^{2p_2Dp_3}.
\end{multline}
The matrix $D_{\m\n}$ differs for the directions tangent to the brane, denoted with indices $a$, $b$, $c$, etc., and directions transverse to the brane, with indices $i$, $j$, $k$, etc. Explicitly,
\be
D_{ab}=\eta_{ab},\qquad D_{ai}=D_{ia}=0,\qquad D_{ij}=-\d_{ij}.
\ee

When writing the result in terms of \(I_{a,b,c,d,e,f}\), we still do not see the manifest exchange symmetry under \(2\leftrightarrow 3\). To observe this symmetry, i.e.\ to show amplitudes with its image under \(2\leftrightarrow 3\) exchanged are the same, we need to go to a minimal set of integrals.  For this, we must understand two sets of relations - first the coefficients of the integrals enjoy identities following from conservation of momentum and on-shell conditions, and second there are relations
among the \(I_{a,b,c,d,e,f}\) themselves that follow from integration by parts.

We now derive the second type of conditions. If we write a polar decomposition $z_i=r_ie^{i\phi_i}$, then we note that the integrand $\widetilde{\mathcal{K}}\mathcal{K}$ depends on $r_2$, $r_3$, and the average angle $\hlf(\phi_2+\phi_3)$, but is independent of the relative angle $\phi_2-\phi_3$.
Therefore we expect three relations from integration
by parts.

The integration by parts from \(\int \frac{\partial }{\partial \bar{z}_2}\frac{\partial }{\partial z_2}\widetilde{\mathcal{K}}\mathcal{K}=0\) (i.e.\ from the $r_2$ integration)
is
\begin{align}\label{uniquelization1}
0=& 2\left(p_1p_2+a+1\right)I_{a,b,c,d,e,f}-2\left(p_2Dp_2+c\right)I_{a+1,b,c-1,d,e,f}\non\\
& \quad +\left(p_2p_3+e\right)\left(I_{a,b,c,d,e,f}+I_{a+1,b,c,d,e-1,f}-I_{a,b+1,c,d,e-1,f}\right)\non\\
& \quad +\left(p_2Dp_3+f\right)\left(I_{a,b,c,d,e,f}-I_{a,b,c,d,e,f-1}+I_{a+1,b+1,c,d,e,f-1}\right).
\end{align}
Similarly from \(\int \frac{\partial }{\partial \bar{z}_3}\frac{\partial }{\partial z_3}\widetilde{\mathcal{K}}\mathcal{K}=0\) (the $r_3$ integration) we have
\begin{align}\label{uniquelization2}
0=& 2\left(p_1p_3+b+1\right)I_{a,b,c,d,e,f}-2\left(p_3Dp_3+d\right)I_{a,b+1,c,d-1,e,f}\non\\
& \quad +\left(p_2p_3+e\right)\left(I_{a,b,c,d,e,f}+I_{a,b+1,c,d,e-1,f}-I_{a+1,b,c,d,e-1,f}\right)\non\\
& \quad +\left(p_2Dp_3+f\right)\left(I_{a,b,c,d,e,f}-I_{a,b,c,d,e,f-1}+I_{a+1,b+1,c,d,e,f-1}\right).
\end{align}
Finally we have a relation from \(\int (\frac{\partial }{\partial \bar{z}_3}\frac{\partial }{\partial z_2}+\frac{\partial }{\partial \bar{z}_2}\frac{\partial
}{\partial z_3})\widetilde{\mathcal{K}}\mathcal{K}=0\) (which corresponds to the non-trivial angular integration)\footnote{Here we already used relations (\ref{uniquelization1}) and (\ref{uniquelization2}) to eliminate \(p_1p_2\) and \(p_1p_3\). }
\begin{align}\label{uniquelization3}
0=& \left(p_2Dp_3+f\right) I_{a,b,c,d,e+1,f} \left(-\bar{z}_3 z_2+\bar{z}_2 z_3\right)^2\non\\
& \quad + \left(p_2p_3+e\right) I_{a,b,c,d,e,f+1} \left(-\bar{z}_3
z_2+\bar{z}_2 z_3\right)^2+I_{a,b,c,d,e,f}\mathcal{Z}_{e,f},
\end{align}
where
\begin{align}
\mathcal{Z}_{e,f}=& -\bar{z}_2 \bar{z}_3 z_2^2+2 \bar{z}_3^2 z_2^2+2 \bar{z}_2 \bar{z}_3^2 z_2^3-3 \bar{z}_3^3 z_2^3-\bar{z}_2^2 z_2 z_3-\bar{z}_3^2
z_2 z_3-\bar{z}_2 \bar{z}_3^2 z_2^2 z_3+2 \bar{z}_3^3 z_2^2 z_3\non\\
& \quad -\bar{z}_2^2 \bar{z}_3^2 z_2^3 z_3+2 \bar{z}_2 \bar{z}_3^3 z_2^3 z_3+2 \bar{z}_2^2
z_3^2-\bar{z}_2 \bar{z}_3 z_3^2+2 \bar{z}_2^3 z_2 z_3^2-\bar{z}_2^2 \bar{z}_3 z_2 z_3^2-\bar{z}_2^3 \bar{z}_3 z_2^2 z_3^2\non\\
& \quad -\bar{z}_2 \bar{z}_3^3 z_2^2
z_3^2-3 \bar{z}_2^3 z_3^3+2 \bar{z}_2^2 \bar{z}_3 z_3^3+2 \bar{z}_2^3 \bar{z}_3 z_2 z_3^3-\bar{z}_2^2 \bar{z}_3^2 z_2 z_3^3.
\end{align}

Next we construct a minimal basis for the integrands using the following strategy:
\begin{enumerate}
\item We observe that the amplitudes is real. If the integrand is not real, we know that the imaginary part must vanish upon integration, so we can delete the imaginary part without changing the integration. 
Our integrand can then be written in terms of the integrals \(I_{a,b,c,d,e,f}\).
\item Conservation of momentum in the presence of the brane (which comes from evaluating the zero mode part of the boson sector of the correlator) implies 
\be
0=p_1+Dp_1+p_2+Dp_2+p_3+Dp_3.
\ee
We use this to eliminate \(\left(p_1\right)_a\). Whenever \(p_1\) appears, it should be contracted with only normal indices. For
example we do allow \(p_1Np_2\) where we define \(N=\frac{1-D}{2}\) for contraction of normal indices ($N_{ij}=\d_{ij}$, all other entries zero), but not $p_1p_2$.
\item We use the relations (\ref{uniquelization1}) and (\ref{uniquelization2}) for \(I_{a,b,c,d,e,f}\)
to eliminate \(p_2Dp_2\) and \(p_3Dp_3\), and if any factor of the following form
\begin{equation}
f I_{a,b,c,d,e+1,f} \left(-\bar{z}_3 z_2+\bar{z}_2 z_3\right)^2+ e I_{a,b,c,d,e,f+1} \left(-\bar{z}_3 z_2+\bar{z}_2 z_3\right)^2+I_{a,b,c,d,e,f}\mathcal{Z}_{e,f},
\end{equation}
appears, we use relation (\ref{uniquelization3}) to rewrite it in terms of \(p_2Dp_3\), \(p_2p_3\).
\end{enumerate}

\section{The Amplitudes}
Obtaining the concrete expressions for the amplitudes is rather challenging, so we used the aid of a computer, evaluating the contractions according to the rules in appendix \ref{app:Sectors} and reducing the integrals to a unique form using the procedure explained above.

For $C^{(p-3)}$, we have verified that the result agrees with the computation of~\cite{Becker:2011ar}.  For this case and also for $C^{(p+5)}$ we have confirmed that the result can be written in a manifestly gauge-invariant form (the latter case will be explained in detail in~\cite{paper2}).

Finally, for all cases we have written the result in a way that makes the symmetry under exchange of the two NS-NS fields manifest.  This is a non-trivial check of the results, since the computation treats the two operators on unequal footing (since they are in different pictures and one operator is in integrated form, while the other one is not).

Since the results are long and elaborate, we list them below without further commentary.  Each result can be split into pieces according to the number of indices of the R-R polarization $C^{(n)}_{\m_1\cdots \m_n}$ which are contracted with the world-volume epsilon tensor $\varepsilon_{a_1\cdots a_{p+1}}$.  The remaining indices are contracted with linear combinations of the NS-NS polarizations $\e_{2\,\m\n}$ and $\e_{3\,\m\n}$ and the three momenta $p_1$, $p_2$, and $p_3$.  Finally, each term in this linear combination multiplies a scalar integral $I_k$, $k=0,\cdots,24$ (or $I_k'$, which is obtained from $I_k$ by interchanging $z_2$ with $z_3$ in the integrand) and these combinations are defined in appendix \ref{app:Integrals}.

\subsection{$C^{(p+5)}$ amplitudes}

\be
\mathcal{A}_{C^{(p+5)}BB}=\mathcal{A}_{C^{(p+5)}BB}^{(4)}+\mathcal{A}_{C^{(p+5)}BB}^{(5)}+\mathcal{A}_{C^{(p+5)}BB}^{(6)}.
\ee

\begin{align}
\mathcal{A}_{C^{(p+5)}BB}^{(4)}=& \frac{2 i^{p (p+1)} \sqrt{2}}{(p+1)!}{\varepsilon}^{b_1...b_{p+1}}
   {{C}^{ijkl}}_{b_1...b_{p+1}}\lp 4 {p_2}_{i} {p_3}_{j} {\left(p_1N\epsilon
   _2\right)}_{k} {\left(p_2\epsilon _3\right)}_{l}I_{9}\right.\non\\
   & \qquad\left. +4 {p_2}_{k} {p_3}_{i}
   {\left(p_1N\epsilon _2\right)}_{l} {\left(p_1N\epsilon _3\right)}_{j}I_{10}-4
   {p_2}_{k} {p_3}_{i} {\left(p_1N\epsilon _2\right)}_{l} {\left(p_2D\epsilon
   _3\right)}_{j}I_{5}\right.\non\\
   & \qquad\left. -2 \left(p_1N\epsilon _3p_2\right) {p_2}_{i} {p_3}_{j} {\epsilon
   _2}_{kl}I_{9}-2 \left(p_1N\epsilon _3Dp_2\right) {p_2}_{i} {p_3}_{j} {\epsilon
   _2}_{kl}I_{5}\right.\non\\
   & \qquad\left. -2 \left(p_1Np_2\right) {p_3}_{j} {\left(p_2\epsilon _3\right)}_{i}
   {\epsilon _2}_{kl}I_{9}-2 \left(p_1Np_3\right) {p_2}_{i} {\left(p_2\epsilon
   _3\right)}_{j} {\epsilon _2}_{kl}I_{9}\right.\non\\
   & \qquad\left. +2 \left(p_2p_3\right) {p_3}_{j}
   {\left(p_1N\epsilon _3\right)}_{i} {\epsilon _2}_{kl}I_{9}+4 \left(p_1Np_2\right)
   {p_3}_{j} {\left(p_1N\epsilon _3\right)}_{i} {\epsilon _2}_{kl}I_{10}\right.\non\\
   & \qquad\left. -2
   \left(p_2Dp_3\right) {p_3}_{j} {\left(p_1N\epsilon _3\right)}_{i} {\epsilon
   _2}_{kl}I_{5}+2 \left(p_2p_3\right) {p_2}_{i} {\left(p_1N\epsilon _3\right)}_{j}
   {\epsilon _2}_{kl}I_{9}\right.\non\\
   & \qquad\left. +2 \left(p_2Dp_3\right) {p_2}_{i} {\left(p_1N\epsilon
   _3\right)}_{j} {\epsilon _2}_{kl}I_{5}-2 \left(p_1Np_2\right) {p_3}_{j}
   {\left(p_2D\epsilon _3\right)}_{i} {\epsilon _2}_{kl}I_{5}\right.\non\\
   & \qquad\left. -2 \left(p_1Np_3\right)
   {p_2}_{i} {\left(p_2D\epsilon _3\right)}_{j} {\epsilon
   _2}_{kl}I_{5}+\left(p_1Np_2\right) \left(p_2p_3\right) {\epsilon _2}_{jk} {\epsilon
   _3}_{li}I_{9}\right.\non\\
   & \qquad\left. -\left(p_1Np_2\right) \left(p_1Np_3\right) {\epsilon _2}_{jk} {\epsilon
   _3}_{li}I_{10}+\left(p_1Np_2\right) \left(p_2Dp_3\right) {\epsilon _2}_{jk}
   {\epsilon _3}_{li}I_{5}\rp\non\\
   & \quad +(2\leftrightarrow 3),
\end{align}

\begin{align}
\mathcal{A}_{C^{(p+5)}BB}^{(5)}=& \frac{2 i^{p (p+1)} \sqrt{2}}{p!}{{C}^{ijklm}}_{b_1...b_p} {\varepsilon}^{ab_1...b_p}\lp 2
   {p_2}_{a} {p_2}_{i} {p_3}_{j} {\left(p_2\epsilon _3\right)}_{m} {\epsilon
   _2}_{kl}I_{9}\right.\non\\
   & \qquad\left. +2 {p_2}_{i} {p_3}_{a} {p_3}_{j} {\left(p_2\epsilon _3\right)}_{m}
   {\epsilon _2}_{kl}I_{9}-4 {p_2}_{a} {p_2}_{i} {p_3}_{j} {\left(p_1N\epsilon
   _3\right)}_{m} {\epsilon _2}_{kl}I_{10}\right.\non\\
   & \qquad\left. +2 {p_2}_{a} {p_2}_{i} {p_3}_{j}
   {\left(p_2D\epsilon _3\right)}_{m} {\epsilon _2}_{kl}I_{5}+2 {p_2}_{i} {p_3}_{a}
   {p_3}_{j} {\left(p_2D\epsilon _3\right)}_{m} {\epsilon
   _2}_{kl}I_{5}\right.\non\\
   & \qquad\left. -\left(p_2p_3\right) {p_2}_{a} {p_2}_{i} {\epsilon _2}_{jk} {\epsilon
   _3}_{lm}I_{9}+2 \left(p_1Np_3\right) {p_2}_{a} {p_2}_{i} {\epsilon _2}_{jk}
   {\epsilon _3}_{lm}I_{10}\right.\non\\
   & \qquad\left. -\left(p_2Dp_3\right) {p_2}_{a} {p_2}_{i} {\epsilon _2}_{jk}
   {\epsilon _3}_{lm}I_{5}+\left(p_2p_3\right) {p_2}_{a} {p_3}_{i} {\epsilon _2}_{jk}
   {\epsilon _3}_{lm}I_{9}\right.\non\\
   & \qquad\left. -\left(p_2Dp_3\right) {p_2}_{a} {p_3}_{i} {\epsilon _2}_{jk}
   {\epsilon _3}_{lm}I_{5}\rp+(2\leftrightarrow 3),
\end{align}

\be
\mathcal{A}_{C^{(p+5)}BB}^{(6)}=\frac{2 i^{p (p+1)} \sqrt{2}}{(p-1)!}{{C}^{ijklmn}}_{b_1...b_{p-1}}
   {\varepsilon}^{abb_1...b_{p-1}}{p_2}_{b} {p_2}_{j} {p_3}_{a} {p_3}_{i} {\epsilon
   _2}_{kl} {\epsilon _3}_{mn}I_{10}+(2\leftrightarrow 3).
\ee

\subsection{$C^{(p+3)}$ amplitudes}

\be
\mathcal{A}_{C^{(p+3)}Bh}=\mathcal{A}_{C^{(p+3)}Bh}^{(2)}+\mathcal{A}_{C^{(p+3)}Bh}^{(3)}+\mathcal
{A}_{C^{(p+3)}Bh}^{(4)}+\mathcal{A}_{C^{(p+3)}Bh}^{(5)}.
\ee


\begin{align}
\mathcal{A}_{C^{(p+3)}Bh}^{(2)}=& \frac{4 i^{p (p+1)} \sqrt{2}}{(p-3)!}{{C}^{ij}}_{b_1...b_{p-3}}
   {\varepsilon}^{abcdb_1...b_{p-3}}\left(2 {p_2}_{d} {p_2}_{j} {p_3}_{a} {p_3}_{i}
   {\left(\epsilon _2\epsilon _3\right)}_{bc}I_{9}\right.\non\\
   & \qquad\left. +2 {p_2}_{d} {p_2}_{j} {p_3}_{a}
   {p_3}_{i} {\left(\epsilon _2D\epsilon _3\right)}_{bc}I_{5}+{p_2}_{d} {p_2}_{i}
   {p_3}_{j} {\left(p_2\epsilon _3\right)}_{c} {\epsilon _2}_{ab}I_{9}\right.\non\\
   & \qquad\left. +{p_2}_{i}
   {p_3}_{d} {p_3}_{j} {\left(p_2\epsilon _3\right)}_{c} {\epsilon _2}_{ab}I_{9}-2
   {p_2}_{d} {p_2}_{i} {p_3}_{j} {\left(p_1N\epsilon _3\right)}_{c} {\epsilon
   _2}_{ab}I_{10}\right.\non\\
   & \qquad\left. +{p_2}_{d} {p_2}_{i} {p_3}_{j} {\left(p_2D\epsilon _3\right)}_{c}
   {\epsilon _2}_{ab}I_{5}+{p_2}_{i} {p_3}_{d} {p_3}_{j} {\left(p_2D\epsilon
   _3\right)}_{c} {\epsilon _2}_{ab}I_{5}\right.\non\\
   & \qquad\left. -\frac{1}{2} \text{tr}\left(D.\epsilon
   _3\right) {p_2}_{d} {p_2}_{j} {p_3}_{a} {p_3}_{i} {\epsilon _2}_{bc}I'_{4}+{p_2}_{d}
   {p_2}_{j} {p_3}_{a} {\left(p_2\epsilon _3\right)}_{i} {\epsilon
   _2}_{bc}I_{9}\right.\non\\
   & \qquad\left. +{p_2}_{d} {p_3}_{a} {p_3}_{j} {\left(p_2\epsilon _3\right)}_{i}
   {\epsilon _2}_{bc}I_{9}-2 {p_2}_{d} {p_2}_{j} {p_3}_{a} {\left(p_1N\epsilon
   _3\right)}_{i} {\epsilon _2}_{bc}I_{10}\right.\non\\
   & \qquad\left. +{p_2}_{d} {p_2}_{j} {p_3}_{a}
   {\left(p_2D\epsilon _3\right)}_{i} {\epsilon _2}_{bc}I_{5}+{p_2}_{d} {p_3}_{a}
   {p_3}_{i} {\left(p_2D\epsilon _3\right)}_{j} {\epsilon _2}_{bc}I_{5}\right.\non\\
   & \qquad\left. +{p_2}_{d}
   {p_2}_{j} {p_3}_{a} {\left(p_3D\epsilon _3\right)}_{i} {\epsilon _2}_{bc}I'_{4}+2
   {p_2}_{d} {p_2}_{i} {p_3}_{a} {\left(p_3\epsilon _2\right)}_{c} {\epsilon
   _3}_{bj}I_{9}\right.\non\\
   & \qquad\left. -2 {p_2}_{a} {p_3}_{d} {p_3}_{i} {\left(p_3\epsilon _2\right)}_{c}
   {\epsilon _3}_{bj}I_{9}-2 {p_2}_{d} {p_3}_{a} {p_3}_{i} {\left(p_2D\epsilon
   _2\right)}_{c} {\epsilon _3}_{bj}I_{4}\right.\non\\
   & \qquad\left. +2 {p_2}_{d} {p_2}_{i} {p_3}_{a}
   {\left(p_3D\epsilon _2\right)}_{c} {\epsilon _3}_{bj}I_{5}+2 {p_2}_{a} {p_3}_{d}
   {p_3}_{i} {\left(p_3D\epsilon _2\right)}_{c} {\epsilon _3}_{bj}I_{5}\right.\non\\
   & \qquad\left. +4 {p_2}_{d}
   {p_3}_{b} {p_3}_{i} {\left(p_1N\epsilon _2\right)}_{a} {\epsilon
   _3}_{cj}I_{10}-\left(p_2p_3\right) {p_2}_{d} {p_2}_{i} {\epsilon _2}_{ab} {\epsilon
   _3}_{cj}I_{9}\right.\non\\
   & \qquad\left. +2 \left(p_1Np_3\right) {p_2}_{d} {p_2}_{i} {\epsilon _2}_{ab}
   {\epsilon _3}_{cj}I_{10}-\left(p_2Dp_3\right) {p_2}_{d} {p_2}_{i} {\epsilon _2}_{ab}
   {\epsilon _3}_{cj}I_{5}\right.\non\\
   & \qquad\left. +\left(p_2p_3\right) {p_2}_{d} {p_3}_{i} {\epsilon _2}_{ab}
   {\epsilon _3}_{cj}I_{9}-\left(p_2Dp_3\right) {p_2}_{d} {p_3}_{i} {\epsilon _2}_{ab}
   {\epsilon _3}_{cj}I_{5}\right.\non\\
   & \qquad\left. -\left(p_2p_3\right) {p_2}_{j} {p_3}_{a} {\epsilon _2}_{bc}
   {\epsilon _3}_{di}I_{9}-\left(p_2Dp_3\right) {p_2}_{j} {p_3}_{a} {\epsilon _2}_{bc}
   {\epsilon _3}_{di}I_{5}\right.\non\\
   & \qquad\left. +\left(p_2p_3\right) {p_3}_{a} {p_3}_{j} {\epsilon _2}_{bc}
   {\epsilon _3}_{di}I_{9}+2 \left(p_1Np_2\right) {p_3}_{a} {p_3}_{j} {\epsilon
   _2}_{bc} {\epsilon _3}_{di}I_{10}\right.\non\\
   & \qquad\left. -\left(p_2Dp_3\right) {p_3}_{a} {p_3}_{j} {\epsilon
   _2}_{bc} {\epsilon _3}_{di}I_{5}\right),
\end{align}

\be
\mathcal{A}_{C^{(p+3)}Bh}^{(3)}=\frac{8 i^{p (p+1)} \sqrt{2}}{(p-4)!}{{C}^{ijk}}_{b_1...b_{p-4}}
   {\varepsilon}^{abcdeb_1...b_{p-4}}{p_2}_{e} {p_2}_{j} {p_3}_{a} {p_3}_{i} {\epsilon
   _2}_{bc} {\epsilon _3}_{dk}I_{10},
\ee

\begin{align}
\mathcal{A}_{C^{(p+3)}Bh}^{(4)}=& \frac{2 i^{p (p+1)} \sqrt{2}}{(p-1)!}{{C}^{ijkl}}_{b_1...b_{p-1}}
   {\varepsilon}^{abb_1...b_{p-1}}\lp -4 {p_2}_{b} {p_2}_{j} {p_3}_{a} {p_3}_{i}
   {\left(\epsilon _2\epsilon _3\right)}_{kl}I_{9}\right.\non\\
   & \qquad\left. +4 {p_2}_{b} {p_2}_{j} {p_3}_{a}
   {p_3}_{i} {\left(\epsilon _2D\epsilon _3\right)}_{kl}I_{5}-2 {p_2}_{b} {p_2}_{i}
   {p_3}_{a} {\left(p_2\epsilon _3\right)}_{l} {\epsilon _2}_{jk}I_{9}\right.\non\\
   & \qquad\left. +2 {p_2}_{a}
   {p_3}_{b} {p_3}_{i} {\left(p_2\epsilon _3\right)}_{l} {\epsilon _2}_{jk}I_{9}+4
   {p_2}_{b} {p_2}_{i} {p_3}_{a} {\left(p_1N\epsilon _3\right)}_{l} {\epsilon
   _2}_{jk}I_{10}\right.\non\\
   & \qquad\left. -2 {p_2}_{b} {p_2}_{i} {p_3}_{a} {\left(p_2D\epsilon _3\right)}_{l}
   {\epsilon _2}_{jk}I_{5}+2 {p_2}_{b} {p_3}_{a} {p_3}_{i} {\left(p_2D\epsilon
   _3\right)}_{l} {\epsilon _2}_{jk}I_{5}\right.\non\\
   & \qquad\left. -2 {p_2}_{b} {p_2}_{i} {p_3}_{a}
   {\left(p_3D\epsilon _3\right)}_{l} {\epsilon _2}_{jk}I'_{4}-tr\left(D\epsilon
   _3\right) {p_2}_{b} {p_2}_{j} {p_3}_{a} {p_3}_{i} {\epsilon _2}_{kl}I'_{4}\right.\non\\
   & \qquad\left. -2
   {p_2}_{b} {p_2}_{j} {p_3}_{i} {\left(p_2\epsilon _3\right)}_{a} {\epsilon
   _2}_{kl}I_{9}-2 {p_2}_{j} {p_3}_{b} {p_3}_{i} {\left(p_2\epsilon _3\right)}_{a}
   {\epsilon _2}_{kl}I_{9}\right.\non\\
   & \qquad\left. +4 {p_2}_{b} {p_2}_{j} {p_3}_{i} {\left(p_1N\epsilon
   _3\right)}_{a} {\epsilon _2}_{kl}I_{10}-2 {p_2}_{b} {p_2}_{j} {p_3}_{i}
   {\left(p_2D\epsilon _3\right)}_{a} {\epsilon _2}_{kl}I_{5}\right.\non\\
   & \qquad\left. -2 {p_2}_{j} {p_3}_{b}
   {p_3}_{i} {\left(p_2D\epsilon _3\right)}_{a} {\epsilon _2}_{kl}I_{5}-4 {p_2}_{b}
   {p_2}_{i} {p_3}_{j} {\left(p_3\epsilon _2\right)}_{l} {\epsilon _3}_{ak}I_{9}\right.\non\\
   & \qquad\left. -4
   {p_2}_{i} {p_3}_{b} {p_3}_{j} {\left(p_3\epsilon _2\right)}_{l} {\epsilon
   _3}_{ak}I_{9}-4 {p_2}_{b} {p_2}_{j} {p_3}_{i} {\left(p_3D\epsilon _2\right)}_{l}
   {\epsilon _3}_{ak}I_{5}\right.\non\\
   & \qquad\left. -4 {p_2}_{j} {p_3}_{b} {p_3}_{i} {\left(p_3D\epsilon
   _2\right)}_{l} {\epsilon _3}_{ak}I_{5}-2 \left(p_2p_3\right) {p_2}_{b} {p_2}_{i}
   {\epsilon _2}_{jk} {\epsilon _3}_{al}I_{9}\right.\non\\
   & \qquad\left. +4 \left(p_1Np_3\right) {p_2}_{b}
   {p_2}_{i} {\epsilon _2}_{jk} {\epsilon _3}_{al}I_{10}-2 \left(p_2Dp_3\right)
   {p_2}_{b} {p_2}_{i} {\epsilon _2}_{jk} {\epsilon _3}_{al}I_{5}\right.\non\\
   & \qquad\left. +2 \left(p_2p_3\right)
   {p_2}_{b} {p_3}_{i} {\epsilon _2}_{jk} {\epsilon _3}_{al}I_{9}-2
   \left(p_2Dp_3\right) {p_2}_{b} {p_3}_{i} {\epsilon _2}_{jk} {\epsilon
   _3}_{al}I_{5}\right.\non\\
   & \qquad\left. -2 \left(p_2p_3\right) {p_2}_{j} {p_3}_{a} {\epsilon _2}_{kl} {\epsilon
   _3}_{bi}I_{9}-2 \left(p_2Dp_3\right) {p_2}_{j} {p_3}_{a} {\epsilon _2}_{kl}
   {\epsilon _3}_{bi}I_{5}\right.\non\\
   & \qquad\left. +2 \left(p_2p_3\right) {p_3}_{a} {p_3}_{j} {\epsilon _2}_{kl}
   {\epsilon _3}_{bi}I_{9}+4 \left(p_1Np_2\right) {p_3}_{a} {p_3}_{j} {\epsilon
   _2}_{kl} {\epsilon _3}_{bi}I_{10}\right.\non\\
   & \qquad\left. -2 \left(p_2Dp_3\right) {p_3}_{a} {p_3}_{j}
   {\epsilon _2}_{kl} {\epsilon _3}_{bi}I_{5}-8 {p_2}_{j} {p_3}_{a} {p_3}_{i}
   {\left(p_1N\epsilon _2\right)}_{l} {\epsilon _3}_{bk}I_{10}\rp,
\end{align}

\be
\mathcal{A}_{C^{(p+3)}Bh}^{(5)}=\frac{8 i^{p (p+1)} \sqrt{2}}{(p-2)!}{{C}^{ijklm}}_{b_1...b_{p-2}}
   {\varepsilon}^{abcb_1...b_{p-2}}{p_2}_{c} {p_2}_{j} {p_3}_{a} {p_3}_{i} {\epsilon
   _2}_{kl} {\epsilon _3}_{bm}I_{10}.
\ee

\subsection{$C^{(p+1)}$ amplitudes}


\be
\mathcal{A}_{C^{(p+1)}BB}=\mathcal{A}_{C^{(p+1)}BB}^{(0)}+\mathcal{A}_{C^{(p+1)}BB}^{(1)}+\mathcal
{A}_{C^{(p+1)}BB}^{(2)}+\mathcal{A}_{C^{(p+1)}BB}^{(3)}+\mathcal{A}_{C^{(p+1)}BB}^{(4)}.
\ee


\begin{align}
\mathcal{A}_{C^{(p+1)}BB}^{(0)}=& \frac{2 i^{p (p+1)} \sqrt{2}}{(p+1)!}{C}_{b_1...b_{p+1}}
   {\varepsilon}^{b_1...b_{p+1}}\lp 2 \left(p_1N\epsilon _2p_3\right) \left(p_1N\epsilon
   _3p_2\right)I_{2}\right.\non\\
   & \qquad\left. -2 \left(p_1N\epsilon _3p_2\right) \left(p_2D\epsilon
   _2p_3\right)I_{20}-2 \left(p_1N\epsilon _2p_3\right) \left(p_2D\epsilon
   _3p_2\right)I_{14}\right.\non\\
   & \qquad\left. -16 \left(p_1N\epsilon _2Dp_3\right) \left(p_1N\epsilon
   _3p_2\right)I_{0}+2 \left(p_1N\epsilon _2Dp_3\right) \left(p_2D\epsilon
   _3p_2\right)I_{15}\right.\non\\
   & \qquad\left. -2 \left(p_1N\epsilon _2Dp_3\right) \left(p_2\epsilon
   _3Dp_3\right)I_{22}+4 \left(p_1N\epsilon _2\epsilon _3p_2\right)I_{23}\right.\non\\
   & \qquad\left. -2
   \left(p_1N\epsilon _2\epsilon _3p_2\right) \left(p_2p_3\right)I_{16}+2
   \left(p_1N\epsilon _2Dp_3\right) \left(p_1N\epsilon _3Dp_2\right)I_{1}\right.\non\\
   & \qquad\left. -2
   \left(p_1N\epsilon _3Dp_2\right) \left(p_2D\epsilon _2Dp_3\right)I_{21}+2
   \left(p_1Np_3\right) \left(p_2D\epsilon _2\epsilon _3p_2\right)I_{20}\right.\non\\
   & \qquad\left. -2
   \left(p_1N\epsilon _2p_3\right) \left(p_2D\epsilon _3Dp_3\right)I_{22}-2
   \left(p_1Np_3\right) \left(p_2D\epsilon _3\epsilon _2p_3\right)I_{14}\right.\non\\
   & \qquad\left. -2
   \left(p_1Np_2\right) \left(p_2\epsilon _3D\epsilon _2p_3\right)I_{14}+2
   \left(p_1N\epsilon _2D\epsilon _3p_2\right) \left(p_2p_3\right)I_{14}\right.\non\\
   & \qquad\left. +2
   \left(p_1N\epsilon _2\epsilon _3Dp_2\right) \left(p_2Dp_3\right)I_{15}-2
   \left(p_1N\epsilon _2\epsilon _3Dp_3\right) \left(p_2p_3\right)I'_{20}\right.\non\\
   & \qquad\left. +2
   \left(p_1N\epsilon _2\epsilon _3Dp_3\right) \left(p_2Dp_3\right)I_{22}+2
   \left(p_1N\epsilon _2\epsilon _3Np_1\right) \left(p_2p_3\right)I_{2}\right.\non\\
   & \qquad\left. -8
   \left(p_1N\epsilon _2\epsilon _3Np_1\right) \left(p_2Dp_3\right)I_{0}-4
   \left(p_1N\epsilon _3D\epsilon _2p_3\right) \left(p_1Np_2\right)I'_{11}\right.\non\\
   & \qquad\left. -4
   \left(p_1N\epsilon _3\epsilon _2Dp_3\right) \left(p_1Np_2\right)I'_{11}-2
   \left(p_1Np_2\right) \left(p_2D\epsilon _3D\epsilon _2p_3\right)I_{15}\right.\non\\
   & \qquad\left. -2
   \left(p_1Np_2\right) \left(p_2D\epsilon _3\epsilon _2Dp_3\right)I_{15}-2
   \left(p_1Np_2\right) \left(p_3D\epsilon _2\epsilon _3Dp_3\right)I_{22}\right.\non\\
   & \qquad\left. -2
   \left(p_1Np_2\right) \left(p_3D\epsilon _3D\epsilon _2p_3\right)I_{22}+4
   \left(p_1N\epsilon _2D\epsilon _3Dp_2\right)I_{24}\right.\non\\
   & \qquad\left. -2 \left(p_1N\epsilon _2D\epsilon
   _3Dp_2\right) \left(p_2Dp_3\right)I_{17}+2 \left(p_1N\epsilon _2D\epsilon
   _3Dp_3\right) \left(p_2p_3\right)I_{22}\right.\non\\
   & \qquad\left. -2 \left(p_1N\epsilon _2D\epsilon
   _3Dp_3\right) \left(p_2Dp_3\right)I'_{21}-8 \left(p_1N\epsilon _2D\epsilon
   _3Np_1\right) \left(p_2p_3\right)I_{0}\right.\non\\
   & \qquad\left. +2 \left(p_1N\epsilon _2D\epsilon
   _3Np_1\right) \left(p_2Dp_3\right)I_{1}+2 \left(p_1Np_3\right) \left(p_2D\epsilon
   _2D\epsilon _3Dp_2\right)I_{21}\right.\non\\
   & \qquad\left. +2 \tr(-\epsilon _2\epsilon _3)
   \left(p_1Np_2\right)I_{23}-\tr(-\epsilon _2\epsilon _3)
   \left(p_1Np_2\right) \left(p_2p_3\right)I_{16}\right.\non\\
   & \qquad\left. +\tr(-\epsilon _2\epsilon
   _3) \left(p_1Np_2\right) \left(p_1Np_3\right)I_{2}-\tr(-\epsilon
   _2\epsilon _3) \left(p_1Np_2\right) \left(p_2Dp_3\right)I_{14}\right.\non\\
   & \qquad\left. -2
   \tr(D\epsilon _2D\epsilon _3) \left(p_1Np_2\right)I_{24}+\tr(D\epsilon
   _2D\epsilon _3) \left(p_1Np_2\right)
   \left(p_2p_3\right)I_{15}\right.\non\\
   & \qquad\left. -\tr(D\epsilon _2D\epsilon _3)
   \left(p_1Np_2\right) \left(p_1Np_3\right)I_{1}\right.\non\\
   & \qquad\left. +\tr(D\epsilon _2D\epsilon
   _3) \left(p_1Np_2\right) \left(p_2Dp_3\right)I_{17}\rp+(2\leftrightarrow 3),
\end{align}



\begin{align}
\mathcal{A}_{C^{(p+1)}BB}^{(1)}=& \frac{2 i^{p (p+1)} \sqrt{2}}{p!}{{C}^{i}}_{b_1...b_p} {\varepsilon}^{ab_1...b_p}\lp -2
   \left(p_2\epsilon _3D\epsilon _2p_3\right) {p_2}_{a} {p_2}_{i}I_{14}\right.\non\\
   & \qquad\left. -2
   \left(p_2\epsilon _3\epsilon _2Dp_3\right) {p_2}_{a} {p_2}_{i}I_{14}+2
   \left(p_3D\epsilon _3\epsilon _2p_3\right) {p_2}_{a} {p_2}_{i}I'_{20}\right.\non\\
   & \qquad\left. -4
   \left(p_1N\epsilon _3D\epsilon _2p_3\right) {p_2}_{a} {p_2}_{i}I'_{11}-4
   \left(p_1N\epsilon _3\epsilon _2Dp_3\right) {p_2}_{a} {p_2}_{i}I'_{11}\right.\non\\
   & \qquad\left. -2
   \left(p_2D\epsilon _3D\epsilon _2p_3\right) {p_2}_{a} {p_2}_{i}I_{15}-2
   \left(p_2D\epsilon _3\epsilon _2Dp_3\right) {p_2}_{a} {p_2}_{i}I_{15}\right.\non\\
   & \qquad\left. -2
   \left(p_3D\epsilon _2\epsilon _3Dp_3\right) {p_2}_{a} {p_2}_{i}I_{22}-2
   \left(p_3D\epsilon _3D\epsilon _2p_3\right) {p_2}_{a} {p_2}_{i}I_{22}\right.\non\\
   & \qquad\left. +2
   \left(p_3D\epsilon _2D\epsilon _3Dp_3\right) {p_2}_{a} {p_2}_{i}I'_{21}+2
   \tr(-\epsilon _2\epsilon _3) {p_2}_{a} {p_2}_{i}I_{23}\right.\non\\
   & \qquad\left. -\tr(-\epsilon
   _2\epsilon _3) \left(p_2p_3\right) {p_2}_{a} {p_2}_{i}I_{16}+2
   \tr(-\epsilon _2\epsilon _3) \left(p_1Np_3\right) {p_2}_{a}
   {p_2}_{i}I_{2}
   \right.\non
\end{align}
\begin{align}
   & \qquad\left.
   -\tr(-\epsilon _2\epsilon _3) \left(p_2Dp_3\right) {p_2}_{a}
   {p_2}_{i}I_{14}-2 \tr(D\epsilon _2D\epsilon _3) {p_2}_{a}
   {p_2}_{i}I_{24}\right.\non\\
   & \qquad\left. +\tr(D\epsilon _2D\epsilon _3) \left(p_2p_3\right)
   {p_2}_{a} {p_2}_{i}I_{15}-2 \tr(D\epsilon _2D\epsilon _3)
   \left(p_1Np_3\right) {p_2}_{a} {p_2}_{i}I_{1}\right.\non\\
   & \qquad\left. +\tr(D\epsilon _2D\epsilon
   _3) \left(p_2Dp_3\right) {p_2}_{a} {p_2}_{i}I_{17}+2 \left(p_1N\epsilon
   _2\epsilon _3p_2\right) {p_2}_{i} {p_3}_{a}I_{2}\right.\non\\
   & \qquad\left. -2 \left(p_1N\epsilon _3\epsilon
   _2p_3\right) {p_2}_{i} {p_3}_{a}I_{2}-2 \left(p_2\epsilon _3D\epsilon _2p_3\right)
   {p_2}_{i} {p_3}_{a}I_{14}-2 \left(p_2\epsilon _3\epsilon _2Dp_3\right) {p_2}_{i}
   {p_3}_{a}I_{14}\right.\non\\
   & \qquad\left. +2 \left(p_3D\epsilon _3\epsilon _2p_3\right) {p_2}_{i}
   {p_3}_{a}I'_{20}+4 \left(p_1N\epsilon _2\epsilon _3Dp_3\right) {p_2}_{i}
   {p_3}_{a}I'_{6}\right.\non\\
   & \qquad\left. -4 \left(p_1N\epsilon _2\epsilon _3Np_1\right) {p_2}_{i}
   {p_3}_{a}I_{9}-2 \left(p_1N\epsilon _3D\epsilon _2p_3\right) {p_2}_{i}
   {p_3}_{a}I'_{11}\right.\non\\
   & \qquad\left. -2 \left(p_1N\epsilon _3\epsilon _2Dp_3\right) {p_2}_{i}
   {p_3}_{a}I'_{11}-2 \left(p_2D\epsilon _3D\epsilon _2p_3\right) {p_2}_{i}
   {p_3}_{a}I_{15}\right.\non\\
   & \qquad\left. -2 \left(p_2D\epsilon _3\epsilon _2Dp_3\right) {p_2}_{i}
   {p_3}_{a}I_{15}-2 \left(p_3D\epsilon _2\epsilon _3Dp_3\right) {p_2}_{i}
   {p_3}_{a}I_{22}\right.\non\\
   & \qquad\left. -2 \left(p_3D\epsilon _3D\epsilon _2p_3\right) {p_2}_{i}
   {p_3}_{a}I_{22}-2 \left(p_1N\epsilon _2D\epsilon _3Dp_2\right) {p_2}_{i}
   {p_3}_{a}I_{1}\right.\non\\
   & \qquad\left. +4 \left(p_1N\epsilon _2D\epsilon _3Dp_3\right) {p_2}_{i}
   {p_3}_{a}I'_{7}+4 \left(p_1N\epsilon _2D\epsilon _3Np_1\right) {p_2}_{i}
   {p_3}_{a}I_{5}\right.\non\\
   & \qquad\left. -2 \left(p_1N\epsilon _3D\epsilon _2Dp_3\right) {p_2}_{i}
   {p_3}_{a}I_{1}+2 \left(p_3D\epsilon _2D\epsilon _3Dp_3\right) {p_2}_{i}
   {p_3}_{a}I'_{21}\right.\non\\
   & \qquad\left. +2 \tr(-\epsilon _2\epsilon _3) {p_2}_{i}
   {p_3}_{a}I_{23}-\tr(-\epsilon _2\epsilon _3) \left(p_2p_3\right) {p_2}_{i}
   {p_3}_{a}I_{16}\right.\non\\
   & \qquad\left. -\tr(-\epsilon _2\epsilon _3) \left(p_2Dp_3\right)
   {p_2}_{i} {p_3}_{a}I_{14}-2 \tr(D\epsilon _2D\epsilon _3) {p_2}_{i}
   {p_3}_{a}I_{24}\right.\non\\
   & \qquad\left. +\tr(D\epsilon _2D\epsilon _3) \left(p_2p_3\right)
   {p_2}_{i} {p_3}_{a}I_{15}+\tr(D\epsilon _2D\epsilon _3)
   \left(p_2Dp_3\right) {p_2}_{i} {p_3}_{a}I_{17}\right.\non\\
   & \qquad\left.
   -2 \left(p_1N\epsilon _3D\epsilon
   _2p_3\right) {p_2}_{a} {p_3}_{i}I'_{11}+2 \left(p_1N\epsilon _3\epsilon
   _2Dp_3\right) {p_2}_{a} {p_3}_{i}I'_{11}\right.\non\\
   & \qquad\left. +2 \left(p_1N\epsilon _2p_3\right) {p_2}_{i}
   {\left(p_2\epsilon _3\right)}_{a}I_{2}+2 \left(p_1N\epsilon _2p_3\right) {p_3}_{i}
   {\left(p_2\epsilon _3\right)}_{a}I_{2}\right.\non\\
   & \qquad\left. +2 \left(p_1N\epsilon _2Dp_3\right) {p_3}_{i}
   {\left(p_2\epsilon _3\right)}_{a}I'_{11}-2 \left(p_1N\epsilon _2p_3\right) {p_2}_{a}
   {\left(p_2\epsilon _3\right)}_{i}I_{2}\right.\non\\
   & \qquad\left. +2 \left(p_2D\epsilon _2p_3\right) {p_2}_{a}
   {\left(p_2\epsilon _3\right)}_{i}I_{20}+2 \left(p_3D\epsilon _2p_3\right) {p_2}_{a}
   {\left(p_2\epsilon _3\right)}_{i}I_{14}\right.\non\\
   & \qquad\left. -2 \left(p_1N\epsilon _2p_3\right) {p_3}_{a}
   {\left(p_2\epsilon _3\right)}_{i}I_{2}+2 \left(p_2D\epsilon _2p_3\right) {p_3}_{a}
   {\left(p_2\epsilon _3\right)}_{i}I_{20}\right.\non\\
   & \qquad\left. +2 \left(p_3D\epsilon _2p_3\right) {p_3}_{a}
   {\left(p_2\epsilon _3\right)}_{i}I_{14}-2 \left(p_1N\epsilon _3Dp_2\right) {p_3}_{i}
   {\left(p_3\epsilon _2\right)}_{a}I'_{11}\right.\non\\
   & \qquad\left. +2 \left(p_1Np_2\right) {\left(p_2\epsilon
   _3\right)}_{i} {\left(p_3\epsilon _2\right)}_{a}I_{2}+2 \left(p_1Np_3\right)
   {\left(p_2\epsilon _3\right)}_{i} {\left(p_3\epsilon _2\right)}_{a}I_{2}\right.\non\\
   & \qquad\left. -2
   \left(p_1N\epsilon _3Dp_2\right) {p_2}_{a} {\left(p_3\epsilon
   _2\right)}_{i}I'_{12}+2 \left(p_2D\epsilon _3Dp_3\right) {p_2}_{a}
   {\left(p_3\epsilon _2\right)}_{i}I_{22}\right.\non\\
   & \qquad\left. -2 \left(p_1N\epsilon _3Dp_2\right) {p_3}_{a}
   {\left(p_3\epsilon _2\right)}_{i}I'_{11}+2 \left(p_2D\epsilon _3Dp_3\right)
   {p_3}_{a} {\left(p_3\epsilon _2\right)}_{i}I_{22}\right.\non\\
   & \qquad\left. +2 \left(p_1Np_2\right)
   \left(p_2p_3\right) {\left(\epsilon _2\epsilon _3\right)}_{ai}I_{2}
-2\left(p_1Np_3\right) \left(p_2p_3\right) {\left(\epsilon _2\epsilon
   _3\right)}_{ai}I_{2}\right.\non\\
   & \qquad\left. -4 \left(p_1Np_2\right) \left(p_1Np_3\right) {\left(\epsilon
   _2\epsilon _3\right)}_{ai}I_{9}-2 \left(p_1Np_3\right) \left(p_2Dp_3\right)
   {\left(\epsilon _2\epsilon _3\right)}_{ai}I'_{11}\right.\non\\
   & \qquad\left. +2 \left(p_1Np_3\right)
   \left(p_2Dp_3\right) {\left(\epsilon _2\epsilon _3\right)}_{ia}I'_{11}+4
   \left(p_1N\epsilon _3p_2\right) {p_3}_{i} {\left(p_1N\epsilon _2\right)}_{a}I_{9}\right.\non\\
   & \qquad\left. +4
   \left(p_1N\epsilon _3Dp_2\right) {p_3}_{i} {\left(p_1N\epsilon _2\right)}_{a}I_{5}+4
   \left(p_1Np_3\right) {\left(p_1N\epsilon _2\right)}_{a} {\left(p_2\epsilon
   _3\right)}_{i}I_{9}\right.\non\\
   & \qquad\left. -4 \left(p_1N\epsilon _3p_2\right) {p_3}_{a} {\left(p_1N\epsilon
   _2\right)}_{i}I_{9}-4 \left(p_2\epsilon _3Dp_3\right) {p_3}_{a} {\left(p_1N\epsilon
   _2\right)}_{i}I'_{6}\right.\non\\
   & \qquad\left. +4 \left(p_1N\epsilon _3Dp_2\right) {p_3}_{a}
   {\left(p_1N\epsilon _2\right)}_{i}I_{5}-4 \left(p_2D\epsilon _3Dp_3\right) {p_3}_{a}
   {\left(p_1N\epsilon _2\right)}_{i}I'_{7}\right.\non\\
   & \qquad\left. -4 \left(p_2p_3\right) {\left(p_1N\epsilon
   _2\right)}_{i} {\left(p_2\epsilon _3\right)}_{a}I_{2}+4 \left(p_1Np_3\right)
   {\left(p_1N\epsilon _2\right)}_{i} {\left(p_2\epsilon _3\right)}_{a}I_{9}\right.\non\\
   & \qquad\left. +4
   \left(p_2p_3\right) {\left(p_1N\epsilon _2\right)}_{i} {\left(p_1N\epsilon
   _3\right)}_{a}I_{9}-4 \left(p_2Dp_3\right) {\left(p_1N\epsilon _2\right)}_{i}
   {\left(p_1N\epsilon _3\right)}_{a}I_{5}\right.\non\\
   & \qquad\left. +4 \left(p_3D\epsilon _2p_3\right) {p_2}_{a}
   {\left(p_1N\epsilon _3\right)}_{i}I'_{11}+4 \left(p_1N\epsilon _3p_2\right)
   {p_3}_{i} {\left(p_2D\epsilon _2\right)}_{a}I_{6}\right.\non\\
    &\qquad\left.
   +4 \left(p_1N\epsilon _3Dp_2\right)
   {p_3}_{i} {\left(p_2D\epsilon _2\right)}_{a}I_{7}+4 \left(p_1Np_3\right)
   {\left(p_2D\epsilon _2\right)}_{a} {\left(p_2\epsilon _3\right)}_{i}I_{6}\right.\non
\end{align}
\begin{align}
   & \qquad\left. -4
   \left(p_2p_3\right) {\left(p_1N\epsilon _3\right)}_{i} {\left(p_2D\epsilon
   _2\right)}_{a}I_{6}-4 \left(p_2Dp_3\right) {\left(p_1N\epsilon _3\right)}_{i}
   {\left(p_2D\epsilon _2\right)}_{a}I_{7}\right.\non\\
   & \qquad\left. -2 \left(p_1N\epsilon _2Dp_3\right) {p_2}_{i}
   {\left(p_2D\epsilon _3\right)}_{a}I_{1}+2 \left(p_1N\epsilon _2p_3\right) {p_3}_{i}
   {\left(p_2D\epsilon _3\right)}_{a}I'_{11}\right.\non\\
   & \qquad\left. +2 \left(p_1N\epsilon _2Dp_3\right)
   {p_3}_{i} {\left(p_2D\epsilon _3\right)}_{a}I_{1}+2 \left(p_1Np_3\right)
   {\left(p_2D\epsilon _3\right)}_{a} {\left(p_3\epsilon _2\right)}_{i}I'_{11}\right.\non\\
   & \qquad\left. -4
   \left(p_1Np_3\right) {\left(p_1N\epsilon _2\right)}_{i} {\left(p_2D\epsilon
   _3\right)}_{a}I_{5}+4 \left(p_2Dp_3\right) {\left(p_1N\epsilon _2\right)}_{i}
   {\left(p_2D\epsilon _3\right)}_{a}I_{1}\right.\non\\
   & \qquad\left. +2 \left(p_3D\epsilon _2p_3\right) {p_2}_{a}
   {\left(p_2D\epsilon _3\right)}_{i}I_{15}-2 \left(p_1N\epsilon _2Dp_3\right)
   {p_2}_{a} {\left(p_2D\epsilon _3\right)}_{i}I_{1}\right.\non\\
   & \qquad\left. +2 \left(p_2D\epsilon _2Dp_3\right)
   {p_2}_{a} {\left(p_2D\epsilon _3\right)}_{i}I_{21}+2 \left(p_3D\epsilon _2p_3\right)
   {p_3}_{a} {\left(p_2D\epsilon _3\right)}_{i}I_{15}\right.\non\\
   & \qquad\left. -2 \left(p_1N\epsilon
   _2Dp_3\right) {p_3}_{a} {\left(p_2D\epsilon _3\right)}_{i}I_{1}+2 \left(p_2D\epsilon
   _2Dp_3\right) {p_3}_{a} {\left(p_2D\epsilon _3\right)}_{i}I_{21}\right.\non\\
   & \qquad\left. -2
   \left(p_1Np_3\right) {\left(p_2D\epsilon _3\right)}_{i} {\left(p_3\epsilon
   _2\right)}_{a}I'_{11}+4 \left(p_1Np_3\right) {\left(p_1N\epsilon _2\right)}_{a}
   {\left(p_2D\epsilon _3\right)}_{i}I_{5}\right.\non\\
   & \qquad\left. +4 \left(p_1Np_3\right) {\left(p_2D\epsilon
   _2\right)}_{a} {\left(p_2D\epsilon _3\right)}_{i}I_{7}+2 \left(p_1Np_3\right)
   {p_2}_{i} {\left(p_2\epsilon _3\epsilon _2\right)}_{a}I_{2}\right.\non\\
   & \qquad\left. -2 \left(p_1Np_2\right)
   {p_3}_{i} {\left(p_2\epsilon _3\epsilon _2\right)}_{a}I_{2}+4 {p_2}_{a}
   {\left(p_2\epsilon _3\epsilon _2\right)}_{i}I_{23}-2 \left(p_2p_3\right) {p_2}_{a}
   {\left(p_2\epsilon _3\epsilon _2\right)}_{i}I_{16}\right.\non\\
   & \qquad\left. +2 \left(p_1Np_3\right) {p_2}_{a}
   {\left(p_2\epsilon _3\epsilon _2\right)}_{i}I_{2}+4 {p_3}_{a} {\left(p_2\epsilon
   _3\epsilon _2\right)}_{i}I_{23}-2 \left(p_2p_3\right) {p_3}_{a} {\left(p_2\epsilon
   _3\epsilon _2\right)}_{i}I_{16}\right.\non\\
   & \qquad\left. -2 \left(p_1Np_2\right) {p_3}_{a} {\left(p_2\epsilon
   _3\epsilon _2\right)}_{i}I_{2}+2 \left(p_1N\epsilon _3p_2\right) {p_3}_{i}
   {\left(p_3D\epsilon _2\right)}_{a}I'_{11}\right.\non\\
   & \qquad\left. +2 \left(p_1Np_3\right) {\left(p_2\epsilon
   _3\right)}_{i} {\left(p_3D\epsilon _2\right)}_{a}I'_{11}+2 \left(p_1Np_2\right)
   {\left(p_2D\epsilon _3\right)}_{i} {\left(p_3D\epsilon _2\right)}_{a}I_{1}\right.\non\\
   & \qquad\left. -2
   \left(p_1Np_3\right) {\left(p_2D\epsilon _3\right)}_{i} {\left(p_3D\epsilon
   _2\right)}_{a}I_{1}
   -2 \left(p_1N\epsilon _3p_2\right) {p_2}_{a} {\left(p_3D\epsilon
   _2\right)}_{i}I'_{12}\right.\non\\
   & \qquad\left. +2 \left(p_2\epsilon _3Dp_3\right) {p_2}_{a}
   {\left(p_3D\epsilon _2\right)}_{i}I_{22}-2 \left(p_1N\epsilon _3p_2\right) {p_3}_{a}
   {\left(p_3D\epsilon _2\right)}_{i}I'_{11}\right.\non\\
   & \qquad\left. +2 \left(p_2\epsilon _3Dp_3\right)
   {p_3}_{a} {\left(p_3D\epsilon _2\right)}_{i}I_{22}+2 \left(p_1Np_3\right)
   {\left(p_2\epsilon _3\right)}_{a} {\left(p_3D\epsilon _2\right)}_{i}I'_{11}\right.\non\\
   & \qquad\left. +2
   \left(p_1Np_3\right) \left(p_2p_3\right) {\left(\epsilon _2D\epsilon
   _3\right)}_{ai}I'_{11}-4 \left(p_1Np_2\right) \left(p_1Np_3\right) {\left(\epsilon
   _2D\epsilon _3\right)}_{ai}I_{5}\right.\non\\
   & \qquad\left. +2 \left(p_1Np_2\right) \left(p_2Dp_3\right)
   {\left(\epsilon _2D\epsilon _3\right)}_{ai}I_{1}+2 \left(p_1Np_3\right)
   \left(p_2Dp_3\right) {\left(\epsilon _2D\epsilon _3\right)}_{ai}I_{1}\right.\non\\
   & \qquad\left. +2
   \left(p_1Np_3\right) \left(p_2p_3\right) {\left(\epsilon _2D\epsilon
   _3\right)}_{ia}I'_{11}+2 \left(p_2p_3\right) {p_2}_{i} {\left(p_1N\epsilon
   _2\epsilon _3\right)}_{a}I_{2}\right.\non\\
   & \qquad\left. -4 \left(p_1Np_3\right) {p_2}_{i} {\left(p_1N\epsilon
   _2\epsilon _3\right)}_{a}I_{9}-2 \left(p_2p_3\right) {p_3}_{i} {\left(p_1N\epsilon
   _2\epsilon _3\right)}_{a}I_{2}\right.\non\\
   & \qquad\left. -2 \left(p_2Dp_3\right) {p_3}_{i} {\left(p_1N\epsilon
   _2\epsilon _3\right)}_{a}I'_{11}+2 \left(p_2p_3\right) {p_2}_{a} {\left(p_1N\epsilon
   _2\epsilon _3\right)}_{i}I_{2}\right.\non\\
   & \qquad\left. -4 \left(p_1Np_3\right) {p_2}_{a} {\left(p_1N\epsilon
   _2\epsilon _3\right)}_{i}I_{9}+2 \left(p_2p_3\right) {p_3}_{a} {\left(p_1N\epsilon
   _2\epsilon _3\right)}_{i}I_{2}\right.\non\\
   & \qquad\left. +2 \left(p_2Dp_3\right) {p_3}_{i} {\left(p_1N\epsilon
   _3\epsilon _2\right)}_{a}I'_{11}+2 \left(p_2Dp_3\right) {p_2}_{a}
   {\left(p_1N\epsilon _3\epsilon _2\right)}_{i}I'_{12}\right.\non\\
   & \qquad\left. +2 \left(p_2Dp_3\right)
   {p_3}_{a} {\left(p_1N\epsilon _3\epsilon _2\right)}_{i}I'_{11}-2 \left(p_2p_3\right)
   {p_2}_{a} {\left(p_2D\epsilon _2\epsilon _3\right)}_{i}I_{20}\right.\non\\
   & \qquad\left. -4 \left(p_1Np_3\right)
   {p_2}_{a} {\left(p_2D\epsilon _2\epsilon _3\right)}_{i}I_{6}-2 \left(p_2p_3\right)
   {p_3}_{a} {\left(p_2D\epsilon _2\epsilon _3\right)}_{i}I_{20}\right.\non\\
   & \qquad\left. +2 \left(p_2Dp_3\right)
   {p_2}_{a} {\left(p_2D\epsilon _3\epsilon _2\right)}_{i}I_{15}+2 \left(p_2Dp_3\right)
   {p_3}_{a} {\left(p_2D\epsilon _3\epsilon _2\right)}_{i}I_{15}\right.\non\\
   & \qquad\left. +2 \left(p_2p_3\right)
   {p_2}_{a} {\left(p_2\epsilon _3D\epsilon _2\right)}_{i}I_{14}+2 \left(p_2p_3\right)
   {p_3}_{a} {\left(p_2\epsilon _3D\epsilon _2\right)}_{i}I_{14}\right.\non\\
   & \qquad\left. -2 \left(p_1Np_3\right)
   {p_2}_{i} {\left(p_3D\epsilon _2\epsilon _3\right)}_{a}I'_{11}+2
   \left(p_1Np_2\right) {p_3}_{i} {\left(p_3D\epsilon _2\epsilon
   _3\right)}_{a}I'_{11}\right.\non\\
   & \qquad\left. -2 \left(p_1Np_3\right) {p_2}_{a} {\left(p_3D\epsilon
   _2\epsilon _3\right)}_{i}I'_{11}-2 \left(p_1Np_2\right) {p_3}_{a}
   {\left(p_3D\epsilon _2\epsilon _3\right)}_{i}I'_{11}\right.\non\\
   & \qquad\left. +2 \left(p_2Dp_3\right)
   {p_2}_{a} {\left(p_3D\epsilon _3\epsilon _2\right)}_{i}I_{22}+2 \left(p_2Dp_3\right)
   {p_3}_{a} {\left(p_3D\epsilon _3\epsilon _2\right)}_{i}I_{22}\right.\non\\
   & \qquad\left. -2 \left(p_1Np_3\right)
   {p_2}_{i} {\left(p_3\epsilon _2D\epsilon _3\right)}_{a}I'_{11}+2
   \left(p_1Np_2\right) {p_3}_{i} {\left(p_3\epsilon _2D\epsilon
   _3\right)}_{a}I'_{11}\right.\non\\
   & \qquad\left. +2 \left(p_1Np_3\right) {p_2}_{a} {\left(p_3\epsilon
   _2D\epsilon _3\right)}_{i}I'_{11}-2 \left(p_1Np_2\right) {p_3}_{a}
   {\left(p_3\epsilon _2D\epsilon _3\right)}_{i}I'_{11}\right.\non\\
   & \qquad\left. +4 \left(p_1Np_3\right)
   {p_2}_{i} {\left(p_1N\epsilon _2D\epsilon _3\right)}_{a}I_{5}-2 \left(p_2Dp_3\right)
   {p_2}_{i} {\left(p_1N\epsilon _2D\epsilon _3\right)}_{a}I_{1}\right.\non
\end{align}
\begin{align}
   & \qquad\left.-2 \left(p_2p_3\right)
   {p_3}_{i} {\left(p_1N\epsilon _2D\epsilon _3\right)}_{a}I'_{11}-2
   \left(p_2Dp_3\right) {p_3}_{i} {\left(p_1N\epsilon _2D\epsilon _3\right)}_{a}I_{1}\right.\non\\
   & \qquad\left. -4
   \left(p_1Np_3\right) {p_2}_{a} {\left(p_1N\epsilon _2D\epsilon _3\right)}_{i}I_{5}+2
   \left(p_2Dp_3\right) {p_2}_{a} {\left(p_1N\epsilon _2D\epsilon _3\right)}_{i}I_{1}\right.\non\\
   & \qquad\left. +2
   \left(p_2Dp_3\right) {p_3}_{a} {\left(p_1N\epsilon _2D\epsilon _3\right)}_{i}I_{1}-2
   \left(p_2p_3\right) {p_3}_{i} {\left(p_1N\epsilon _3D\epsilon
   _2\right)}_{a}I'_{11}\right.\non\\
   & \qquad\left. +2 \left(p_2p_3\right) {p_2}_{a} {\left(p_1N\epsilon
   _3D\epsilon _2\right)}_{i}I'_{12}+2 \left(p_2p_3\right) {p_3}_{a}
   {\left(p_1N\epsilon _3D\epsilon _2\right)}_{i}I'_{11}\right.\non\\
   & \qquad\left. -4 \left(p_1Np_3\right)
   {p_2}_{a} {\left(p_2D\epsilon _2D\epsilon _3\right)}_{i}I_{7}-2 \left(p_2Dp_3\right)
   {p_2}_{a} {\left(p_2D\epsilon _2D\epsilon _3\right)}_{i}I_{21}\right.\non\\
   & \qquad\left. -2
   \left(p_2Dp_3\right) {p_3}_{a} {\left(p_2D\epsilon _2D\epsilon
   _3\right)}_{i}I_{21}+2 \left(p_1Np_3\right) {p_2}_{i} {\left(p_2D\epsilon
   _3D\epsilon _2\right)}_{a}I_{1}\right.\non\\
   & \qquad\left. -2 \left(p_1Np_2\right) {p_3}_{i} {\left(p_2D\epsilon
   _3D\epsilon _2\right)}_{a}I_{1}+4 {p_2}_{a} {\left(p_2D\epsilon _3D\epsilon
   _2\right)}_{i}I_{24}\right.\non\\
   & \qquad\left. +2 \left(p_1Np_3\right) {p_2}_{a} {\left(p_2D\epsilon
   _3D\epsilon _2\right)}_{i}I_{1}-2 \left(p_2Dp_3\right) {p_2}_{a} {\left(p_2D\epsilon
   _3D\epsilon _2\right)}_{i}I_{17}\right.\non\\
   & \qquad\left. +4 {p_3}_{a} {\left(p_2D\epsilon _3D\epsilon
   _2\right)}_{i}I_{24}+2 \left(p_1Np_2\right) {p_3}_{a} {\left(p_2D\epsilon
   _3D\epsilon _2\right)}_{i}I_{1}\right.\non\\
   & \qquad\left. -2 \left(p_2Dp_3\right) {p_3}_{a} {\left(p_2D\epsilon
   _3D\epsilon _2\right)}_{i}I_{17}+2 \left(p_2p_3\right) {p_2}_{a} {\left(p_3D\epsilon
   _3D\epsilon _2\right)}_{i}I_{22}\right.\non\\
   & \qquad\left. +2 \left(p_2p_3\right) {p_3}_{a} {\left(p_3D\epsilon
   _3D\epsilon _2\right)}_{i}I_{22}\rp,
\end{align}



\begin{align}
\mathcal{A}_{C^{(p+1)}BB}^{(2)}=& \frac{2 i^{p (p+1)} \sqrt{2}}{(p-1)!}{{C}^{ij}}_{b_1...b_{p-1}}
   {\varepsilon}^{abb_1...b_{p-1}}\lp\tr(-\epsilon _2\epsilon _3) {p_2}_{b}
   {p_2}_{j} {p_3}_{a} {p_3}_{i}I_{2}\right.\non\\
   & \qquad\left. -\tr(D\epsilon _2D\epsilon _3) {p_2}_{b}
   {p_2}_{j} {p_3}_{a} {p_3}_{i}I_{1}+2 {p_2}_{b} {p_2}_{j} {\left(p_2\epsilon
   _3\right)}_{i} {\left(p_3\epsilon _2\right)}_{a}I_{2}\right.\non\\
   & \qquad\left. +2 {p_2}_{j} {p_3}_{b}
   {\left(p_2\epsilon _3\right)}_{i} {\left(p_3\epsilon _2\right)}_{a}I_{2}+2 {p_2}_{b}
   {p_3}_{j} {\left(p_2\epsilon _3\right)}_{i} {\left(p_3\epsilon _2\right)}_{a}I_{2}\right.\non\\
   & \qquad\left. +2
   {p_3}_{b} {p_3}_{j} {\left(p_2\epsilon _3\right)}_{i} {\left(p_3\epsilon
   _2\right)}_{a}I_{2}-2 \left(p_2p_3\right) {p_2}_{b} {p_2}_{i} {\left(\epsilon
   _2\epsilon _3\right)}_{aj}I_{2}\right.\non\\
   & \qquad\left. +4 \left(p_1Np_3\right) {p_2}_{b} {p_2}_{i}
   {\left(\epsilon _2\epsilon _3\right)}_{aj}I_{9}+2 \left(p_2p_3\right) {p_2}_{b}
   {p_3}_{i} {\left(\epsilon _2\epsilon _3\right)}_{aj}I_{2}\right.\non\\
   & \qquad\left. +2 \left(p_2Dp_3\right)
   {p_2}_{b} {p_3}_{i} {\left(\epsilon _2\epsilon _3\right)}_{aj}I'_{11}-2
   \left(p_2p_3\right) {p_2}_{j} {p_3}_{a} {\left(\epsilon _2\epsilon
   _3\right)}_{bi}I_{2}\right.\non\\
   & \qquad\left. -2 \left(p_2p_3\right) {p_3}_{a} {p_3}_{i} {\left(\epsilon
   _2\epsilon _3\right)}_{bj}I_{2}-4 \left(p_1Np_2\right) {p_3}_{a} {p_3}_{i}
   {\left(\epsilon _2\epsilon _3\right)}_{bj}I_{9}\right.\non\\
   & \qquad\left. -2 \left(p_2Dp_3\right) {p_3}_{a}
   {p_3}_{i} {\left(\epsilon _2\epsilon _3\right)}_{bj}I'_{11}+4 \left(p_2Dp_3\right)
   {p_2}_{b} {p_3}_{a} {\left(\epsilon _2\epsilon _3\right)}_{ij}I_{0}\right.\non\\
   & \qquad\left. -2
   \left(p_2Dp_3\right) {p_2}_{b} {p_3}_{i} {\left(\epsilon _2\epsilon
   _3\right)}_{ja}I'_{11}+2 \left(p_2Dp_3\right) {p_3}_{a} {p_3}_{i} {\left(\epsilon
   _2\epsilon _3\right)}_{jb}I'_{11}\right.\non\\
   & \qquad\left. -4 {p_2}_{b} {p_3}_{i} {\left(p_1N\epsilon
   _2\right)}_{a} {\left(p_2\epsilon _3\right)}_{j}I_{9}-4 {p_3}_{b} {p_3}_{i}
   {\left(p_1N\epsilon _2\right)}_{a} {\left(p_2\epsilon _3\right)}_{j}I_{9}\right.\non\\
   & \qquad\left. +4
   {p_3}_{b} {p_3}_{j} {\left(p_1N\epsilon _2\right)}_{i} {\left(p_2\epsilon
   _3\right)}_{a}I_{9}+4 {p_2}_{i} {p_3}_{a} {\left(p_1N\epsilon _2\right)}_{j}
   {\left(p_2\epsilon _3\right)}_{b}I_{9}\right.\non\\
   & \qquad\left. -8 {p_2}_{i} {p_3}_{a} {\left(p_1N\epsilon
   _2\right)}_{j} {\left(p_1N\epsilon _3\right)}_{b}I_{10}-4 {p_2}_{b} {p_3}_{i}
   {\left(p_2D\epsilon _2\right)}_{a} {\left(p_2\epsilon _3\right)}_{j}I_{6}\right.\non\\
   & \qquad\left. -4
   {p_3}_{b} {p_3}_{i} {\left(p_2D\epsilon _2\right)}_{a} {\left(p_2\epsilon
   _3\right)}_{j}I_{6}-4 {p_2}_{b} {p_3}_{i} {\left(p_1N\epsilon _3\right)}_{j}
   {\left(p_2D\epsilon _2\right)}_{a}I_{4}\right.\non\\
   & \qquad\left. -2 {p_2}_{b} {p_3}_{i} {\left(p_2D\epsilon
   _3\right)}_{a} {\left(p_3\epsilon _2\right)}_{j}I'_{11}-2 {p_3}_{b} {p_3}_{i}
   {\left(p_2D\epsilon _3\right)}_{a} {\left(p_3\epsilon _2\right)}_{j}I'_{11}\right.\non\\
   & \qquad\left. +4
   {p_3}_{b} {p_3}_{i} {\left(p_1N\epsilon _2\right)}_{j} {\left(p_2D\epsilon
   _3\right)}_{a}I_{5}+4 {p_2}_{i} {p_3}_{a} {\left(p_1N\epsilon _2\right)}_{j}
   {\left(p_2D\epsilon _3\right)}_{b}I_{5}\right.\non\\
   & \qquad\left. +2 {p_2}_{b} {p_3}_{i} {\left(p_2D\epsilon
   _3\right)}_{j} {\left(p_3\epsilon _2\right)}_{a}I'_{11}+2 {p_3}_{b} {p_3}_{i}
   {\left(p_2D\epsilon _3\right)}_{j} {\left(p_3\epsilon _2\right)}_{a}I'_{11}\right.\non\\
   & \qquad\left. +8
   {p_2}_{b} {p_3}_{a} {\left(p_2D\epsilon _3\right)}_{j} {\left(p_3\epsilon
   _2\right)}_{i}I_{0}-4 {p_2}_{b} {p_3}_{i} {\left(p_1N\epsilon _2\right)}_{a}
   {\left(p_2D\epsilon _3\right)}_{j}I_{5}\right.\non\\
   & \qquad\left. -4 {p_3}_{b} {p_3}_{i} {\left(p_1N\epsilon
   _2\right)}_{a} {\left(p_2D\epsilon _3\right)}_{j}I_{5}-4 {p_2}_{b} {p_3}_{i}
   {\left(p_2D\epsilon _2\right)}_{a} {\left(p_2D\epsilon _3\right)}_{j}I_{7}\right.\non
\end{align}
\begin{align}
   & \qquad\left. -4
   {p_3}_{b} {p_3}_{i} {\left(p_2D\epsilon _2\right)}_{a} {\left(p_2D\epsilon
   _3\right)}_{j}I_{7}-2 {p_2}_{b} {p_2}_{j} {p_3}_{i} {\left(p_2\epsilon _3\epsilon
   _2\right)}_{a}I_{2}-2 {p_2}_{j} {p_3}_{b} {p_3}_{i} {\left(p_2\epsilon _3\epsilon
   _2\right)}_{a}I_{2}\right.\non\\
   & \qquad\left. +2 {p_2}_{b} {p_2}_{i} {p_3}_{a} {\left(p_2\epsilon _3\epsilon
   _2\right)}_{j}I_{2}+2 {p_2}_{b} {p_3}_{a} {p_3}_{i} {\left(p_2\epsilon _3\epsilon
   _2\right)}_{j}I_{2}-2 {p_2}_{b} {p_3}_{i} {\left(p_2\epsilon _3\right)}_{j}
   {\left(p_3D\epsilon _2\right)}_{a}I'_{11}\right.\non\\
   & \qquad\left. -2 {p_3}_{b} {p_3}_{i} {\left(p_2\epsilon
   _3\right)}_{j} {\left(p_3D\epsilon _2\right)}_{a}I'_{11}+2 {p_2}_{b} {p_2}_{j}
   {\left(p_2D\epsilon _3\right)}_{i} {\left(p_3D\epsilon _2\right)}_{a}I_{1}\right.\non\\
   & \qquad\left. +2
   {p_2}_{j} {p_3}_{b} {\left(p_2D\epsilon _3\right)}_{i} {\left(p_3D\epsilon
   _2\right)}_{a}I_{1}+2 {p_2}_{b} {p_3}_{i} {\left(p_2D\epsilon _3\right)}_{j}
   {\left(p_3D\epsilon _2\right)}_{a}I_{1}\right.\non\\
   & \qquad\left. +2 {p_3}_{b} {p_3}_{i} {\left(p_2D\epsilon
   _3\right)}_{j} {\left(p_3D\epsilon _2\right)}_{a}I_{1}+2 {p_2}_{b} {p_3}_{j}
   {\left(p_2\epsilon _3\right)}_{a} {\left(p_3D\epsilon _2\right)}_{i}I'_{11}\right.\non\\
   & \qquad\left. +2
   {p_3}_{b} {p_3}_{j} {\left(p_2\epsilon _3\right)}_{a} {\left(p_3D\epsilon
   _2\right)}_{i}I'_{11}+4 \left(p_1Np_3\right) {p_2}_{b} {p_2}_{i} {\left(\epsilon
   _2D\epsilon _3\right)}_{aj}I_{5}\right.\non\\
   & \qquad\left. -2 \left(p_2Dp_3\right) {p_2}_{b} {p_2}_{i}
   {\left(\epsilon _2D\epsilon _3\right)}_{aj}I_{1}-2 \left(p_2p_3\right) {p_2}_{b}
   {p_3}_{i} {\left(\epsilon _2D\epsilon _3\right)}_{aj}I'_{11}\right.\non\\
   & \qquad\left. -2 \left(p_2Dp_3\right)
   {p_2}_{b} {p_3}_{i} {\left(\epsilon _2D\epsilon _3\right)}_{aj}I_{1}-2
   \left(p_2Dp_3\right) {p_2}_{j} {p_3}_{a} {\left(\epsilon _2D\epsilon
   _3\right)}_{bi}I_{1}\right.\non\\
   & \qquad\left. +2 \left(p_2p_3\right) {p_3}_{a} {p_3}_{i} {\left(\epsilon
   _2D\epsilon _3\right)}_{bj}I'_{11}-4 \left(p_1Np_2\right) {p_3}_{a} {p_3}_{i}
   {\left(\epsilon _2D\epsilon _3\right)}_{bj}I_{5}\right.\non\\
   & \qquad\left. +2 \left(p_2Dp_3\right) {p_3}_{a}
   {p_3}_{i} {\left(\epsilon _2D\epsilon _3\right)}_{bj}I_{1}-2 \left(p_2p_3\right)
   {p_2}_{b} {p_3}_{i} {\left(\epsilon _2D\epsilon _3\right)}_{ja}I'_{11}\right.\non\\
   & \qquad\left. +2
   \left(p_2p_3\right) {p_3}_{a} {p_3}_{i} {\left(\epsilon _2D\epsilon
   _3\right)}_{jb}I'_{11}-4 \left(p_2p_3\right) {p_2}_{b} {p_3}_{a} {\left(\epsilon
   _2D\epsilon _3\right)}_{ji}I_{0}\right.\non\\
   & \qquad\left. -4 {p_2}_{j} {p_3}_{a} {p_3}_{i} {\left(p_1N\epsilon
   _2\epsilon _3\right)}_{b}I_{9}-4 {p_2}_{b} {p_3}_{a} {p_3}_{i} {\left(p_1N\epsilon
   _2\epsilon _3\right)}_{j}I_{9}\right.\non\\
   & \qquad\left. -4 {p_2}_{b} {p_3}_{a} {p_3}_{i} {\left(p_2D\epsilon
   _2\epsilon _3\right)}_{j}I_{6}+2 {p_2}_{b} {p_2}_{j} {p_3}_{i} {\left(p_3D\epsilon
   _2\epsilon _3\right)}_{a}I'_{11}+2 {p_2}_{j} {p_3}_{b} {p_3}_{i} {\left(p_3D\epsilon
   _2\epsilon _3\right)}_{a}I'_{11}\right.\non\\
   & \qquad\left. +2 {p_2}_{b} {p_2}_{i} {p_3}_{a} {\left(p_3D\epsilon
   _2\epsilon _3\right)}_{j}I'_{11}-2 {p_2}_{b} {p_3}_{a} {p_3}_{i} {\left(p_3D\epsilon
   _2\epsilon _3\right)}_{j}I'_{11}+2 {p_2}_{b} {p_2}_{j} {p_3}_{i} {\left(p_3\epsilon
   _2D\epsilon _3\right)}_{a}I'_{11}\right.\non\\
   & \qquad\left. +2 {p_2}_{j} {p_3}_{b} {p_3}_{i} {\left(p_3\epsilon
   _2D\epsilon _3\right)}_{a}I'_{11}+2 {p_2}_{b} {p_2}_{i} {p_3}_{a} {\left(p_3\epsilon
   _2D\epsilon _3\right)}_{j}I'_{11}+2 {p_2}_{b} {p_3}_{a} {p_3}_{i} {\left(p_3\epsilon
   _2D\epsilon _3\right)}_{j}I'_{11}\right.\non\\
   & \qquad\left. +4 {p_2}_{j} {p_3}_{a} {p_3}_{i}
   {\left(p_1N\epsilon _2D\epsilon _3\right)}_{b}I_{5}-4 {p_2}_{b} {p_3}_{a} {p_3}_{i}
   {\left(p_1N\epsilon _2D\epsilon _3\right)}_{j}I_{5}\right.\non\\
   & \qquad\left. -4 {p_2}_{b} {p_3}_{a} {p_3}_{i}
   {\left(p_2D\epsilon _2D\epsilon _3\right)}_{j}I_{7}-2 {p_2}_{b} {p_2}_{j} {p_3}_{i}
   {\left(p_2D\epsilon _3D\epsilon _2\right)}_{a}I_{1}\right.\non\\
   & \qquad\left. -2 {p_2}_{j} {p_3}_{b} {p_3}_{i}
   {\left(p_2D\epsilon _3D\epsilon _2\right)}_{a}I_{1}-2 {p_2}_{b} {p_2}_{i} {p_3}_{a}
   {\left(p_2D\epsilon _3D\epsilon _2\right)}_{j}I_{1}\right.\non\\
   & \qquad\left. +2 {p_2}_{b} {p_3}_{a} {p_3}_{i}
   {\left(p_2D\epsilon _3D\epsilon _2\right)}_{j}I_{1}+2 \left(p_1N\epsilon
   _3p_2\right) {p_2}_{j} {p_3}_{i} {\epsilon _2}_{ab}I_{9}\right.\non\\
   & \qquad\left. +2 \left(p_1N\epsilon
   _3Dp_2\right) {p_2}_{j} {p_3}_{i} {\epsilon _2}_{ab}I_{5}-2 \left(p_1Np_2\right)
   {p_3}_{j} {\left(p_2\epsilon _3\right)}_{i} {\epsilon _2}_{ab}I_{9}\right.\non\\
   & \qquad\left. -2
   \left(p_1Np_3\right) {p_2}_{i} {\left(p_2\epsilon _3\right)}_{j} {\epsilon
   _2}_{ab}I_{9}+2 \left(p_2p_3\right) {p_3}_{j} {\left(p_1N\epsilon _3\right)}_{i}
   {\epsilon _2}_{ab}I_{9}\right.\non\\
   & \qquad\left. +4 \left(p_1Np_2\right) {p_3}_{j} {\left(p_1N\epsilon
   _3\right)}_{i} {\epsilon _2}_{ab}I_{10}-2 \left(p_2Dp_3\right) {p_3}_{j}
   {\left(p_1N\epsilon _3\right)}_{i} {\epsilon _2}_{ab}I_{5}\right.\non\\
   & \qquad\left. +2 \left(p_2p_3\right)
   {p_2}_{i} {\left(p_1N\epsilon _3\right)}_{j} {\epsilon _2}_{ab}I_{9}+2
   \left(p_2Dp_3\right) {p_2}_{i} {\left(p_1N\epsilon _3\right)}_{j} {\epsilon
   _2}_{ab}I_{5}\right.\non\\
   & \qquad\left. -2 \left(p_1Np_2\right) {p_3}_{j} {\left(p_2D\epsilon _3\right)}_{i}
   {\epsilon _2}_{ab}I_{5}-2 \left(p_1Np_3\right) {p_2}_{i} {\left(p_2D\epsilon
   _3\right)}_{j} {\epsilon _2}_{ab}I_{5}\right.\non\\
   & \qquad\left. +2 \left(p_2\epsilon _3Dp_3\right) {p_2}_{b}
   {p_3}_{a} {\epsilon _2}_{ij}I'_{6}+2 \left(p_2D\epsilon _3Dp_3\right) {p_2}_{b}
   {p_3}_{a} {\epsilon _2}_{ij}I'_{7}\right.\non\\
   & \qquad\left. -2 \left(p_1N\epsilon _3p_2\right) {p_2}_{a}
   {p_3}_{b} {\epsilon _2}_{ij}I_{9}+2 \left(p_1N\epsilon _3Dp_2\right) {p_2}_{a}
   {p_3}_{b} {\epsilon _2}_{ij}I_{5}\right.\non\\
   & \qquad\left. +2 \left(p_2p_3\right) {p_2}_{b} {\left(p_2\epsilon
   _3\right)}_{a} {\epsilon _2}_{ij}I_{2}-2 \left(p_1Np_3\right) {p_2}_{b}
   {\left(p_2\epsilon _3\right)}_{a} {\epsilon _2}_{ij}I_{9}\right.\non\\
   & \qquad\left. +2 \left(p_2p_3\right)
   {p_3}_{b} {\left(p_2\epsilon _3\right)}_{a} {\epsilon _2}_{ij}I_{2}+2
   \left(p_1Np_2\right) {p_3}_{b} {\left(p_2\epsilon _3\right)}_{a} {\epsilon
   _2}_{ij}I_{9}\right.\non\\
   & \qquad\left. -2 \left(p_2p_3\right) {p_2}_{b} {\left(p_1N\epsilon _3\right)}_{a}
   {\epsilon _2}_{ij}I_{9}+2 \left(p_2Dp_3\right) {p_2}_{b} {\left(p_1N\epsilon
   _3\right)}_{a} {\epsilon _2}_{ij}I_{5}\right.\non\\
   & \qquad\left. +2 \left(p_1Np_3\right) {p_2}_{b}
   {\left(p_2D\epsilon _3\right)}_{a} {\epsilon _2}_{ij}I_{5}-2 \left(p_2Dp_3\right)
   {p_2}_{b} {\left(p_2D\epsilon _3\right)}_{a} {\epsilon _2}_{ij}I_{1}\right.\non\\
   & \qquad\left. +2
   \left(p_1Np_2\right) {p_3}_{b} {\left(p_2D\epsilon _3\right)}_{a} {\epsilon
   _2}_{ij}I_{5}-2 \left(p_2Dp_3\right) {p_3}_{b} {\left(p_2D\epsilon _3\right)}_{a}
   {\epsilon _2}_{ij}I_{1}\right.\non
\end{align}
\begin{align}
   & \qquad\left. +2 \left(p_2p_3\right) {p_2}_{b} {\left(p_3D\epsilon
   _3\right)}_{a} {\epsilon _2}_{ij}I'_{6}+2 \left(p_2Dp_3\right) {p_2}_{b}
   {\left(p_3D\epsilon _3\right)}_{a} {\epsilon _2}_{ij}I'_{7}\right.\non\\
   & \qquad\left. -2 \left(p_2p_3\right)
   {p_3}_{a} {\left(p_1N\epsilon _3\right)}_{b} {\epsilon _2}_{ji}I_{9}-4
   \left(p_1Np_2\right) {p_3}_{a} {\left(p_1N\epsilon _3\right)}_{b} {\epsilon
   _2}_{ji}I_{10}\right.\non\\
   & \qquad\left. +2 \left(p_2Dp_3\right) {p_3}_{a} {\left(p_1N\epsilon _3\right)}_{b}
   {\epsilon _2}_{ji}I_{5}+2 \left(p_2p_3\right) {p_3}_{a} {\left(p_3D\epsilon
   _3\right)}_{b} {\epsilon _2}_{ji}I'_{6}\right.\non\\
   & \qquad\left. +2 \left(p_1Np_2\right) {p_3}_{a}
   {\left(p_3D\epsilon _3\right)}_{b} {\epsilon _2}_{ji}I'_{4}+2 \left(p_2Dp_3\right)
   {p_3}_{a} {\left(p_3D\epsilon _3\right)}_{b} {\epsilon
   _2}_{ji}I'_{7}\right.\non\\
   & \qquad\left. +\left(p_1Np_2\right) \left(p_2p_3\right) {\epsilon _2}_{ji} {\epsilon
   _3}_{ab}I_{9}-\left(p_1Np_3\right) \left(p_2p_3\right) {\epsilon _2}_{ji} {\epsilon
   _3}_{ab}I_{9}\right.\non\\
   & \qquad\left. -2 \left(p_1Np_2\right) \left(p_1Np_3\right) {\epsilon _2}_{ji}
   {\epsilon _3}_{ab}I_{10}+\left(p_1Np_2\right) \left(p_2Dp_3\right) {\epsilon
   _2}_{ji} {\epsilon _3}_{ab}I_{5}\right.\non\\
   & \qquad\left. +\left(p_1Np_3\right) \left(p_2Dp_3\right) {\epsilon
   _2}_{ji} {\epsilon _3}_{ab}I_{5}+2 \left(p_3D\epsilon _2p_3\right) {p_2}_{b}
   {p_3}_{a} {\epsilon _3}_{ij}I'_{11})+(2\leftrightarrow 3\rp,
\end{align}


\begin{align}
\mathcal{A}_{C^{(p+1)}BB}^{(3)}=& \frac{2 i^{p (p+1)} \sqrt{2}}{(p-2)!}{{C}^{ijk}}_{b_1...b_{p-2}}
   {\varepsilon}^{abcb_1...b_{p-2}}\lp 4 {p_2}_{c} {p_2}_{j} {p_3}_{a} {p_3}_{i}
   {\left(\epsilon _2\epsilon _3\right)}_{bk}I_{9}\right.\non\\
   & \qquad\left. +4 {p_2}_{c} {p_2}_{j} {p_3}_{a}
   {p_3}_{i} {\left(\epsilon _2D\epsilon _3\right)}_{bk}I_{5}+2 {p_2}_{c} {p_2}_{k}
   {p_3}_{i} {\left(p_2\epsilon _3\right)}_{j} {\epsilon _2}_{ab}I_{9}\right.\non\\
   & \qquad\left. +2 {p_2}_{k}
   {p_3}_{c} {p_3}_{i} {\left(p_2\epsilon _3\right)}_{j} {\epsilon _2}_{ab}I_{9}-4
   {p_2}_{c} {p_2}_{k} {p_3}_{i} {\left(p_1N\epsilon _3\right)}_{j} {\epsilon
   _2}_{ab}I_{10}\right.\non\\
   & \qquad\left. +2 {p_2}_{c} {p_2}_{k} {p_3}_{i} {\left(p_2D\epsilon _3\right)}_{j}
   {\epsilon _2}_{ab}I_{5}+2 {p_2}_{k} {p_3}_{c} {p_3}_{i} {\left(p_2D\epsilon
   _3\right)}_{j} {\epsilon _2}_{ab}I_{5}\right.\non\\
   & \qquad\left. +2 {p_2}_{c} {p_3}_{b} {p_3}_{i}
   {\left(p_2\epsilon _3\right)}_{a} {\epsilon _2}_{jk}I_{9}-2 {p_2}_{c} {p_2}_{i}
   {p_3}_{a} {\left(p_2\epsilon _3\right)}_{b} {\epsilon _2}_{jk}I_{9}\right.\non\\
   & \qquad\left. +4 {p_2}_{c}
   {p_2}_{i} {p_3}_{a} {\left(p_1N\epsilon _3\right)}_{b} {\epsilon _2}_{jk}I_{10}+2
   {p_2}_{b} {p_3}_{c} {p_3}_{i} {\left(p_2D\epsilon _3\right)}_{a} {\epsilon
   _2}_{jk}I_{5}\right.\non\\
   & \qquad\left. -2 {p_2}_{c} {p_2}_{i} {p_3}_{a} {\left(p_2D\epsilon _3\right)}_{b}
   {\epsilon _2}_{jk}I_{5}-2 {p_2}_{c} {p_2}_{i} {p_3}_{a} {\left(p_3D\epsilon
   _3\right)}_{b} {\epsilon _2}_{jk}I'_{4}\right.\non\\
   & \qquad\left. -\left(p_2p_3\right) {p_2}_{c} {p_2}_{i}
   {\epsilon _2}_{jk} {\epsilon _3}_{ab}I_{9}+2 \left(p_1Np_3\right) {p_2}_{c}
   {p_2}_{i} {\epsilon _2}_{jk} {\epsilon _3}_{ab}I_{10}\right.\non\\
   & \qquad\left. -\left(p_2Dp_3\right) {p_2}_{c}
   {p_2}_{i} {\epsilon _2}_{jk} {\epsilon _3}_{ab}I_{5}+\left(p_2p_3\right) {p_2}_{c}
   {p_3}_{i} {\epsilon _2}_{jk} {\epsilon _3}_{ab}I_{9}\right.\non\\
   & \qquad\left. -\left(p_2Dp_3\right) {p_2}_{c}
   {p_3}_{i} {\epsilon _2}_{jk} {\epsilon _3}_{ab}I_{5}-\left(p_2p_3\right) {p_2}_{j}
   {p_3}_{a} {\epsilon _2}_{ki} {\epsilon _3}_{bc}I_{9}\right.\non\\
   & \qquad\left. -\left(p_2Dp_3\right) {p_2}_{j}
   {p_3}_{a} {\epsilon _2}_{ki} {\epsilon _3}_{bc}I_{5}+\left(p_2p_3\right) {p_3}_{a}
   {p_3}_{j} {\epsilon _2}_{ki} {\epsilon _3}_{bc}I_{9}\right.\non\\
   & \qquad\left. +2 \left(p_1Np_2\right)
   {p_3}_{a} {p_3}_{j} {\epsilon _2}_{ki} {\epsilon _3}_{bc}I_{10}-\left(p_2Dp_3\right)
   {p_3}_{a} {p_3}_{j} {\epsilon _2}_{ki} {\epsilon _3}_{bc}I_{5})+(2\leftrightarrow 3\rp,
\end{align}


\be
\mathcal{A}_{C^{(p+1)}BB}^{(4)}=\frac{4 i^{p (p+1)} \sqrt{2}}{(p-3)!}{{C}^{ijkl}}_{b_1...b_{p-3}}
   {\varepsilon}^{abcdb_1...b_{p-3}}{p_2}_{d} {p_2}_{j} {p_3}_{a} {p_3}_{i} {\epsilon
   _2}_{kl} {\epsilon _3}_{bc}I_{10}+(2\leftrightarrow 3).
\ee


\be
\mathcal{A}_{C^{(p+1)}hh}=\mathcal{A}_{C^{(p+1)}hh}^{(0)}+\mathcal{A}_{C^{(p+1)}hh}^{(1)}+\mathcal
{A}_{C^{(p+1)}hh}^{(2)}+\mathcal{A}_{C^{(p+1)}hh}^{(3)}+\mathcal{A}_{C^{(p+1)}hh}^{(4)}.
\ee


\begin{align}
\mathcal{A}_{C^{(p+1)}hh}^{(0)}=& \frac{2 i^{p (p+1)} \sqrt{2}}{(p+1)!}{C}_{b_1...b_{p+1}}
   {\varepsilon}^{b_1...b_{p+1}}\lp -2 \left(p_1N\epsilon _2p_3\right) \left(p_1N\epsilon
   _3p_2\right)I_{2}\right.\non\\
   & \qquad\left. +2 \left(p_1N\epsilon _3p_2\right) \left(p_2D\epsilon
   _2p_3\right)I_{20}+2 \left(p_1N\epsilon _2p_3\right) \left(p_2D\epsilon
   _3p_2\right)I_{14}\right.\non\\
   & \qquad\left. -2 \left(p_1N\epsilon _2Dp_2\right) \left(p_2\epsilon
   _3p_2\right)I_{20}-8 \left(p_1N\epsilon _2Dp_2\right) \left(p_1N\epsilon
   _3p_2\right)I_{6}\right.\non\\
   & \qquad\left. -4 \left(p_1N\epsilon _2Dp_2\right) \left(p_2\epsilon
   _3Dp_3\right)I_{18}-2 \left(p_1N\epsilon _2Dp_3\right) \left(p_2\epsilon
   _3p_2\right)I_{14}\right.\non\\
   & \qquad\left. +8 \left(p_1N\epsilon _2Dp_3\right) \left(p_1N\epsilon
   _3p_2\right)I_{8}-2 \left(p_1N\epsilon _2Dp_3\right) \left(p_2D\epsilon
   _3p_2\right)I_{15}\right.\non\\
   & \qquad\left. -2 \left(p_1N\epsilon _2Dp_3\right) \left(p_2\epsilon
   _3Dp_3\right)I_{22}+2 \left(p_1N\epsilon _2Np_1\right) \left(p_2\epsilon
   _3p_2\right)I_{2}\right.\non\\
   & \qquad\left. -8 \left(p_1N\epsilon _2Np_1\right) \left(p_1N\epsilon
   _3p_2\right)I_{9}+4 \left(p_1N\epsilon _2Np_1\right) \left(p_2\epsilon
   _3Dp_3\right)I'_{6}\right.\non\\
   & \qquad\left. -4 \left(p_1N\epsilon _2\epsilon _3p_2\right)I_{23}+2
   \left(p_1N\epsilon _2\epsilon _3p_2\right) \left(p_2p_3\right)I_{16}\right.\non\\
   & \qquad\left. -8
   \left(p_1N\epsilon _2Dp_2\right) \left(p_1N\epsilon _3Dp_2\right)I_{7}+2
   \left(p_1N\epsilon _2Dp_3\right) \left(p_1N\epsilon _3Dp_2\right)I_{1}\right.\non\\
   & \qquad\left. -8
   \left(p_1N\epsilon _2Np_1\right) \left(p_1N\epsilon _3Dp_2\right)I_{5}+4
   \left(p_1N\epsilon _3Dp_3\right) \left(p_3D\epsilon _2p_3\right)I_{22}\right.\non\\
   & \qquad\left. +4
   \left(p_1N\epsilon _2Dp_2\right) \left(p_1N\epsilon _3Dp_3\right)I_{3}-4
   \left(p_1N\epsilon _2Np_1\right) \left(p_1N\epsilon _3Dp_3\right)I'_{4}\right.\non\\
   & \qquad\left. +4
   \left(p_1N\epsilon _3Np_1\right) \left(p_3D\epsilon _2p_3\right)I'_{11}+4
   \left(p_1N\epsilon _2Np_1\right) \left(p_1N\epsilon _3Np_1\right)I_{10}\right.\non\\
   & \qquad\left. -2
   \left(p_1N\epsilon _3Dp_2\right) \left(p_2D\epsilon _2Dp_3\right)I_{21}-4
   \left(p_1N\epsilon _3Dp_3\right) \left(p_2D\epsilon _2Dp_3\right)I_{19}\right.\non\\
   & \qquad\left. -4
   \left(p_1N\epsilon _3Np_1\right) \left(p_2D\epsilon _2Dp_3\right)I_{7}-2
   \left(p_1Np_3\right) \left(p_2D\epsilon _2\epsilon _3p_2\right)I_{20}\right.\non\\
   & \qquad\left. +2
   \left(p_1N\epsilon _2p_3\right) \left(p_2D\epsilon _3Dp_2\right)I_{15}-2
   \left(p_1N\epsilon _2Dp_2\right) \left(p_2D\epsilon _3Dp_2\right)I_{21}\right.\non\\
   & \qquad\left. +2
   \left(p_1N\epsilon _2Np_1\right) \left(p_2D\epsilon _3Dp_2\right)I_{1}+2
   \left(p_1N\epsilon _2p_3\right) \left(p_2D\epsilon _3Dp_3\right)I_{22}\right.\non\\
   & \qquad\left. +2
   \left(p_1Np_3\right) \left(p_2D\epsilon _3\epsilon _2p_3\right)I_{14}-2
   \left(p_1Np_2\right) \left(p_2\epsilon _3D\epsilon _2p_3\right)I_{14}\right.\non\\
   & \qquad\left. +2
   \left(p_1N\epsilon _2D\epsilon _3p_2\right) \left(p_2p_3\right)I_{14}-2
   \left(p_1N\epsilon _2\epsilon _3Dp_2\right) \left(p_2Dp_3\right)I_{15}\right.\non\\
   & \qquad\left. +2
   \left(p_1N\epsilon _2\epsilon _3Dp_3\right) \left(p_2p_3\right)I'_{20}-2
   \left(p_1N\epsilon _2\epsilon _3Dp_3\right) \left(p_2Dp_3\right)I_{22}\right.\non\\
   & \qquad\left. -2
   \left(p_1N\epsilon _2\epsilon _3Np_1\right) \left(p_2p_3\right)I_{2}+8
   \left(p_1N\epsilon _2\epsilon _3Np_1\right) \left(p_2Dp_3\right)I_{0}\right.\non\\
   & \qquad\left. -4
   \left(p_1N\epsilon _3D\epsilon _2p_3\right) \left(p_1Np_2\right)I'_{11}+4
   \left(p_1N\epsilon _3\epsilon _2Dp_3\right) \left(p_1Np_2\right)I'_{11}\right.\non\\
   & \qquad\left. -2
   \left(p_1Np_2\right) \left(p_2D\epsilon _3D\epsilon _2p_3\right)I_{15}+2
   \left(p_1Np_2\right) \left(p_2D\epsilon _3\epsilon _2Dp_3\right)I_{15}\right.\non\\
   & \qquad\left. +2
   \left(p_1Np_2\right) \left(p_3D\epsilon _2\epsilon _3Dp_3\right)I_{22}-2
   \left(p_1Np_2\right) \left(p_3D\epsilon _3D\epsilon _2p_3\right)I_{22}\right.\non\\
   & \qquad\left. +4
   \left(p_1N\epsilon _2D\epsilon _3Dp_2\right)I_{24}-2 \left(p_1N\epsilon _2D\epsilon
   _3Dp_2\right) \left(p_2Dp_3\right)I_{17}\right.\non\\
   & \qquad\left. +2 \left(p_1N\epsilon _2D\epsilon
   _3Dp_3\right) \left(p_2p_3\right)I_{22}-2 \left(p_1N\epsilon _2D\epsilon
   _3Dp_3\right) \left(p_2Dp_3\right)I'_{21}\right.\non\\
   & \qquad\left. -8 \left(p_1N\epsilon _2D\epsilon
   _3Np_1\right) \left(p_2p_3\right)I_{0}+2 \left(p_1N\epsilon _2D\epsilon
   _3Np_1\right) \left(p_2Dp_3\right)I_{1}\right.\non\\
   & \qquad\left. +2 \left(p_1Np_3\right) \left(p_2D\epsilon
   _2D\epsilon _3Dp_2\right)I_{21}+\tr(D\epsilon _2) \left(p_1Np_2\right)
   \left(p_2\epsilon _3p_2\right)I_{20}\right.\non\\
   & \qquad\left. +\tr(D\epsilon _2)
   \left(p_1Np_3\right) \left(p_2\epsilon _3p_2\right)I_{20}-2 \tr(D\epsilon
   _2) \left(p_1N\epsilon _3p_2\right) \left(p_2p_3\right)I_{20}\right.\non\\
   & \qquad\left. +4
   \tr(D\epsilon _2) \left(p_1N\epsilon _3p_2\right)
   \left(p_1Np_2\right)I_{6}+2 \tr(D\epsilon _2) \left(p_1Np_2\right)
   \left(p_2\epsilon _3Dp_3\right)I_{18}\right.\non\\
   & \qquad\left. +4 \tr(D\epsilon _2)
   \left(p_1N\epsilon _3Dp_2\right) \left(p_1Np_2\right)I_{7}\right.\non\\
   & \qquad\left. +2 \tr(D\epsilon
   _2) \left(p_1N\epsilon _3Dp_2\right) \left(p_2Dp_3\right)I_{21}\right.\non\\
   & \qquad\left. -2
   \tr(D\epsilon _2) \left(p_1N\epsilon _3Dp_3\right)
   \left(p_2p_3\right)I_{18}\right.\non
\end{align}
\begin{align}
   & \qquad\left. -4 \tr(D\epsilon _2) \left(p_1N\epsilon
   _3Dp_3\right) \left(p_1Np_2\right)I_{3}+2 \tr(D\epsilon _2)
   \left(p_1N\epsilon _3Dp_3\right) \left(p_2Dp_3\right)I_{19}\right.\non\\
   & \qquad\left. -2 \tr(D\epsilon
   _2) \left(p_1N\epsilon _3Np_1\right) \left(p_2p_3\right)I_{6}+2
   \tr(D\epsilon _2) \left(p_1N\epsilon _3Np_1\right)
   \left(p_1Np_2\right)I_{4}\right.\non\\
   & \qquad\left. +2 \tr(D\epsilon _2) \left(p_1N\epsilon
   _3Np_1\right) \left(p_2Dp_3\right)I_{7}+\tr(D\epsilon _2)
   \left(p_1Np_2\right) \left(p_2D\epsilon _3Dp_2\right)I_{21}\right.\non\\
   & \qquad\left. -\tr(D\epsilon
   _2) \left(p_1Np_3\right) \left(p_2D\epsilon _3Dp_2\right)I_{21}+2
   \tr(D\epsilon _2) \left(p_1Np_2\right) \left(p_2D\epsilon
   _3Dp_3\right)I_{19}\right.\non\\
   & \qquad\left. -2 \tr(D\epsilon _3) \left(p_1Np_3\right)
   \left(p_3D\epsilon _2p_3\right)I_{22}-\tr(D\epsilon _2) \tr(D\epsilon
   _3) \left(p_1Np_2\right) \left(p_2p_3\right)I_{18}\right.\non\\
   & \qquad\left. +\tr(D\epsilon _2)
   \tr(D\epsilon _3) \left(p_1Np_2\right)
   \left(p_1Np_3\right)I_{3}-\tr(D\epsilon _2) \tr(D\epsilon _3)
   \left(p_1Np_2\right) \left(p_2Dp_3\right)I_{19}\right.\non\\
   & \qquad\left. +2 \tr(\epsilon _2\epsilon
   _3) \left(p_1Np_2\right)I_{23}-\tr(\epsilon _2\epsilon _3)
   \left(p_1Np_2\right) \left(p_2p_3\right)I_{16}\right.\non\\
   & \qquad\left. +\tr(\epsilon _2\epsilon _3)
   \left(p_1Np_2\right) \left(p_1Np_3\right)I_{2}-\tr(\epsilon _2\epsilon _3)
   \left(p_1Np_2\right) \left(p_2Dp_3\right)I_{14}\right.\non\\
   & \qquad\left. -2 \tr(D\epsilon _2D\epsilon
   _3) \left(p_1Np_2\right)I_{24}+\tr(D\epsilon _2D\epsilon _3)
   \left(p_1Np_2\right) \left(p_2p_3\right)I_{15}\right.\non\\
   & \qquad\left. -\tr(D\epsilon _2D\epsilon
   _3) \left(p_1Np_2\right) \left(p_1Np_3\right)I_{1}+\tr(D\epsilon
   _2D\epsilon _3) \left(p_1Np_2\right) \left(p_2Dp_3\right)I_{17}\rp\non\\
   & \quad +(2\leftrightarrow 3),
\end{align}


\begin{align}
\mathcal{A}_{C^{(p+1)}hh}^{(1)}=& \frac{2 i^{p (p+1)} \sqrt{2}}{p!}{{C}^{i}}_{b_1...b_p} {\varepsilon}^{ab_1...b_p}\lp -2
   \left(p_2\epsilon _3D\epsilon _2p_3\right) {p_2}_{a} {p_2}_{i}I_{14}\right.\non\\
   & \qquad\left. +2
   \left(p_2\epsilon _3\epsilon _2Dp_3\right) {p_2}_{a} {p_2}_{i}I_{14}-2
   \left(p_3D\epsilon _3\epsilon _2p_3\right) {p_2}_{a} {p_2}_{i}I'_{20}\right.\non\\
   & \qquad\left. -4
   \left(p_1N\epsilon _3D\epsilon _2p_3\right) {p_2}_{a} {p_2}_{i}I'_{11}+4
   \left(p_1N\epsilon _3\epsilon _2Dp_3\right) {p_2}_{a} {p_2}_{i}I'_{11}\right.\non\\
   & \qquad\left. -2
   \left(p_2D\epsilon _3D\epsilon _2p_3\right) {p_2}_{a} {p_2}_{i}I_{15}+2
   \left(p_2D\epsilon _3\epsilon _2Dp_3\right) {p_2}_{a} {p_2}_{i}I_{15}\right.\non\\
   & \qquad\left. +2
   \left(p_3D\epsilon _2\epsilon _3Dp_3\right) {p_2}_{a} {p_2}_{i}I_{22}-2
   \left(p_3D\epsilon _3D\epsilon _2p_3\right) {p_2}_{a} {p_2}_{i}I_{22}\right.\non\\
   & \qquad\left. +2
   \left(p_3D\epsilon _2D\epsilon _3Dp_3\right) {p_2}_{a}
   {p_2}_{i}I'_{21}+\tr(D\epsilon _2) \left(p_2\epsilon _3p_2\right)
   {p_2}_{a} {p_2}_{i}I_{20}\right.\non\\
   & \qquad\left. +4 \tr(D\epsilon _2) \left(p_1N\epsilon
   _3p_2\right) {p_2}_{a} {p_2}_{i}I_{6}+2 \tr(D\epsilon _2)
   \left(p_2\epsilon _3Dp_3\right) {p_2}_{a} {p_2}_{i}I_{18}\right.\non\\
   & \qquad\left. +4 \tr(D\epsilon
   _2) \left(p_1N\epsilon _3Dp_2\right) {p_2}_{a} {p_2}_{i}I_{7}-4
   \tr(D\epsilon _2) \left(p_1N\epsilon _3Dp_3\right) {p_2}_{a}
   {p_2}_{i}I_{3}\right.\non\\
   & \qquad\left. +2 \tr(D\epsilon _2) \left(p_1N\epsilon _3Np_1\right)
   {p_2}_{a} {p_2}_{i}I_{4}+\tr(D\epsilon _2) \left(p_2D\epsilon
   _3Dp_2\right) {p_2}_{a} {p_2}_{i}I_{21}\right.\non\\
   & \qquad\left. +2 \tr(D\epsilon _2)
   \left(p_2D\epsilon _3Dp_3\right) {p_2}_{a} {p_2}_{i}I_{19}+\tr(D\epsilon
   _3) \left(p_3\epsilon _2p_3\right) {p_2}_{a}
   {p_2}_{i}I'_{20}\right.\non\\
   & \qquad\left. -\tr(D\epsilon _3) \left(p_3D\epsilon _2Dp_3\right)
   {p_2}_{a} {p_2}_{i}I'_{21}-\tr(D\epsilon _2) \tr(D\epsilon _3)
   \left(p_2p_3\right) {p_2}_{a} {p_2}_{i}I_{18}\right.\non\\
   & \qquad\left. +2 \tr(D\epsilon _2)
   \tr(D\epsilon _3) \left(p_1Np_3\right) {p_2}_{a}
   {p_2}_{i}I_{3}\right.\non\\
   & \qquad\left. -\tr(D\epsilon _2) \tr(D\epsilon _3)
   \left(p_2Dp_3\right) {p_2}_{a} {p_2}_{i}I_{19}+2 \tr(\epsilon _2\epsilon
   _3) {p_2}_{a} {p_2}_{i}I_{23}\right.\non\\
   & \qquad\left. -\tr(\epsilon _2\epsilon _3)
   \left(p_2p_3\right) {p_2}_{a} {p_2}_{i}I_{16}+2 \tr(\epsilon _2\epsilon
   _3) \left(p_1Np_3\right) {p_2}_{a} {p_2}_{i}I_{2}\right.\non\\
   & \qquad\left. -\tr(\epsilon _2\epsilon
   _3) \left(p_2Dp_3\right) {p_2}_{a} {p_2}_{i}I_{14}-2 \tr(D\epsilon
   _2D\epsilon _3) {p_2}_{a} {p_2}_{i}I_{24}\right.\non\\
   & \qquad\left. +\tr(D\epsilon _2D\epsilon
   _3) \left(p_2p_3\right) {p_2}_{a} {p_2}_{i}I_{15}-2 \tr(D\epsilon
   _2D\epsilon _3) \left(p_1Np_3\right) {p_2}_{a}
   {p_2}_{i}I_{1}\right.\non\\
   & \qquad\left. +\tr(D\epsilon _2D\epsilon _3) \left(p_2Dp_3\right)
   {p_2}_{a} {p_2}_{i}I_{17}-2 \left(p_1N\epsilon _2\epsilon _3p_2\right) {p_2}_{i}
   {p_3}_{a}I_{2}\right.\non\\
   & \qquad\left. +2 \left(p_1N\epsilon _3\epsilon _2p_3\right) {p_2}_{i}
   {p_3}_{a}I_{2}-2 \left(p_2\epsilon _3D\epsilon _2p_3\right) {p_2}_{i}
   {p_3}_{a}I_{14}\right.\non\\
   & \qquad\left. +2 \left(p_2\epsilon _3\epsilon _2Dp_3\right) {p_2}_{i}
   {p_3}_{a}I_{14}-2 \left(p_3D\epsilon _3\epsilon _2p_3\right) {p_2}_{i}
   {p_3}_{a}I'_{20}\right.\non\\
   & \qquad\left. -4 \left(p_1N\epsilon _2\epsilon _3Dp_3\right) {p_2}_{i}
   {p_3}_{a}I'_{6}+4 \left(p_1N\epsilon _2\epsilon _3Np_1\right) {p_2}_{i}
   {p_3}_{a}I_{9}\right.\non
\end{align}
\begin{align}
   & \qquad\left. -2 \left(p_1N\epsilon _3D\epsilon _2p_3\right) {p_2}_{i}
   {p_3}_{a}I'_{11}+2 \left(p_1N\epsilon _3\epsilon _2Dp_3\right) {p_2}_{i}
   {p_3}_{a}I'_{11}\right.\non\\
   & \qquad\left. -2 \left(p_2D\epsilon _3D\epsilon _2p_3\right) {p_2}_{i}
   {p_3}_{a}I_{15}+2 \left(p_2D\epsilon _3\epsilon _2Dp_3\right) {p_2}_{i}
   {p_3}_{a}I_{15}\right.\non\\
   & \qquad\left. +2 \left(p_3D\epsilon _2\epsilon _3Dp_3\right) {p_2}_{i}
   {p_3}_{a}I_{22}-2 \left(p_3D\epsilon _3D\epsilon _2p_3\right) {p_2}_{i}
   {p_3}_{a}I_{22}\right.\non\\
   & \qquad\left. -2 \left(p_1N\epsilon _2D\epsilon _3Dp_2\right) {p_2}_{i}
   {p_3}_{a}I_{1}+4 \left(p_1N\epsilon _2D\epsilon _3Dp_3\right) {p_2}_{i}
   {p_3}_{a}I'_{7}\right.\non\\
   & \qquad\left. +4 \left(p_1N\epsilon _2D\epsilon _3Np_1\right) {p_2}_{i}
   {p_3}_{a}I_{5}-2 \left(p_1N\epsilon _3D\epsilon _2Dp_3\right) {p_2}_{i}
   {p_3}_{a}I_{1}\right.\non\\
   & \qquad\left. +2 \left(p_3D\epsilon _2D\epsilon _3Dp_3\right) {p_2}_{i}
   {p_3}_{a}I'_{21}+\tr(D\epsilon _2) \left(p_2\epsilon _3p_2\right)
   {p_2}_{i} {p_3}_{a}I_{20}\right.\non\\
   & \qquad\left. +2 \tr(D\epsilon _2) \left(p_1N\epsilon
   _3p_2\right) {p_2}_{i} {p_3}_{a}I_{6}+2 \tr(D\epsilon _2)
   \left(p_2\epsilon _3Dp_3\right) {p_2}_{i} {p_3}_{a}I_{18}\right.\non\\
   & \qquad\left. +2 \tr(D\epsilon
   _2) \left(p_1N\epsilon _3Dp_2\right) {p_2}_{i}
   {p_3}_{a}I_{7}+\tr(D\epsilon _2) \left(p_2D\epsilon _3Dp_2\right)
   {p_2}_{i} {p_3}_{a}I_{21}\right.\non\\
   & \qquad\left. +2 \tr(D\epsilon _2) \left(p_2D\epsilon
   _3Dp_3\right) {p_2}_{i} {p_3}_{a}I_{19}+\tr(D\epsilon _3)
   \left(p_3\epsilon _2p_3\right) {p_2}_{i} {p_3}_{a}I'_{20}\right.\non\\
   & \qquad\left. +2 \tr(D\epsilon
   _3) \left(p_1N\epsilon _2p_3\right) {p_2}_{i} {p_3}_{a}I'_{6}-2
   \tr(D\epsilon _3) \left(p_1N\epsilon _2Dp_3\right) {p_2}_{i}
   {p_3}_{a}I'_{7}\right.\non\\
   & \qquad\left. -\tr(D\epsilon _3) \left(p_3D\epsilon _2Dp_3\right)
   {p_2}_{i} {p_3}_{a}I'_{21}-\tr(D\epsilon _2) \tr(D\epsilon _3)
   \left(p_2p_3\right) {p_2}_{i} {p_3}_{a}I_{18}\right.\non\\
   & \qquad\left. -\tr(D\epsilon _2)
   \tr(D\epsilon _3) \left(p_2Dp_3\right) {p_2}_{i} {p_3}_{a}I_{19}+2
   tr\left(\epsilon _2\epsilon _3\right) {p_2}_{i} {p_3}_{a}I_{23}\right.\non\\
   & \qquad\left. -\tr(\epsilon
   _2\epsilon _3) \left(p_2p_3\right) {p_2}_{i} {p_3}_{a}I_{16}-\tr(\epsilon
   _2\epsilon _3) \left(p_2Dp_3\right) {p_2}_{i} {p_3}_{a}I_{14}\right.\non\\
   & \qquad\left. -2
   \tr(D\epsilon _2D\epsilon _3) {p_2}_{i} {p_3}_{a}I_{24}+\tr(D\epsilon
   _2D\epsilon _3) \left(p_2p_3\right) {p_2}_{i}
   {p_3}_{a}I_{15}\right.\non\\
   & \qquad\left. +\tr(D\epsilon _2D\epsilon _3) \left(p_2Dp_3\right)
   {p_2}_{i} {p_3}_{a}I_{17}-2 \left(p_1N\epsilon _3D\epsilon _2p_3\right) {p_2}_{a}
   {p_3}_{i}I'_{11}\right.\non\\
   & \qquad\left. -2 \left(p_1N\epsilon _3\epsilon _2Dp_3\right) {p_2}_{a}
   {p_3}_{i}I'_{11}-2 \tr(D\epsilon _3) \left(p_3D\epsilon _2p_3\right)
   {p_2}_{a} {p_3}_{i}I_{22}\right.\non\\
   & \qquad\left. -2 \tr(D\epsilon _3) \left(p_3D\epsilon
   _2p_3\right) {p_3}_{a} {p_3}_{i}I_{22}+2 \left(p_1N\epsilon _2p_3\right) {p_2}_{i}
   {\left(p_2\epsilon _3\right)}_{a}I_{2}\right.\non\\
   & \qquad\left. +2 \tr(D\epsilon _2)
   \left(p_1Np_3\right) {p_2}_{i} {\left(p_2\epsilon _3\right)}_{a}I_{6}+2
   \left(p_1N\epsilon _2p_3\right) {p_3}_{i} {\left(p_2\epsilon _3\right)}_{a}I_{2}\right.\non\\
   & \qquad\left. +4
   \left(p_1N\epsilon _2Dp_2\right) {p_3}_{i} {\left(p_2\epsilon _3\right)}_{a}I_{6}+2
   \left(p_1N\epsilon _2Dp_3\right) {p_3}_{i} {\left(p_2\epsilon
   _3\right)}_{a}I'_{11}\right.\non\\
   & \qquad\left. +4 \left(p_1N\epsilon _2Np_1\right) {p_3}_{i}
   {\left(p_2\epsilon _3\right)}_{a}I_{9}-2 \tr(D\epsilon _2)
   \left(p_1Np_2\right) {p_3}_{i} {\left(p_2\epsilon _3\right)}_{a}I_{6}\right.\non\\
   & \qquad\left. -2
   \left(p_1N\epsilon _2p_3\right) {p_2}_{a} {\left(p_2\epsilon _3\right)}_{i}I_{2}+2
   \left(p_2D\epsilon _2p_3\right) {p_2}_{a} {\left(p_2\epsilon _3\right)}_{i}I_{20}\right.\non\\
   & \qquad\left. +2
   \left(p_3D\epsilon _2p_3\right) {p_2}_{a} {\left(p_2\epsilon _3\right)}_{i}I_{14}-2
   \left(p_3D\epsilon _2Dp_3\right) {p_2}_{a} {\left(p_2\epsilon _3\right)}_{i}I_{15}\right.\non\\
   & \qquad\left. -2
   \tr(D\epsilon _2) \left(p_2p_3\right) {p_2}_{a} {\left(p_2\epsilon
   _3\right)}_{i}I_{20}-2 \tr(D\epsilon _2) \left(p_1Np_3\right) {p_2}_{a}
   {\left(p_2\epsilon _3\right)}_{i}I_{6}\right.\non\\
   & \qquad\left. -2 \left(p_1N\epsilon _2p_3\right) {p_3}_{a}
   {\left(p_2\epsilon _3\right)}_{i}I_{2}+2 \left(p_2D\epsilon _2p_3\right) {p_3}_{a}
   {\left(p_2\epsilon _3\right)}_{i}I_{20}\right.\non\\
   & \qquad\left. +2 \left(p_3D\epsilon _2p_3\right) {p_3}_{a}
   {\left(p_2\epsilon _3\right)}_{i}I_{14}-4 \left(p_1N\epsilon _2Dp_2\right) {p_3}_{a}
   {\left(p_2\epsilon _3\right)}_{i}I_{6}\right.\non\\
   & \qquad\left. -4 \left(p_1N\epsilon _2Np_1\right) {p_3}_{a}
   {\left(p_2\epsilon _3\right)}_{i}I_{9}-2 \left(p_3D\epsilon _2Dp_3\right) {p_3}_{a}
   {\left(p_2\epsilon _3\right)}_{i}I_{15}\right.\non\\
   & \qquad\left. -2 \tr(D\epsilon _2)
   \left(p_2p_3\right) {p_3}_{a} {\left(p_2\epsilon _3\right)}_{i}I_{20}+2
   \tr(D\epsilon _2) \left(p_1Np_2\right) {p_3}_{a} {\left(p_2\epsilon
   _3\right)}_{i}I_{6}\right.\non\\
   & \qquad\left. +2 \left(p_1N\epsilon _3Dp_2\right) {p_3}_{i} {\left(p_3\epsilon
   _2\right)}_{a}I'_{11}-2 \left(p_1Np_2\right) {\left(p_2\epsilon _3\right)}_{i}
   {\left(p_3\epsilon _2\right)}_{a}I_{2}\right.\non\\
   & \qquad\left. -2 \left(p_1Np_3\right) {\left(p_2\epsilon
   _3\right)}_{i} {\left(p_3\epsilon _2\right)}_{a}I_{2}+2 \left(p_1N\epsilon
   _3Dp_2\right) {p_2}_{a} {\left(p_3\epsilon _2\right)}_{i}I'_{13}\right.\non\\
   & \qquad\left. +2
   \left(p_2D\epsilon _3Dp_3\right) {p_2}_{a} {\left(p_3\epsilon _2\right)}_{i}I_{22}+2
   \left(p_1N\epsilon _3Dp_2\right) {p_3}_{a} {\left(p_3\epsilon
   _2\right)}_{i}I'_{11}\right.\non\\
   & \qquad\left. +2 \left(p_2D\epsilon _3Dp_3\right) {p_3}_{a}
   {\left(p_3\epsilon _2\right)}_{i}I_{22}-2 \left(p_1Np_2\right) \left(p_2p_3\right)
   {\left(\epsilon _2\epsilon _3\right)}_{ai}I_{2}\right.\non\\
   & \qquad\left. +2 \left(p_1Np_3\right)
   \left(p_2p_3\right) {\left(\epsilon _2\epsilon _3\right)}_{ai}I_{2}+4
   \left(p_1Np_2\right) \left(p_1Np_3\right) {\left(\epsilon _2\epsilon
   _3\right)}_{ai}I_{9}\right.\non\\
   & \qquad\left. +2 \left(p_1Np_3\right) \left(p_2Dp_3\right) {\left(\epsilon
   _2\epsilon _3\right)}_{ai}I'_{11}-2 \left(p_1Np_3\right) \left(p_2Dp_3\right)
   {\left(\epsilon _2\epsilon _3\right)}_{ia}I'_{11}\right.\non\\
   & \qquad\left. -2 \left(p_2\epsilon _3p_2\right)
   {p_2}_{i} {\left(p_1N\epsilon _2\right)}_{a}I_{2}+8 \left(p_1N\epsilon _3p_2\right)
   {p_2}_{i} {\left(p_1N\epsilon _2\right)}_{a}I_{9}\right.\non
\end{align}
\begin{align}
   & \qquad\left. -4 \left(p_2\epsilon _3Dp_3\right)
   {p_2}_{i} {\left(p_1N\epsilon _2\right)}_{a}I'_{6}+8 \left(p_1N\epsilon
   _3Dp_2\right) {p_2}_{i} {\left(p_1N\epsilon _2\right)}_{a}I_{5}\right.\non\\
   & \qquad\left. +4 \left(p_1N\epsilon
   _3Dp_3\right) {p_2}_{i} {\left(p_1N\epsilon _2\right)}_{a}I'_{4}-8
   \left(p_1N\epsilon _3Np_1\right) {p_2}_{i} {\left(p_1N\epsilon
   _2\right)}_{a}I_{10}\right.\non\\
   & \qquad\left. -2 \left(p_2D\epsilon _3Dp_2\right) {p_2}_{i}
   {\left(p_1N\epsilon _2\right)}_{a}I_{1}+4 \left(p_2D\epsilon _3Dp_3\right) {p_2}_{i}
   {\left(p_1N\epsilon _2\right)}_{a}I'_{7}\right.\non\\
   & \qquad\left. +2 \tr(D\epsilon _3)
   \left(p_2p_3\right) {p_2}_{i} {\left(p_1N\epsilon _2\right)}_{a}I'_{6}-2
   \tr(D\epsilon _3) \left(p_1Np_3\right) {p_2}_{i} {\left(p_1N\epsilon
   _2\right)}_{a}I'_{4}\right.\non\\
   & \qquad\left. -2 \tr(D\epsilon _3) \left(p_2Dp_3\right) {p_2}_{i}
   {\left(p_1N\epsilon _2\right)}_{a}I'_{7}-2 \left(p_2\epsilon _3p_2\right) {p_3}_{i}
   {\left(p_1N\epsilon _2\right)}_{a}I_{2}\right.\non\\
   & \qquad\left. +4 \left(p_1N\epsilon _3p_2\right) {p_3}_{i}
   {\left(p_1N\epsilon _2\right)}_{a}I_{9}-4 \left(p_1N\epsilon _3Dp_2\right) {p_3}_{i}
   {\left(p_1N\epsilon _2\right)}_{a}I_{5}\right.\non\\
   & \qquad\left. +2 \left(p_2D\epsilon _3Dp_2\right) {p_3}_{i}
   {\left(p_1N\epsilon _2\right)}_{a}I_{1}-2 \tr(D\epsilon _3)
   \left(p_2p_3\right) {p_3}_{i} {\left(p_1N\epsilon _2\right)}_{a}I'_{6}\right.\non\\
   & \qquad\left. -2
   \tr(D\epsilon _3) \left(p_2Dp_3\right) {p_3}_{i} {\left(p_1N\epsilon
   _2\right)}_{a}I'_{7}+4 \left(p_2p_3\right) {\left(p_1N\epsilon _2\right)}_{a}
   {\left(p_2\epsilon _3\right)}_{i}I_{2}\right.\non\\
   & \qquad\left. -4 \left(p_1Np_3\right) {\left(p_1N\epsilon
   _2\right)}_{a} {\left(p_2\epsilon _3\right)}_{i}I_{9}+2 \left(p_2\epsilon
   _3p_2\right) {p_2}_{a} {\left(p_1N\epsilon _2\right)}_{i}I_{2}\right.\non\\
   & \qquad\left. -8 \left(p_1N\epsilon
   _3p_2\right) {p_2}_{a} {\left(p_1N\epsilon _2\right)}_{i}I_{9}+4 \left(p_2\epsilon
   _3Dp_3\right) {p_2}_{a} {\left(p_1N\epsilon _2\right)}_{i}I'_{6}\right.\non\\
   & \qquad\left. -8
   \left(p_1N\epsilon _3Dp_2\right) {p_2}_{a} {\left(p_1N\epsilon _2\right)}_{i}I_{5}-4
   \left(p_1N\epsilon _3Dp_3\right) {p_2}_{a} {\left(p_1N\epsilon
   _2\right)}_{i}I'_{4}\right.\non\\
   & \qquad\left. +8 \left(p_1N\epsilon _3Np_1\right) {p_2}_{a}
   {\left(p_1N\epsilon _2\right)}_{i}I_{10}+2 \left(p_2D\epsilon _3Dp_2\right)
   {p_2}_{a} {\left(p_1N\epsilon _2\right)}_{i}I_{1}\right.\non\\
   & \qquad\left. -4 \left(p_2D\epsilon _3Dp_3\right)
   {p_2}_{a} {\left(p_1N\epsilon _2\right)}_{i}I'_{7}-2 \tr(D\epsilon _3)
   \left(p_2p_3\right) {p_2}_{a} {\left(p_1N\epsilon _2\right)}_{i}I'_{6}\right.\non\\
   & \qquad\left. +2
   \tr(D\epsilon _3) \left(p_1Np_3\right) {p_2}_{a} {\left(p_1N\epsilon
   _2\right)}_{i}I'_{4}+2 \tr(D\epsilon _3) \left(p_2Dp_3\right) {p_2}_{a}
   {\left(p_1N\epsilon _2\right)}_{i}I'_{7}\right.\non\\
   & \qquad\left. +2 \left(p_2\epsilon _3p_2\right) {p_3}_{a}
   {\left(p_1N\epsilon _2\right)}_{i}I_{2}-4 \left(p_1N\epsilon _3p_2\right) {p_3}_{a}
   {\left(p_1N\epsilon _2\right)}_{i}I_{9}\right.\non\\
   & \qquad\left. +4 \left(p_2\epsilon _3Dp_3\right) {p_3}_{a}
   {\left(p_1N\epsilon _2\right)}_{i}I'_{6}-4 \left(p_1N\epsilon _3Dp_2\right)
   {p_3}_{a} {\left(p_1N\epsilon _2\right)}_{i}I_{5}\right.\non\\
   & \qquad\left. +2 \left(p_2D\epsilon _3Dp_2\right)
   {p_3}_{a} {\left(p_1N\epsilon _2\right)}_{i}I_{1}-4 \left(p_2D\epsilon _3Dp_3\right)
   {p_3}_{a} {\left(p_1N\epsilon _2\right)}_{i}I'_{7}\right.\non\\
   & \qquad\left. -2 \tr(D\epsilon _3)
   \left(p_2p_3\right) {p_3}_{a} {\left(p_1N\epsilon _2\right)}_{i}I'_{6}+2
   \tr(D\epsilon _3) \left(p_2Dp_3\right) {p_3}_{a} {\left(p_1N\epsilon
   _2\right)}_{i}I'_{7}\right.\non\\
   & \qquad\left. -4 \left(p_1Np_3\right) {\left(p_1N\epsilon _2\right)}_{i}
   {\left(p_2\epsilon _3\right)}_{a}I_{9}-4 \left(p_3D\epsilon _2p_3\right) {p_3}_{i}
   {\left(p_1N\epsilon _3\right)}_{a}I'_{11}\right.\non\\
   & \qquad\left. +4 \left(p_2p_3\right) {\left(p_1N\epsilon
   _2\right)}_{i} {\left(p_1N\epsilon _3\right)}_{a}I_{9}+4 \left(p_2Dp_3\right)
   {\left(p_1N\epsilon _2\right)}_{i} {\left(p_1N\epsilon _3\right)}_{a}I_{5}\right.\non\\
   & \qquad\left. +4
   \left(p_3D\epsilon _2p_3\right) {p_2}_{a} {\left(p_1N\epsilon
   _3\right)}_{i}I'_{11}+4 \left(p_3D\epsilon _2p_3\right) {p_3}_{a}
   {\left(p_1N\epsilon _3\right)}_{i}I'_{11}\right.\non\\
   & \qquad\left. -2 \left(p_2\epsilon _3p_2\right) {p_2}_{a}
   {\left(p_2D\epsilon _2\right)}_{i}I_{20}-8 \left(p_1N\epsilon _3p_2\right) {p_2}_{a}
   {\left(p_2D\epsilon _2\right)}_{i}I_{6}\right.\non\\
   & \qquad\left. -4 \left(p_2\epsilon _3Dp_3\right) {p_2}_{a}
   {\left(p_2D\epsilon _2\right)}_{i}I_{18}-8 \left(p_1N\epsilon _3Dp_2\right)
   {p_2}_{a} {\left(p_2D\epsilon _2\right)}_{i}I_{7}\right.\non\\
   & \qquad\left. +8 \left(p_1N\epsilon _3Dp_3\right)
   {p_2}_{a} {\left(p_2D\epsilon _2\right)}_{i}I_{3}-4 \left(p_1N\epsilon _3Np_1\right)
   {p_2}_{a} {\left(p_2D\epsilon _2\right)}_{i}I_{4}\right.\non\\
   & \qquad\left. -2 \left(p_2D\epsilon _3Dp_2\right)
   {p_2}_{a} {\left(p_2D\epsilon _2\right)}_{i}I_{21}-4 \left(p_2D\epsilon
   _3Dp_3\right) {p_2}_{a} {\left(p_2D\epsilon _2\right)}_{i}I_{19}\right.\non\\
   & \qquad\left. +2 \tr(D\epsilon
   _3) \left(p_2p_3\right) {p_2}_{a} {\left(p_2D\epsilon _2\right)}_{i}I_{18}-4
   \tr(D\epsilon _3) \left(p_1Np_3\right) {p_2}_{a} {\left(p_2D\epsilon
   _2\right)}_{i}I_{3}\right.\non\\
   & \qquad\left. +2 \tr(D\epsilon _3) \left(p_2Dp_3\right) {p_2}_{a}
   {\left(p_2D\epsilon _2\right)}_{i}I_{19}-2 \left(p_2\epsilon _3p_2\right) {p_3}_{a}
   {\left(p_2D\epsilon _2\right)}_{i}I_{20}\right.\non\\
   & \qquad\left. -4 \left(p_1N\epsilon _3p_2\right) {p_3}_{a}
   {\left(p_2D\epsilon _2\right)}_{i}I_{6}-4 \left(p_2\epsilon _3Dp_3\right) {p_3}_{a}
   {\left(p_2D\epsilon _2\right)}_{i}I_{18}\right.\non\\
   & \qquad\left. -4 \left(p_1N\epsilon _3Dp_2\right)
   {p_3}_{a} {\left(p_2D\epsilon _2\right)}_{i}I_{7}-2 \left(p_2D\epsilon _3Dp_2\right)
   {p_3}_{a} {\left(p_2D\epsilon _2\right)}_{i}I_{21}\right.\non\\
   & \qquad\left. -4 \left(p_2D\epsilon
   _3Dp_3\right) {p_3}_{a} {\left(p_2D\epsilon _2\right)}_{i}I_{19}+2 \tr(D\epsilon
   _3) \left(p_2p_3\right) {p_3}_{a} {\left(p_2D\epsilon _2\right)}_{i}I_{18}\right.\non\\
   & \qquad\left. +2
   \tr(D\epsilon _3) \left(p_2Dp_3\right) {p_3}_{a} {\left(p_2D\epsilon
   _2\right)}_{i}I_{19}-4 \left(p_1Np_3\right) {\left(p_2D\epsilon _2\right)}_{i}
   {\left(p_2\epsilon _3\right)}_{a}I_{6}\right.\non\\
   & \qquad\left. +4 \left(p_2p_3\right) {\left(p_1N\epsilon
   _3\right)}_{a} {\left(p_2D\epsilon _2\right)}_{i}I_{6}+4 \left(p_2Dp_3\right)
   {\left(p_1N\epsilon _3\right)}_{a} {\left(p_2D\epsilon _2\right)}_{i}I_{7}\right.\non\\
   & \qquad\left. +2
   \left(p_1N\epsilon _2Dp_3\right) {p_2}_{i} {\left(p_2D\epsilon _3\right)}_{a}I_{1}+2
   tr\left(D\epsilon _2\right) \left(p_1Np_3\right) {p_2}_{i} {\left(p_2D\epsilon
   _3\right)}_{a}I_{7}\right.\non\\
   & \qquad\left. -2 \left(p_1N\epsilon _2p_3\right) {p_3}_{i} {\left(p_2D\epsilon
   _3\right)}_{a}I'_{11}+4 \left(p_1N\epsilon _2Dp_2\right) {p_3}_{i}
   {\left(p_2D\epsilon _3\right)}_{a}I_{7}\right.\non
\end{align}
\begin{align}
   & \qquad\left.-2 \left(p_1N\epsilon _2Dp_3\right) {p_3}_{i}
   {\left(p_2D\epsilon _3\right)}_{a}I_{1}+4 \left(p_1N\epsilon _2Np_1\right) {p_3}_{i}
   {\left(p_2D\epsilon _3\right)}_{a}I_{5}\right.\non\\
   & \qquad\left. -2 \tr(D\epsilon _2)
   \left(p_1Np_2\right) {p_3}_{i} {\left(p_2D\epsilon _3\right)}_{a}I_{7}+2
   \left(p_1Np_3\right) {\left(p_2D\epsilon _3\right)}_{a} {\left(p_3\epsilon
   _2\right)}_{i}I'_{11}\right.\non\\
   & \qquad\left. -4 \left(p_1Np_3\right) {\left(p_1N\epsilon _2\right)}_{i}
   {\left(p_2D\epsilon _3\right)}_{a}I_{5}-4 \left(p_1Np_3\right) {\left(p_2D\epsilon
   _2\right)}_{i} {\left(p_2D\epsilon _3\right)}_{a}I_{7}\right.\non\\
   & \qquad\left. -2 \left(p_3\epsilon
   _2p_3\right) {p_2}_{a} {\left(p_2D\epsilon _3\right)}_{i}I_{14}+2 \left(p_3D\epsilon
   _2p_3\right) {p_2}_{a} {\left(p_2D\epsilon _3\right)}_{i}I_{15}\right.\non\\
   & \qquad\left. +2 \left(p_1N\epsilon
   _2Dp_3\right) {p_2}_{a} {\left(p_2D\epsilon _3\right)}_{i}I_{1}-2 \left(p_2D\epsilon
   _2Dp_3\right) {p_2}_{a} {\left(p_2D\epsilon _3\right)}_{i}I_{21}\right.\non\\
   & \qquad\left. +2 \tr(D\epsilon
   _2) \left(p_1Np_3\right) {p_2}_{a} {\left(p_2D\epsilon _3\right)}_{i}I_{7}+2
   \tr(D\epsilon _2) \left(p_2Dp_3\right) {p_2}_{a} {\left(p_2D\epsilon
   _3\right)}_{i}I_{21}\right.\non\\
   & \qquad\left. -2 \left(p_3\epsilon _2p_3\right) {p_3}_{a} {\left(p_2D\epsilon
   _3\right)}_{i}I_{14}+2 \left(p_3D\epsilon _2p_3\right) {p_3}_{a} {\left(p_2D\epsilon
   _3\right)}_{i}I_{15}\right.\non\\
   & \qquad\left. -4 \left(p_1N\epsilon _2Dp_2\right) {p_3}_{a}
   {\left(p_2D\epsilon _3\right)}_{i}I_{7}+2 \left(p_1N\epsilon _2Dp_3\right) {p_3}_{a}
   {\left(p_2D\epsilon _3\right)}_{i}I_{1}\right.\non\\
   & \qquad\left. -4 \left(p_1N\epsilon _2Np_1\right) {p_3}_{a}
   {\left(p_2D\epsilon _3\right)}_{i}I_{5}-2 \left(p_2D\epsilon _2Dp_3\right) {p_3}_{a}
   {\left(p_2D\epsilon _3\right)}_{i}I_{21}\right.\non\\
   & \qquad\left. +2 \tr(D\epsilon _2)
   \left(p_1Np_2\right) {p_3}_{a} {\left(p_2D\epsilon _3\right)}_{i}I_{7}+2
   \tr(D\epsilon _2) \left(p_2Dp_3\right) {p_3}_{a} {\left(p_2D\epsilon
   _3\right)}_{i}I_{21}\right.\non\\
   & \qquad\left. -2 \left(p_1Np_3\right) {\left(p_2D\epsilon _3\right)}_{i}
   {\left(p_3\epsilon _2\right)}_{a}I'_{11}+4 \left(p_1Np_3\right) {\left(p_1N\epsilon
   _2\right)}_{a} {\left(p_2D\epsilon _3\right)}_{i}I_{5}\right.\non\\
   & \qquad\left. -4 \left(p_2Dp_3\right)
   {\left(p_1N\epsilon _2\right)}_{a} {\left(p_2D\epsilon _3\right)}_{i}I_{1}-2
   \left(p_1Np_3\right) {p_2}_{i} {\left(p_2\epsilon _3\epsilon _2\right)}_{a}I_{2}\right.\non\\
   & \qquad\left. +2
   \left(p_1Np_2\right) {p_3}_{i} {\left(p_2\epsilon _3\epsilon _2\right)}_{a}I_{2}-4
   {p_2}_{a} {\left(p_2\epsilon _3\epsilon _2\right)}_{i}I_{23}+2 \left(p_2p_3\right)
   {p_2}_{a} {\left(p_2\epsilon _3\epsilon _2\right)}_{i}I_{16}\right.\non\\
   & \qquad\left. -2 \left(p_1Np_3\right)
   {p_2}_{a} {\left(p_2\epsilon _3\epsilon _2\right)}_{i}I_{2}-4 {p_3}_{a}
   {\left(p_2\epsilon _3\epsilon _2\right)}_{i}I_{23}+2 \left(p_2p_3\right) {p_3}_{a}
   {\left(p_2\epsilon _3\epsilon _2\right)}_{i}I_{16}\right.\non\\
   & \qquad\left. +2 \left(p_1Np_2\right) {p_3}_{a}
   {\left(p_2\epsilon _3\epsilon _2\right)}_{i}I_{2}+2 \left(p_1N\epsilon _3p_2\right)
   {p_3}_{i} {\left(p_3D\epsilon _2\right)}_{a}I'_{11}\right.\non\\
   & \qquad\left. -2 \left(p_1Np_3\right)
   {\left(p_2\epsilon _3\right)}_{i} {\left(p_3D\epsilon _2\right)}_{a}I'_{11}+2
   \left(p_1Np_2\right) {\left(p_2D\epsilon _3\right)}_{i} {\left(p_3D\epsilon
   _2\right)}_{a}I_{1}\right.\non\\
   & \qquad\left. -2 \left(p_1Np_3\right) {\left(p_2D\epsilon _3\right)}_{i}
   {\left(p_3D\epsilon _2\right)}_{a}I_{1}-2 \left(p_1N\epsilon _3p_2\right) {p_2}_{a}
   {\left(p_3D\epsilon _2\right)}_{i}I'_{13}\right.\non\\
   & \qquad\left. -2 \left(p_2\epsilon _3Dp_3\right)
   {p_2}_{a} {\left(p_3D\epsilon _2\right)}_{i}I_{22}-2 \left(p_1N\epsilon _3p_2\right)
   {p_3}_{a} {\left(p_3D\epsilon _2\right)}_{i}I'_{11}\right.\non\\
   & \qquad\left. -2 \left(p_2\epsilon
   _3Dp_3\right) {p_3}_{a} {\left(p_3D\epsilon _2\right)}_{i}I_{22}-2
   \left(p_1Np_3\right) {\left(p_2\epsilon _3\right)}_{a} {\left(p_3D\epsilon
   _2\right)}_{i}I'_{11}\right.\non\\
   & \qquad\left. +4 \left(p_3D\epsilon _2p_3\right) {p_2}_{a}
   {\left(p_3D\epsilon _3\right)}_{i}I_{22}+4 \left(p_3D\epsilon _2p_3\right) {p_3}_{a}
   {\left(p_3D\epsilon _3\right)}_{i}I_{22}\right.\non\\
   & \qquad\left. +2 \left(p_1Np_3\right) \left(p_2p_3\right)
   {\left(\epsilon _2D\epsilon _3\right)}_{ai}I'_{11}-4 \left(p_1Np_2\right)
   \left(p_1Np_3\right) {\left(\epsilon _2D\epsilon _3\right)}_{ai}I_{5}\right.\non\\
   & \qquad\left. +2
   \left(p_1Np_2\right) \left(p_2Dp_3\right) {\left(\epsilon _2D\epsilon
   _3\right)}_{ai}I_{1}+2 \left(p_1Np_3\right) \left(p_2Dp_3\right) {\left(\epsilon
   _2D\epsilon _3\right)}_{ai}I_{1}\right.\non\\
   & \qquad\left. +2 \left(p_1Np_3\right) \left(p_2p_3\right)
   {\left(\epsilon _2D\epsilon _3\right)}_{ia}I'_{11}-2 \left(p_2p_3\right) {p_2}_{i}
   {\left(p_1N\epsilon _2\epsilon _3\right)}_{a}I_{2}\right.\non\\
   & \qquad\left. +4 \left(p_1Np_3\right) {p_2}_{i}
   {\left(p_1N\epsilon _2\epsilon _3\right)}_{a}I_{9}+2 \left(p_2p_3\right) {p_3}_{i}
   {\left(p_1N\epsilon _2\epsilon _3\right)}_{a}I_{2}\right.\non\\
   & \qquad\left. +2 \left(p_2Dp_3\right) {p_3}_{i}
   {\left(p_1N\epsilon _2\epsilon _3\right)}_{a}I'_{11}-2 \left(p_2p_3\right) {p_2}_{a}
   {\left(p_1N\epsilon _2\epsilon _3\right)}_{i}I_{2}\right.\non\\
   & \qquad\left. +4 \left(p_1Np_3\right) {p_2}_{a}
   {\left(p_1N\epsilon _2\epsilon _3\right)}_{i}I_{9}-2 \left(p_2p_3\right) {p_3}_{a}
   {\left(p_1N\epsilon _2\epsilon _3\right)}_{i}I_{2}\right.\non\\
   & \qquad\left. -2 \left(p_2Dp_3\right) {p_3}_{i}
   {\left(p_1N\epsilon _3\epsilon _2\right)}_{a}I'_{11}-2 \left(p_2Dp_3\right)
   {p_2}_{a} {\left(p_1N\epsilon _3\epsilon _2\right)}_{i}I'_{12}\right.\non\\
   & \qquad\left. -2
   \left(p_2Dp_3\right) {p_3}_{a} {\left(p_1N\epsilon _3\epsilon
   _2\right)}_{i}I'_{11}+2 \left(p_2p_3\right) {p_2}_{a} {\left(p_2D\epsilon _2\epsilon
   _3\right)}_{i}I_{20}\right.\non\\
   & \qquad\left. +4 \left(p_1Np_3\right) {p_2}_{a} {\left(p_2D\epsilon _2\epsilon
   _3\right)}_{i}I_{6}+2 \left(p_2p_3\right) {p_3}_{a} {\left(p_2D\epsilon _2\epsilon
   _3\right)}_{i}I_{20}\right.\non\\
   & \qquad\left. -2 \left(p_2Dp_3\right) {p_2}_{a} {\left(p_2D\epsilon _3\epsilon
   _2\right)}_{i}I_{15}-2 \left(p_2Dp_3\right) {p_3}_{a} {\left(p_2D\epsilon _3\epsilon
   _2\right)}_{i}I_{15}\right.\non\\
   & \qquad\left. +2 \left(p_2p_3\right) {p_2}_{a} {\left(p_2\epsilon _3D\epsilon
   _2\right)}_{i}I_{14}+2 \left(p_2p_3\right) {p_3}_{a} {\left(p_2\epsilon _3D\epsilon
   _2\right)}_{i}I_{14}\right.\non\\
   & \qquad\left. +2 \left(p_1Np_3\right) {p_2}_{i} {\left(p_3D\epsilon _2\epsilon
   _3\right)}_{a}I'_{11}-2 \left(p_1Np_2\right) {p_3}_{i} {\left(p_3D\epsilon
   _2\epsilon _3\right)}_{a}I'_{11}\right.\non\\
   & \qquad\left. +2 \left(p_1Np_3\right) {p_2}_{a}
   {\left(p_3D\epsilon _2\epsilon _3\right)}_{i}I'_{11}+2 \left(p_1Np_2\right)
   {p_3}_{a} {\left(p_3D\epsilon _2\epsilon _3\right)}_{i}I'_{11}\right.\non\\
   & \qquad\left. -2
   \left(p_2Dp_3\right) {p_2}_{a} {\left(p_3D\epsilon _3\epsilon _2\right)}_{i}I_{22}-2
   \left(p_2Dp_3\right) {p_3}_{a} {\left(p_3D\epsilon _3\epsilon _2\right)}_{i}I_{22}\right.\non
\end{align}
\begin{align}
   & \qquad\left. -2
   \left(p_1Np_3\right) {p_2}_{i} {\left(p_3\epsilon _2D\epsilon
   _3\right)}_{a}I'_{11}+2 \left(p_1Np_2\right) {p_3}_{i} {\left(p_3\epsilon
   _2D\epsilon _3\right)}_{a}I'_{11}\right.\non\\
   & \qquad\left. +2 \left(p_1Np_3\right) {p_2}_{a}
   {\left(p_3\epsilon _2D\epsilon _3\right)}_{i}I'_{11}-2 \left(p_1Np_2\right)
   {p_3}_{a} {\left(p_3\epsilon _2D\epsilon _3\right)}_{i}I'_{11}\right.\non\\
   & \qquad\left. +4
   \left(p_1Np_3\right) {p_2}_{i} {\left(p_1N\epsilon _2D\epsilon _3\right)}_{a}I_{5}-2
   \left(p_2Dp_3\right) {p_2}_{i} {\left(p_1N\epsilon _2D\epsilon _3\right)}_{a}I_{1}\right.\non\\
   & \qquad\left. -2
   \left(p_2p_3\right) {p_3}_{i} {\left(p_1N\epsilon _2D\epsilon
   _3\right)}_{a}I'_{11}-2 \left(p_2Dp_3\right) {p_3}_{i} {\left(p_1N\epsilon
   _2D\epsilon _3\right)}_{a}I_{1}\right.\non\\
   & \qquad\left. -4 \left(p_1Np_3\right) {p_2}_{a} {\left(p_1N\epsilon
   _2D\epsilon _3\right)}_{i}I_{5}+2 \left(p_2Dp_3\right) {p_2}_{a} {\left(p_1N\epsilon
   _2D\epsilon _3\right)}_{i}I_{1}\right.\non\\
   & \qquad\left. +2 \left(p_2Dp_3\right) {p_3}_{a} {\left(p_1N\epsilon
   _2D\epsilon _3\right)}_{i}I_{1}-2 \left(p_2p_3\right) {p_3}_{i} {\left(p_1N\epsilon
   _3D\epsilon _2\right)}_{a}I'_{11}\right.\non\\
   & \qquad\left. +2 \left(p_2p_3\right) {p_2}_{a}
   {\left(p_1N\epsilon _3D\epsilon _2\right)}_{i}I'_{12}+2 \left(p_2p_3\right)
   {p_3}_{a} {\left(p_1N\epsilon _3D\epsilon _2\right)}_{i}I'_{11}\right.\non\\
   & \qquad\left. -4
   \left(p_1Np_3\right) {p_2}_{a} {\left(p_2D\epsilon _2D\epsilon _3\right)}_{i}I_{7}-2
   \left(p_2Dp_3\right) {p_2}_{a} {\left(p_2D\epsilon _2D\epsilon
   _3\right)}_{i}I_{21}\right.\non\\
   & \qquad\left. -2 \left(p_2Dp_3\right) {p_3}_{a} {\left(p_2D\epsilon
   _2D\epsilon _3\right)}_{i}I_{21}+2 \left(p_1Np_3\right) {p_2}_{i}
   {\left(p_2D\epsilon _3D\epsilon _2\right)}_{a}I_{1}\right.\non\\
   & \qquad\left. -2 \left(p_1Np_2\right) {p_3}_{i}
   {\left(p_2D\epsilon _3D\epsilon _2\right)}_{a}I_{1}+4 {p_2}_{a} {\left(p_2D\epsilon
   _3D\epsilon _2\right)}_{i}I_{24}\right.\non\\
   & \qquad\left. +2 \left(p_1Np_3\right) {p_2}_{a}
   {\left(p_2D\epsilon _3D\epsilon _2\right)}_{i}I_{1}-2 \left(p_2Dp_3\right) {p_2}_{a}
   {\left(p_2D\epsilon _3D\epsilon _2\right)}_{i}I_{17}\right.\non\\
   & \qquad\left. +4 {p_3}_{a} {\left(p_2D\epsilon
   _3D\epsilon _2\right)}_{i}I_{24}+2 \left(p_1Np_2\right) {p_3}_{a}
   {\left(p_2D\epsilon _3D\epsilon _2\right)}_{i}I_{1}\right.\non\\
   & \qquad\left. -2 \left(p_2Dp_3\right) {p_3}_{a}
   {\left(p_2D\epsilon _3D\epsilon _2\right)}_{i}I_{17}+2 \left(p_2p_3\right) {p_2}_{a}
   {\left(p_3D\epsilon _3D\epsilon _2\right)}_{i}I_{22}\right.\non\\
   & \qquad\left. +2 \left(p_2p_3\right) {p_3}_{a}
   {\left(p_3D\epsilon _3D\epsilon _2\right)}_{i}I_{22}+2 \left(p_1Np_2\right)
   \left(p_2\epsilon _3p_2\right) {\epsilon _2}_{ai}I_{2}\right.\non\\
   & \qquad\left. +2 \left(p_1Np_3\right)
   \left(p_2\epsilon _3p_2\right) {\epsilon _2}_{ai}I_{2}-4 \left(p_1N\epsilon
   _3p_2\right) \left(p_2p_3\right) {\epsilon _2}_{ai}I_{2}\right.\non\\
   & \qquad\left. -8 \left(p_1N\epsilon
   _3p_2\right) \left(p_1Np_2\right) {\epsilon _2}_{ai}I_{9}+4 \left(p_1Np_2\right)
   \left(p_2\epsilon _3Dp_3\right) {\epsilon _2}_{ai}I'_{6}\right.\non\\
   & \qquad\left. -8 \left(p_1N\epsilon
   _3Dp_2\right) \left(p_1Np_2\right) {\epsilon _2}_{ai}I_{5}+4 \left(p_1N\epsilon
   _3Dp_2\right) \left(p_2Dp_3\right) {\epsilon _2}_{ai}I_{1}\right.\non\\
   & \qquad\left. -4 \left(p_1N\epsilon
   _3Dp_3\right) \left(p_2p_3\right) {\epsilon _2}_{ai}I'_{6}-4 \left(p_1N\epsilon
   _3Dp_3\right) \left(p_1Np_2\right) {\epsilon _2}_{ai}I'_{4}\right.\non\\
   & \qquad\left. -4 \left(p_1N\epsilon
   _3Dp_3\right) \left(p_2Dp_3\right) {\epsilon _2}_{ai}I'_{7}+4 \left(p_1N\epsilon
   _3Np_1\right) \left(p_2p_3\right) {\epsilon _2}_{ai}I_{9}\right.\non\\
   & \qquad\left. +8 \left(p_1N\epsilon
   _3Np_1\right) \left(p_1Np_2\right) {\epsilon _2}_{ai}I_{10}-4 \left(p_1N\epsilon
   _3Np_1\right) \left(p_2Dp_3\right) {\epsilon _2}_{ai}I_{5}\right.\non\\
   & \qquad\left. +2 \left(p_1Np_2\right)
   \left(p_2D\epsilon _3Dp_2\right) {\epsilon _2}_{ai}I_{1}-2 \left(p_1Np_3\right)
   \left(p_2D\epsilon _3Dp_2\right) {\epsilon _2}_{ai}I_{1}\right.\non\\
   & \qquad\left. -4 \left(p_1Np_2\right)
   \left(p_2D\epsilon _3Dp_3\right) {\epsilon _2}_{ai}I'_{7}-2 \tr(D\epsilon
   _3) \left(p_1Np_2\right) \left(p_2p_3\right) {\epsilon _2}_{ai}I'_{6}\right.\non\\
   & \qquad\left. +2
   \tr(D\epsilon _3) \left(p_1Np_3\right) \left(p_2p_3\right) {\epsilon
   _2}_{ai}I'_{6}+2 \tr(D\epsilon _3) \left(p_1Np_2\right)
   \left(p_1Np_3\right) {\epsilon _2}_{ai}I'_{4}\right.\non\\
   & \qquad\left. +2 \tr(D\epsilon _3)
   \left(p_1Np_2\right) \left(p_2Dp_3\right) {\epsilon _2}_{ai}I'_{7}+2
   \tr(D\epsilon _3) \left(p_1Np_3\right) \left(p_2Dp_3\right) {\epsilon
   _2}_{ai}I'_{7}\right.\non\\
   & \qquad\left. +4 \left(p_1Np_3\right) \left(p_3D\epsilon _2p_3\right) {\epsilon
   _3}_{ai}I'_{11}\rp+(2\leftrightarrow 3),
\end{align}


\begin{align}
\mathcal{A}_{C^{(p+1)}hh}^{(2)}=& \frac{2 i^{p (p+1)} \sqrt{2}}{(p-1)!}{{C}^{ij}}_{b_1...b_{p-1}}
   {\varepsilon}^{abb_1...b_{p-1}}\lp\tr(D\epsilon _2) \tr(D\epsilon
   _3) {p_2}_{b} {p_2}_{j} {p_3}_{a} {p_3}_{i}I_{3}\right.\non\\
   & \qquad\left. +\tr(\epsilon _2\epsilon
   _3) {p_2}_{b} {p_2}_{j} {p_3}_{a} {p_3}_{i}I_{2}-\tr(D\epsilon _2D\epsilon
   _3) {p_2}_{b} {p_2}_{j} {p_3}_{a} {p_3}_{i}I_{1}\right.\non\\
   & \qquad\left. -2 \tr(D\epsilon _2)
   {p_2}_{b} {p_2}_{j} {p_3}_{i} {\left(p_2\epsilon _3\right)}_{a}I_{6}-2
   \tr(D\epsilon _2) {p_2}_{j} {p_3}_{b} {p_3}_{i} {\left(p_2\epsilon
   _3\right)}_{a}I_{6}\right.\non\\
   & \qquad\left. +2 \tr(D\epsilon _2) {p_2}_{b} {p_2}_{j} {p_3}_{a}
   {\left(p_2\epsilon _3\right)}_{i}I_{6}+2 \tr(D\epsilon _2) {p_2}_{b}
   {p_3}_{a} {p_3}_{j} {\left(p_2\epsilon _3\right)}_{i}I_{6}\right.\non\\
   & \qquad\left. -2 {p_2}_{b} {p_2}_{j}
   {\left(p_2\epsilon _3\right)}_{i} {\left(p_3\epsilon _2\right)}_{a}I_{2}-2 {p_2}_{j}
   {p_3}_{b} {\left(p_2\epsilon _3\right)}_{i} {\left(p_3\epsilon _2\right)}_{a}I_{2}\right.\non
\end{align}
\begin{align}
   & \qquad\left. -2
   {p_2}_{b} {p_3}_{j} {\left(p_2\epsilon _3\right)}_{i} {\left(p_3\epsilon
   _2\right)}_{a}I_{2}-2 {p_3}_{b} {p_3}_{j} {\left(p_2\epsilon _3\right)}_{i}
   {\left(p_3\epsilon _2\right)}_{a}I_{2}+2 \left(p_2p_3\right) {p_2}_{b} {p_2}_{i}
   {\left(\epsilon _2\epsilon _3\right)}_{aj}I_{2}\right.\non\\
   & \qquad\left. -4 \left(p_1Np_3\right) {p_2}_{b}
   {p_2}_{i} {\left(\epsilon _2\epsilon _3\right)}_{aj}I_{9}-2 \left(p_2p_3\right)
   {p_2}_{b} {p_3}_{i} {\left(\epsilon _2\epsilon _3\right)}_{aj}I_{2}\right.\non\\
   & \qquad\left. -2
   \left(p_2Dp_3\right) {p_2}_{b} {p_3}_{i} {\left(\epsilon _2\epsilon
   _3\right)}_{aj}I'_{11}+2 \left(p_2p_3\right) {p_2}_{j} {p_3}_{a} {\left(\epsilon
   _2\epsilon _3\right)}_{bi}I_{2}\right.\non\\
   & \qquad\left. +2 \left(p_2p_3\right) {p_3}_{a} {p_3}_{i}
   {\left(\epsilon _2\epsilon _3\right)}_{bj}I_{2}+4 \left(p_1Np_2\right) {p_3}_{a}
   {p_3}_{i} {\left(\epsilon _2\epsilon _3\right)}_{bj}I_{9}\right.\non\\
   & \qquad\left. +2 \left(p_2Dp_3\right)
   {p_3}_{a} {p_3}_{i} {\left(\epsilon _2\epsilon _3\right)}_{bj}I'_{11}-4
   \left(p_2Dp_3\right) {p_2}_{b} {p_3}_{a} {\left(\epsilon _2\epsilon
   _3\right)}_{ij}I_{0}\right.\non\\
   & \qquad\left. +2 \left(p_2Dp_3\right) {p_2}_{b} {p_3}_{i} {\left(\epsilon
   _2\epsilon _3\right)}_{ja}I'_{11}-2 \left(p_2Dp_3\right) {p_3}_{a} {p_3}_{i}
   {\left(\epsilon _2\epsilon _3\right)}_{jb}I'_{11}\right.\non\\
   & \qquad\left. -4 {p_2}_{i} {p_3}_{j}
   {\left(p_1N\epsilon _2\right)}_{a} {\left(p_2\epsilon _3\right)}_{b}I_{9}+4
   {p_3}_{b} {p_3}_{i} {\left(p_1N\epsilon _2\right)}_{a} {\left(p_2\epsilon
   _3\right)}_{j}I_{9}\right.\non\\
   & \qquad\left. -2 \tr(D\epsilon _3) {p_2}_{j} {p_3}_{a} {p_3}_{i}
   {\left(p_1N\epsilon _2\right)}_{b}I'_{4}+4 {p_2}_{j} {p_3}_{a} {\left(p_1N\epsilon
   _2\right)}_{b} {\left(p_2\epsilon _3\right)}_{i}I_{9}\right.\non\\
   & \qquad\left. -2 \tr(D\epsilon _3)
   {p_2}_{b} {p_3}_{a} {p_3}_{j} {\left(p_1N\epsilon _2\right)}_{i}I'_{4}-4 {p_2}_{b}
   {p_3}_{j} {\left(p_1N\epsilon _2\right)}_{i} {\left(p_2\epsilon
   _3\right)}_{a}I_{9}\right.\non\\
   & \qquad\left. -4 {p_3}_{b} {p_3}_{j} {\left(p_1N\epsilon _2\right)}_{i}
   {\left(p_2\epsilon _3\right)}_{a}I_{9}+4 {p_2}_{b} {p_3}_{a} {\left(p_1N\epsilon
   _2\right)}_{i} {\left(p_2\epsilon _3\right)}_{j}I_{9}\right.\non\\
   & \qquad\left. +4 {p_2}_{j} {p_3}_{i}
   {\left(p_1N\epsilon _2\right)}_{b} {\left(p_1N\epsilon _3\right)}_{a}I_{10}+8
   {p_2}_{b} {p_3}_{j} {\left(p_1N\epsilon _2\right)}_{i} {\left(p_1N\epsilon
   _3\right)}_{a}I_{10}\right.\non\\
   & \qquad\left. +4 {p_2}_{b} {p_3}_{a} {\left(p_1N\epsilon _2\right)}_{j}
   {\left(p_1N\epsilon _3\right)}_{i}I_{10}+4 \tr(D\epsilon _3) {p_2}_{b}
   {p_3}_{a} {p_3}_{j} {\left(p_2D\epsilon _2\right)}_{i}I_{3}\right.\non\\
   & \qquad\left. -4 {p_2}_{b} {p_3}_{j}
   {\left(p_2D\epsilon _2\right)}_{i} {\left(p_2\epsilon _3\right)}_{a}I_{6}-4
   {p_3}_{b} {p_3}_{j} {\left(p_2D\epsilon _2\right)}_{i} {\left(p_2\epsilon
   _3\right)}_{a}I_{6}\right.\non\\
   & \qquad\left. +4 {p_2}_{b} {p_3}_{a} {\left(p_2D\epsilon _2\right)}_{i}
   {\left(p_2\epsilon _3\right)}_{j}I_{6}-4 {p_2}_{b} {p_3}_{j} {\left(p_1N\epsilon
   _3\right)}_{a} {\left(p_2D\epsilon _2\right)}_{i}I_{4}\right.\non\\
   & \qquad\left. +4 {p_2}_{b} {p_3}_{a}
   {\left(p_1N\epsilon _3\right)}_{j} {\left(p_2D\epsilon _2\right)}_{i}I_{4}-2
   \tr(D\epsilon _2) {p_2}_{b} {p_2}_{j} {p_3}_{i} {\left(p_2D\epsilon
   _3\right)}_{a}I_{7}\right.\non\\
   & \qquad\left. -2 \tr(D\epsilon _2) {p_2}_{j} {p_3}_{b} {p_3}_{i}
   {\left(p_2D\epsilon _3\right)}_{a}I_{7}-2 {p_2}_{b} {p_3}_{i} {\left(p_2D\epsilon
   _3\right)}_{a} {\left(p_3\epsilon _2\right)}_{j}I'_{11}\right.\non\\
   & \qquad\left. -2 {p_3}_{b} {p_3}_{i}
   {\left(p_2D\epsilon _3\right)}_{a} {\left(p_3\epsilon _2\right)}_{j}I'_{11}-4
   {p_2}_{j} {p_3}_{i} {\left(p_1N\epsilon _2\right)}_{b} {\left(p_2D\epsilon
   _3\right)}_{a}I_{5}\right.\non\\
   & \qquad\left. -4 {p_2}_{b} {p_3}_{j} {\left(p_1N\epsilon _2\right)}_{i}
   {\left(p_2D\epsilon _3\right)}_{a}I_{5}+4 {p_3}_{b} {p_3}_{i} {\left(p_1N\epsilon
   _2\right)}_{j} {\left(p_2D\epsilon _3\right)}_{a}I_{5}\right.\non\\
   & \qquad\left. -4 {p_2}_{b} {p_3}_{j}
   {\left(p_2D\epsilon _2\right)}_{i} {\left(p_2D\epsilon _3\right)}_{a}I_{7}-4
   {p_3}_{b} {p_3}_{j} {\left(p_2D\epsilon _2\right)}_{i} {\left(p_2D\epsilon
   _3\right)}_{a}I_{7}\right.\non\\
   & \qquad\left. +2 \tr(D\epsilon _2) {p_2}_{b} {p_2}_{j} {p_3}_{a}
   {\left(p_2D\epsilon _3\right)}_{i}I_{7}+4 {p_2}_{j} {p_3}_{a} {\left(p_1N\epsilon
   _2\right)}_{b} {\left(p_2D\epsilon _3\right)}_{i}I_{5}\right.\non\\
   & \qquad\left. -4 {p_2}_{b} {p_3}_{a}
   {\left(p_1N\epsilon _2\right)}_{j} {\left(p_2D\epsilon _3\right)}_{i}I_{5}-4
   {p_2}_{b} {p_3}_{a} {\left(p_2D\epsilon _2\right)}_{j} {\left(p_2D\epsilon
   _3\right)}_{i}I_{7}\right.\non\\
   & \qquad\left. +2 \tr(D\epsilon _2) {p_2}_{b} {p_3}_{a} {p_3}_{i}
   {\left(p_2D\epsilon _3\right)}_{j}I_{7}+2 {p_2}_{b} {p_3}_{i} {\left(p_2D\epsilon
   _3\right)}_{j} {\left(p_3\epsilon _2\right)}_{a}I'_{11}\right.\non\\
   & \qquad\left. +2 {p_3}_{b} {p_3}_{i}
   {\left(p_2D\epsilon _3\right)}_{j} {\left(p_3\epsilon _2\right)}_{a}I'_{11}+4
   {p_2}_{b} {p_3}_{a} {\left(p_2D\epsilon _3\right)}_{j} {\left(p_3\epsilon
   _2\right)}_{i}I_{8}\right.\non\\
   & \qquad\left. -4 {p_3}_{b} {p_3}_{i} {\left(p_1N\epsilon _2\right)}_{a}
   {\left(p_2D\epsilon _3\right)}_{j}I_{5}+2 {p_2}_{b} {p_2}_{j} {p_3}_{i}
   {\left(p_2\epsilon _3\epsilon _2\right)}_{a}I_{2}+2 {p_2}_{j} {p_3}_{b} {p_3}_{i}
   {\left(p_2\epsilon _3\epsilon _2\right)}_{a}I_{2}\right.\non\\
   & \qquad\left. -2 {p_2}_{b} {p_2}_{i} {p_3}_{a}
   {\left(p_2\epsilon _3\epsilon _2\right)}_{j}I_{2}-2 {p_2}_{b} {p_3}_{a} {p_3}_{i}
   {\left(p_2\epsilon _3\epsilon _2\right)}_{j}I_{2}+2 {p_2}_{b} {p_3}_{i}
   {\left(p_2\epsilon _3\right)}_{j} {\left(p_3D\epsilon _2\right)}_{a}I'_{11}\right.\non\\
   & \qquad\left. +2
   {p_3}_{b} {p_3}_{i} {\left(p_2\epsilon _3\right)}_{j} {\left(p_3D\epsilon
   _2\right)}_{a}I'_{11}+2 {p_2}_{b} {p_2}_{j} {\left(p_2D\epsilon _3\right)}_{i}
   {\left(p_3D\epsilon _2\right)}_{a}I_{1}\right.\non\\
   & \qquad\left. +2 {p_2}_{j} {p_3}_{b} {\left(p_2D\epsilon
   _3\right)}_{i} {\left(p_3D\epsilon _2\right)}_{a}I_{1}+2 {p_2}_{b} {p_3}_{i}
   {\left(p_2D\epsilon _3\right)}_{j} {\left(p_3D\epsilon _2\right)}_{a}I_{1}\right.\non\\
   & \qquad\left. +2
   {p_3}_{b} {p_3}_{i} {\left(p_2D\epsilon _3\right)}_{j} {\left(p_3D\epsilon
   _2\right)}_{a}I_{1}-2 {p_2}_{b} {p_3}_{j} {\left(p_2\epsilon _3\right)}_{a}
   {\left(p_3D\epsilon _2\right)}_{i}I'_{11}\right.\non\\
   & \qquad\left. -2 {p_3}_{b} {p_3}_{j} {\left(p_2\epsilon
   _3\right)}_{a} {\left(p_3D\epsilon _2\right)}_{i}I'_{11}+4 {p_2}_{b} {p_3}_{a}
   {\left(p_2D\epsilon _2\right)}_{j} {\left(p_3D\epsilon _3\right)}_{i}I_{3}\right.\non\\
   & \qquad\left. +4
   \left(p_1Np_3\right) {p_2}_{b} {p_2}_{i} {\left(\epsilon _2D\epsilon
   _3\right)}_{aj}I_{5}-2 \left(p_2Dp_3\right) {p_2}_{b} {p_2}_{i} {\left(\epsilon
   _2D\epsilon _3\right)}_{aj}I_{1}\right.\non\\
   & \qquad\left. -2 \left(p_2p_3\right) {p_2}_{b} {p_3}_{i}
   {\left(\epsilon _2D\epsilon _3\right)}_{aj}I'_{11}-2 \left(p_2Dp_3\right) {p_2}_{b}
   {p_3}_{i} {\left(\epsilon _2D\epsilon _3\right)}_{aj}I_{1}\right.\non
\end{align}
\begin{align}
   & \qquad\left. -2 \left(p_2Dp_3\right)
   {p_2}_{j} {p_3}_{a} {\left(\epsilon _2D\epsilon _3\right)}_{bi}I_{1}+2
   \left(p_2p_3\right) {p_3}_{a} {p_3}_{i} {\left(\epsilon _2D\epsilon
   _3\right)}_{bj}I'_{11}\right.\non\\
   & \qquad\left. -4 \left(p_1Np_2\right) {p_3}_{a} {p_3}_{i} {\left(\epsilon
   _2D\epsilon _3\right)}_{bj}I_{5}+2 \left(p_2Dp_3\right) {p_3}_{a} {p_3}_{i}
   {\left(\epsilon _2D\epsilon _3\right)}_{bj}I_{1}\right.\non\\
   & \qquad\left. -2 \left(p_2p_3\right) {p_2}_{b}
   {p_3}_{i} {\left(\epsilon _2D\epsilon _3\right)}_{ja}I'_{11}+2 \left(p_2p_3\right)
   {p_3}_{a} {p_3}_{i} {\left(\epsilon _2D\epsilon _3\right)}_{jb}I'_{11}\right.\non\\
   & \qquad\left. -4
   \left(p_2p_3\right) {p_2}_{b} {p_3}_{a} {\left(\epsilon _2D\epsilon
   _3\right)}_{ji}I_{0}+4 {p_2}_{j} {p_3}_{a} {p_3}_{i} {\left(p_1N\epsilon _2\epsilon
   _3\right)}_{b}I_{9}\right.\non\\
   & \qquad\left. +4 {p_2}_{b} {p_3}_{a} {p_3}_{i} {\left(p_1N\epsilon _2\epsilon
   _3\right)}_{j}I_{9}+4 {p_2}_{b} {p_3}_{a} {p_3}_{i} {\left(p_2D\epsilon _2\epsilon
   _3\right)}_{j}I_{6}-2 {p_2}_{b} {p_2}_{j} {p_3}_{i} {\left(p_3D\epsilon _2\epsilon
   _3\right)}_{a}I'_{11}\right.\non\\
   & \qquad\left. -2 {p_2}_{j} {p_3}_{b} {p_3}_{i} {\left(p_3D\epsilon _2\epsilon
   _3\right)}_{a}I'_{11}-2 {p_2}_{b} {p_2}_{i} {p_3}_{a} {\left(p_3D\epsilon _2\epsilon
   _3\right)}_{j}I'_{11}+2 {p_2}_{b} {p_3}_{a} {p_3}_{i} {\left(p_3D\epsilon _2\epsilon
   _3\right)}_{j}I'_{11}\right.\non\\
   & \qquad\left. +2 {p_2}_{b} {p_2}_{j} {p_3}_{i} {\left(p_3\epsilon _2D\epsilon
   _3\right)}_{a}I'_{11}+2 {p_2}_{j} {p_3}_{b} {p_3}_{i} {\left(p_3\epsilon _2D\epsilon
   _3\right)}_{a}I'_{11}+2 {p_2}_{b} {p_2}_{i} {p_3}_{a} {\left(p_3\epsilon _2D\epsilon
   _3\right)}_{j}I'_{11}\right.\non\\
   & \qquad\left. +2 {p_2}_{b} {p_3}_{a} {p_3}_{i} {\left(p_3\epsilon _2D\epsilon
   _3\right)}_{j}I'_{11}+4 {p_2}_{j} {p_3}_{a} {p_3}_{i} {\left(p_1N\epsilon
   _2D\epsilon _3\right)}_{b}I_{5}\right.\non\\
   & \qquad\left. -4 {p_2}_{b} {p_3}_{a} {p_3}_{i} {\left(p_1N\epsilon
   _2D\epsilon _3\right)}_{j}I_{5}-4 {p_2}_{b} {p_3}_{a} {p_3}_{i} {\left(p_2D\epsilon
   _2D\epsilon _3\right)}_{j}I_{7}\right.\non\\
   & \qquad\left. -2 {p_2}_{b} {p_2}_{j} {p_3}_{i} {\left(p_2D\epsilon
   _3D\epsilon _2\right)}_{a}I_{1}-2 {p_2}_{j} {p_3}_{b} {p_3}_{i} {\left(p_2D\epsilon
   _3D\epsilon _2\right)}_{a}I_{1}\right.\non\\
   & \qquad\left. -2 {p_2}_{b} {p_2}_{i} {p_3}_{a} {\left(p_2D\epsilon
   _3D\epsilon _2\right)}_{j}I_{1}+2 {p_2}_{b} {p_3}_{a} {p_3}_{i} {\left(p_2D\epsilon
   _3D\epsilon _2\right)}_{j}I_{1}\right.\non\\
   & \qquad\left. +4 \left(p_1Np_2\right) {p_3}_{j} {\left(p_2\epsilon
   _3\right)}_{b} {\epsilon _2}_{ai}I_{9}-4 \left(p_2p_3\right) {p_2}_{b}
   {\left(p_2\epsilon _3\right)}_{j} {\epsilon _2}_{ai}I_{2}\right.\non\\
   & \qquad\left. +4 \left(p_1Np_3\right)
   {p_2}_{b} {\left(p_2\epsilon _3\right)}_{j} {\epsilon _2}_{ai}I_{9}-4
   \left(p_2p_3\right) {p_3}_{b} {\left(p_2\epsilon _3\right)}_{j} {\epsilon
   _2}_{ai}I_{2}\right.\non\\
   & \qquad\left. -4 \left(p_1Np_2\right) {p_3}_{b} {\left(p_2\epsilon _3\right)}_{j}
   {\epsilon _2}_{ai}I_{9}-4 \left(p_2p_3\right) {p_3}_{j} {\left(p_1N\epsilon
   _3\right)}_{b} {\epsilon _2}_{ai}I_{9}\right.\non\\
   & \qquad\left. -8 \left(p_1Np_2\right) {p_3}_{j}
   {\left(p_1N\epsilon _3\right)}_{b} {\epsilon _2}_{ai}I_{10}+4 \left(p_2Dp_3\right)
   {p_3}_{j} {\left(p_1N\epsilon _3\right)}_{b} {\epsilon _2}_{ai}I_{5}\right.\non\\
   & \qquad\left. +4
   \left(p_2p_3\right) {p_2}_{b} {\left(p_1N\epsilon _3\right)}_{j} {\epsilon
   _2}_{ai}I_{9}-4 \left(p_2Dp_3\right) {p_2}_{b} {\left(p_1N\epsilon _3\right)}_{j}
   {\epsilon _2}_{ai}I_{5}\right.\non\\
   & \qquad\left. +4 \left(p_1Np_2\right) {p_3}_{j} {\left(p_2D\epsilon
   _3\right)}_{b} {\epsilon _2}_{ai}I_{5}-4 \left(p_2p_3\right) {p_2}_{b}
   {\left(p_3D\epsilon _3\right)}_{j} {\epsilon _2}_{ai}I'_{6}\right.\non\\
   & \qquad\left. -4 \left(p_2Dp_3\right)
   {p_2}_{b} {\left(p_3D\epsilon _3\right)}_{j} {\epsilon _2}_{ai}I'_{7}-2
   \left(p_2\epsilon _3p_2\right) {p_2}_{b} {p_2}_{i} {\epsilon _2}_{aj}I_{2}\right.\non\\
   & \qquad\left. +8
   \left(p_1N\epsilon _3p_2\right) {p_2}_{b} {p_2}_{i} {\epsilon _2}_{aj}I_{9}-4
   \left(p_2\epsilon _3Dp_3\right) {p_2}_{b} {p_2}_{i} {\epsilon _2}_{aj}I'_{6}\right.\non\\
   & \qquad\left. +8
   \left(p_1N\epsilon _3Dp_2\right) {p_2}_{b} {p_2}_{i} {\epsilon _2}_{aj}I_{5}+4
   \left(p_1N\epsilon _3Dp_3\right) {p_2}_{b} {p_2}_{i} {\epsilon _2}_{aj}I'_{4}\right.\non\\
   & \qquad\left. -8
   \left(p_1N\epsilon _3Np_1\right) {p_2}_{b} {p_2}_{i} {\epsilon _2}_{aj}I_{10}-2
   \left(p_2D\epsilon _3Dp_2\right) {p_2}_{b} {p_2}_{i} {\epsilon _2}_{aj}I_{1}\right.\non\\
   & \qquad\left. +4
   \left(p_2D\epsilon _3Dp_3\right) {p_2}_{b} {p_2}_{i} {\epsilon _2}_{aj}I'_{7}+2
   \tr(D\epsilon _3) \left(p_2p_3\right) {p_2}_{b} {p_2}_{i} {\epsilon
   _2}_{aj}I'_{6}\right.\non\\
   & \qquad\left. -2 \tr(D\epsilon _3) \left(p_1Np_3\right) {p_2}_{b}
   {p_2}_{i} {\epsilon _2}_{aj}I'_{4}-2 \tr(D\epsilon _3)
   \left(p_2Dp_3\right) {p_2}_{b} {p_2}_{i} {\epsilon _2}_{aj}I'_{7}\right.\non\\
   & \qquad\left. -2
   \left(p_2\epsilon _3p_2\right) {p_2}_{i} {p_3}_{b} {\epsilon _2}_{aj}I_{2}+4
   \left(p_1N\epsilon _3p_2\right) {p_2}_{i} {p_3}_{b} {\epsilon _2}_{aj}I_{9}-4
   \left(p_2\epsilon _3Dp_3\right) {p_2}_{i} {p_3}_{b} {\epsilon _2}_{aj}I'_{6}\right.\non\\
   & \qquad\left. +4
   \left(p_1N\epsilon _3Dp_2\right) {p_2}_{i} {p_3}_{b} {\epsilon _2}_{aj}I_{5}-2
   \left(p_2D\epsilon _3Dp_2\right) {p_2}_{i} {p_3}_{b} {\epsilon _2}_{aj}I_{1}\right.\non\\
   & \qquad\left. +4
   \left(p_2D\epsilon _3Dp_3\right) {p_2}_{i} {p_3}_{b} {\epsilon _2}_{aj}I'_{7}+2
   \tr(D\epsilon _3) \left(p_2p_3\right) {p_2}_{i} {p_3}_{b} {\epsilon
   _2}_{aj}I'_{6}\right.\non\\
   & \qquad\left. -2 \tr(D\epsilon _3) \left(p_2Dp_3\right) {p_2}_{i}
   {p_3}_{b} {\epsilon _2}_{aj}I'_{7}-2 \left(p_2\epsilon _3p_2\right) {p_2}_{b}
   {p_3}_{i} {\epsilon _2}_{aj}I_{2}\right.\non\\
   & \qquad\left. +4 \left(p_1N\epsilon _3p_2\right) {p_2}_{b}
   {p_3}_{i} {\epsilon _2}_{aj}I_{9}-4 \left(p_1N\epsilon _3Dp_2\right) {p_2}_{b}
   {p_3}_{i} {\epsilon _2}_{aj}I_{5}\right.\non\\
   & \qquad\left. +2 \left(p_2D\epsilon _3Dp_2\right) {p_2}_{b}
   {p_3}_{i} {\epsilon _2}_{aj}I_{1}-2 \tr(D\epsilon _3) \left(p_2p_3\right)
   {p_2}_{b} {p_3}_{i} {\epsilon _2}_{aj}I'_{6}\right.\non\\
   & \qquad\left. -2 \tr(D\epsilon _3)
   \left(p_2Dp_3\right) {p_2}_{b} {p_3}_{i} {\epsilon _2}_{aj}I'_{7}-2
   \left(p_2\epsilon _3p_2\right) {p_3}_{b} {p_3}_{i} {\epsilon _2}_{aj}I_{2}\right.\non\\
   & \qquad\left. +2
   \left(p_2D\epsilon _3Dp_2\right) {p_3}_{b} {p_3}_{i} {\epsilon _2}_{aj}I_{1}+4
   \left(p_1Np_3\right) {p_2}_{i} {\left(p_2\epsilon _3\right)}_{b} {\epsilon
   _2}_{aj}I_{9}\right.\non\\
   & \qquad\left. -4 \left(p_2p_3\right) {p_2}_{i} {\left(p_1N\epsilon _3\right)}_{b}
   {\epsilon _2}_{aj}I_{9}-4 \left(p_2Dp_3\right) {p_2}_{i} {\left(p_1N\epsilon
   _3\right)}_{b} {\epsilon _2}_{aj}I_{5}\right.\non
\end{align}
\begin{align}
   & \qquad\left. +4 \left(p_1Np_3\right) {p_2}_{i}
   {\left(p_2D\epsilon _3\right)}_{b} {\epsilon _2}_{aj}I_{5}+4 \left(p_1Np_3\right)
   {p_2}_{b} {\left(p_2D\epsilon _3\right)}_{i} {\epsilon _2}_{aj}I_{5}\right.\non\\
   & \qquad\left. -4
   \left(p_2Dp_3\right) {p_2}_{b} {\left(p_2D\epsilon _3\right)}_{i} {\epsilon
   _2}_{aj}I_{1}+4 \left(p_1Np_2\right) {p_3}_{b} {\left(p_2D\epsilon _3\right)}_{i}
   {\epsilon _2}_{aj}I_{5}\right.\non\\
   & \qquad\left. -4 \left(p_2Dp_3\right) {p_3}_{b} {\left(p_2D\epsilon
   _3\right)}_{i} {\epsilon _2}_{aj}I_{1}+2 \tr(D\epsilon _3)
   \left(p_2p_3\right) {p_3}_{a} {p_3}_{i} {\epsilon _2}_{bj}I'_{6}\right.\non\\
   & \qquad\left. +2 \tr(D\epsilon
   _3) \left(p_1Np_2\right) {p_3}_{a} {p_3}_{i} {\epsilon _2}_{bj}I'_{4}+2
   \tr(D\epsilon _3) \left(p_2Dp_3\right) {p_3}_{a} {p_3}_{i} {\epsilon
   _2}_{bj}I'_{7}\right.\non\\
   & \qquad\left. +4 \left(p_2p_3\right) {p_3}_{a} {\left(p_1N\epsilon _3\right)}_{i}
   {\epsilon _2}_{bj}I_{9}+8 \left(p_1Np_2\right) {p_3}_{a} {\left(p_1N\epsilon
   _3\right)}_{i} {\epsilon _2}_{bj}I_{10}\right.\non\\
   & \qquad\left. -4 \left(p_2Dp_3\right) {p_3}_{a}
   {\left(p_1N\epsilon _3\right)}_{i} {\epsilon _2}_{bj}I_{5}-4 \left(p_2p_3\right)
   {p_3}_{a} {\left(p_3D\epsilon _3\right)}_{i} {\epsilon _2}_{bj}I'_{6}\right.\non\\
   & \qquad\left. -4
   \left(p_1Np_2\right) {p_3}_{a} {\left(p_3D\epsilon _3\right)}_{i} {\epsilon
   _2}_{bj}I'_{4}-4 \left(p_2Dp_3\right) {p_3}_{a} {\left(p_3D\epsilon _3\right)}_{i}
   {\epsilon _2}_{bj}I'_{7}\right.\non\\
   & \qquad\left. -4 \left(p_3D\epsilon _2p_3\right) {p_2}_{b} {p_3}_{i}
   {\epsilon _3}_{aj}I'_{11}-4 \left(p_3D\epsilon _2p_3\right) {p_3}_{b} {p_3}_{i}
   {\epsilon _3}_{aj}I'_{11}\right.\non\\
   & \qquad\left. +4 \left(p_1Np_2\right) \left(p_2p_3\right) {\epsilon
   _2}_{aj} {\epsilon _3}_{bi}I_{9}-4 \left(p_1Np_2\right) \left(p_1Np_3\right)
   {\epsilon _2}_{aj} {\epsilon _3}_{bi}I_{10}\right.\non\\
   & \qquad\left. +4 \left(p_1Np_2\right)
   \left(p_2Dp_3\right) {\epsilon _2}_{aj} {\epsilon _3}_{bi}I_{5}\rp+(2\leftrightarrow 3),
\end{align}


\begin{align}
\mathcal{A}_{C^{(p+1)}hh}^{(3)}=& \frac{4 i^{p (p+1)} \sqrt{2}}{(p-2)!}{{C}^{ijk}}_{b_1...b_{p-2}}
   {\varepsilon}^{abcb_1...b_{p-2}}\lp -2 {p_2}_{c} {p_2}_{j} {p_3}_{a} {p_3}_{i}
   {\left(\epsilon _2\epsilon _3\right)}_{bk}I_{9}\right.\non\\
   & \qquad\left. +2 {p_2}_{c} {p_2}_{j} {p_3}_{a}
   {p_3}_{i} {\left(\epsilon _2D\epsilon _3\right)}_{bk}I_{5}-2 {p_2}_{c} {p_2}_{i}
   {p_3}_{j} {\left(p_2\epsilon _3\right)}_{b} {\epsilon _2}_{ak}I_{9}\right.\non\\
   & \qquad\left. -2 {p_2}_{i}
   {p_3}_{c} {p_3}_{j} {\left(p_2\epsilon _3\right)}_{b} {\epsilon _2}_{ak}I_{9}+4
   {p_2}_{c} {p_2}_{i} {p_3}_{j} {\left(p_1N\epsilon _3\right)}_{b} {\epsilon
   _2}_{ak}I_{10}\right.\non\\
   & \qquad\left. -2 {p_2}_{c} {p_2}_{i} {p_3}_{j} {\left(p_2D\epsilon _3\right)}_{b}
   {\epsilon _2}_{ak}I_{5}-2 {p_2}_{i} {p_3}_{c} {p_3}_{j} {\left(p_2D\epsilon
   _3\right)}_{b} {\epsilon _2}_{ak}I_{5}\right.\non\\
   & \qquad\left. +2 {p_2}_{c} {p_2}_{i} {p_3}_{a}
   {\left(p_2\epsilon _3\right)}_{k} {\epsilon _2}_{bj}I_{9}-2 {p_2}_{a} {p_3}_{c}
   {p_3}_{i} {\left(p_2\epsilon _3\right)}_{k} {\epsilon _2}_{bj}I_{9}\right.\non\\
   & \qquad\left. -4 {p_2}_{c}
   {p_2}_{i} {p_3}_{a} {\left(p_1N\epsilon _3\right)}_{k} {\epsilon _2}_{bj}I_{10}+2
   {p_2}_{c} {p_2}_{i} {p_3}_{a} {\left(p_2D\epsilon _3\right)}_{k} {\epsilon
   _2}_{bj}I_{5}\right.\non\\
   & \qquad\left. -2 {p_2}_{c} {p_3}_{a} {p_3}_{i} {\left(p_2D\epsilon _3\right)}_{k}
   {\epsilon _2}_{bj}I_{5}+2 {p_2}_{c} {p_2}_{i} {p_3}_{a} {\left(p_3D\epsilon
   _3\right)}_{k} {\epsilon _2}_{bj}I'_{4}\right.\non\\
   & \qquad\left. -\tr(D\epsilon _3) {p_2}_{c}
   {p_2}_{j} {p_3}_{a} {p_3}_{i} {\epsilon _2}_{bk}I'_{4}-2 \left(p_2p_3\right)
   {p_2}_{c} {p_2}_{i} {\epsilon _2}_{aj} {\epsilon _3}_{bk}I_{9}\right.\non\\
   & \qquad\left. +4
   \left(p_1Np_3\right) {p_2}_{c} {p_2}_{i} {\epsilon _2}_{aj} {\epsilon
   _3}_{bk}I_{10}-2 \left(p_2Dp_3\right) {p_2}_{c} {p_2}_{i} {\epsilon _2}_{aj}
   {\epsilon _3}_{bk}I_{5}\right.\non\\
   & \qquad\left. +2 \left(p_2p_3\right) {p_2}_{c} {p_3}_{i} {\epsilon _2}_{aj}
   {\epsilon _3}_{bk}I_{9}-2 \left(p_2Dp_3\right) {p_2}_{c} {p_3}_{i} {\epsilon
   _2}_{aj} {\epsilon _3}_{bk}I_{5}\rp+(2\leftrightarrow 3),
\end{align}


\be
\mathcal{A}_{C^{(p+1)}hh}^{(4)}=\frac{8 i^{p (p+1)} \sqrt{2}}{(p-3)!}{{C}^{ijkl}}_{b_1...b_{p-3}}
   {\varepsilon}^{abcdb_1...b_{p-3}}{p_2}_{d} {p_2}_{j} {p_3}_{a} {p_3}_{i} {\epsilon
   _2}_{bk} {\epsilon _3}_{cl}I_{10}+(2\leftrightarrow 3).
\ee

\subsection{$C^{(p-1)}$ amplitudes}


\be
\mathcal{A}_{C^{(p-1)}Bh}=\mathcal{A}_{C^{(p-1)}Bh}^{(0)}+\mathcal{A}_{C^{(p-1)}Bh}^{(1)}+\mathcal
{A}_{C^{(p-1)}Bh}^{(2)}+\mathcal{A}_{C^{(p-1)}Bh}^{(3)}.
\ee


\begin{align}
\mathcal{A}_{C^{(p-1)}Bh}^{(0)}=& \frac{2 i^{p (p+1)} \sqrt{2}}{(p-1)!}{C}_{b_1...b_{p-1}}
   {\varepsilon}^{abb_1...b_{p-1}}\lp 2 \left(p_1N\epsilon _2\epsilon _3p_2\right)
   {p_2}_{b} {p_3}_{a}I_{2}\right.\non\\
   & \qquad\left. +2 \left(p_1N\epsilon _3\epsilon _2p_3\right) {p_2}_{b}
   {p_3}_{a}I_{2}-2 \left(p_2D\epsilon _2\epsilon _3p_2\right) {p_2}_{b}
   {p_3}_{a}I_{20}\right.\non\\
   & \qquad\left. -2 \left(p_2D\epsilon _3\epsilon _2p_3\right) {p_2}_{b}
   {p_3}_{a}I_{14}+2 \left(p_2\epsilon _3\epsilon _2Dp_3\right) {p_2}_{b}
   {p_3}_{a}I_{14}\right.\non\\
   & \qquad\left. +2 \left(p_1N\epsilon _2D\epsilon _3p_2\right) {p_2}_{b}
   {p_3}_{a}I_{11}+2 \left(p_1N\epsilon _2\epsilon _3Dp_2\right) {p_2}_{b}
   {p_3}_{a}I_{11}\right.\non\\
   & \qquad\left. +4 \left(p_1N\epsilon _2\epsilon _3Dp_3\right) {p_2}_{b}
   {p_3}_{a}I'_{6}-4 \left(p_1N\epsilon _2\epsilon _3Np_1\right) {p_2}_{b}
   {p_3}_{a}I_{9}\right.\non\\
   & \qquad\left. -2 \left(p_1N\epsilon _3D\epsilon _2p_3\right) {p_2}_{b}
   {p_3}_{a}I'_{11}+4 \left(p_1N\epsilon _3\epsilon _2Dp_2\right) {p_2}_{b}
   {p_3}_{a}I_{6}\right.\non\\
   & \qquad\left. +2 \left(p_1N\epsilon _3\epsilon _2Dp_3\right) {p_2}_{b}
   {p_3}_{a}I'_{11}+2 \left(p_2D\epsilon _2D\epsilon _3p_2\right) {p_2}_{b}
   {p_3}_{a}I'_{22}\right.\non\\
   & \qquad\left. +2 \left(p_2D\epsilon _2\epsilon _3Dp_2\right) {p_2}_{b}
   {p_3}_{a}I'_{22}-4 \left(p_2D\epsilon _2\epsilon _3Dp_3\right) {p_2}_{b}
   {p_3}_{a}I_{18}\right.\non\\
   & \qquad\left. -2 \left(p_2D\epsilon _3D\epsilon _2p_3\right) {p_2}_{b}
   {p_3}_{a}I_{15}+2 \left(p_2\epsilon _3D\epsilon _2Dp_3\right) {p_2}_{b}
   {p_3}_{a}I_{15}\right.\non\\
   & \qquad\left. -2 \left(p_3D\epsilon _2\epsilon _3Dp_3\right) {p_2}_{b}
   {p_3}_{a}I_{22}+2 \left(p_1N\epsilon _2D\epsilon _3Dp_2\right) {p_2}_{b}
   {p_3}_{a}I_{1}\right.\non\\
   & \qquad\left. -4 \left(p_1N\epsilon _2D\epsilon _3Dp_3\right) {p_2}_{b}
   {p_3}_{a}I'_{7}-4 \left(p_1N\epsilon _2D\epsilon _3Np_1\right) {p_2}_{b}
   {p_3}_{a}I_{5}\right.\non\\
   & \qquad\left. +4 \left(p_1N\epsilon _3D\epsilon _2Dp_2\right) {p_2}_{b}
   {p_3}_{a}I_{7}-2 \left(p_1N\epsilon _3D\epsilon _2Dp_3\right) {p_2}_{b}
   {p_3}_{a}I_{1}\right.\non\\
   & \qquad\left. -2 \left(p_2D\epsilon _2D\epsilon _3Dp_2\right) {p_2}_{b}
   {p_3}_{a}I_{21}-4 \left(p_2D\epsilon _2D\epsilon _3Dp_3\right) {p_2}_{b}
   {p_3}_{a}I_{19}\right.\non\\
   & \qquad\left. -2 \tr(D\epsilon _3) \left(p_1N\epsilon _2p_3\right)
   {p_2}_{b} {p_3}_{a}I'_{6}+2 \tr(D\epsilon _3) \left(p_2D\epsilon
   _2p_3\right) {p_2}_{b} {p_3}_{a}I_{18}\right.\non\\
   & \qquad\left. +2 \tr(D\epsilon _3)
   \left(p_3D\epsilon _2p_3\right) {p_2}_{b} {p_3}_{a}I_{22}+2 \tr(D\epsilon
   _3) \left(p_1N\epsilon _2Dp_3\right) {p_2}_{b} {p_3}_{a}I'_{7}\right.\non\\
   & \qquad\left. +2
   \tr(D\epsilon _3) \left(p_2D\epsilon _2Dp_3\right) {p_2}_{b}
   {p_3}_{a}I_{19}+2 \left(p_3D\epsilon _3\epsilon _2p_3\right) {p_2}_{a}
   {p_3}_{b}I'_{20}\right.\non\\
   & \qquad\left. +2 \left(p_3D\epsilon _3D\epsilon _2p_3\right) {p_2}_{a}
   {p_3}_{b}I_{22}+2 \left(p_3D\epsilon _2D\epsilon _3Dp_3\right) {p_2}_{a}
   {p_3}_{b}I'_{21}\right.\non\\
   & \qquad\left. -2 \left(p_1N\epsilon _2p_3\right) {p_2}_{b} {\left(p_2\epsilon
   _3\right)}_{a}I_{2}+2 \left(p_2D\epsilon _2p_3\right) {p_2}_{b} {\left(p_2\epsilon
   _3\right)}_{a}I_{20}\right.\non\\
   & \qquad\left. +2 \left(p_3D\epsilon _2p_3\right) {p_2}_{b} {\left(p_2\epsilon
   _3\right)}_{a}I_{14}-2 \left(p_1N\epsilon _2Dp_3\right) {p_2}_{b} {\left(p_2\epsilon
   _3\right)}_{a}I_{11}\right.\non\\
   & \qquad\left. -2 \left(p_2D\epsilon _2Dp_3\right) {p_2}_{b} {\left(p_2\epsilon
   _3\right)}_{a}I'_{22}-2 \left(p_1N\epsilon _2p_3\right) {p_3}_{b} {\left(p_2\epsilon
   _3\right)}_{a}I_{2}\right.\non\\
   & \qquad\left. +2 \left(p_2D\epsilon _2p_3\right) {p_3}_{b} {\left(p_2\epsilon
   _3\right)}_{a}I_{20}+2 \left(p_3D\epsilon _2p_3\right) {p_3}_{b} {\left(p_2\epsilon
   _3\right)}_{a}I_{14}\right.\non\\
   & \qquad\left. -2 \left(p_2D\epsilon _2Dp_3\right) {p_3}_{b} {\left(p_2\epsilon
   _3\right)}_{a}I'_{22}+2 \left(p_1N\epsilon _2Dp_3\right) {p_3}_{a}
   {\left(p_2\epsilon _3\right)}_{b}I_{12}\right.\non\\
   & \qquad\left. -2 \left(p_1N\epsilon _3p_2\right) {p_2}_{b}
   {\left(p_3\epsilon _2\right)}_{a}I_{2}+2 \left(p_2D\epsilon _3p_2\right) {p_2}_{b}
   {\left(p_3\epsilon _2\right)}_{a}I_{14}\right.\non\\
   & \qquad\left. +2 \left(p_2\epsilon _3Dp_3\right) {p_2}_{b}
   {\left(p_3\epsilon _2\right)}_{a}I'_{20}+2 \left(p_1N\epsilon _3Dp_2\right)
   {p_2}_{b} {\left(p_3\epsilon _2\right)}_{a}I'_{13}\right.\non\\
   & \qquad\left. -4 \left(p_1N\epsilon
   _3Dp_3\right) {p_2}_{b} {\left(p_3\epsilon _2\right)}_{a}I'_{6}+4 \left(p_1N\epsilon
   _3Np_1\right) {p_2}_{b} {\left(p_3\epsilon _2\right)}_{a}I_{9}\right.\non\\
   & \qquad\left. +2 \left(p_2D\epsilon
   _3Dp_2\right) {p_2}_{b} {\left(p_3\epsilon _2\right)}_{a}I_{15}+2 \left(p_2D\epsilon
   _3Dp_3\right) {p_2}_{b} {\left(p_3\epsilon _2\right)}_{a}I_{22}\right.\non\\
   & \qquad\left. -2 \tr(D\epsilon
   _3) \left(p_2p_3\right) {p_2}_{b} {\left(p_3\epsilon _2\right)}_{a}I'_{20}+2
   \tr(D\epsilon _3) \left(p_1Np_3\right) {p_2}_{b} {\left(p_3\epsilon
   _2\right)}_{a}I'_{6}\right.\non\\
   & \qquad\left. -2 \left(p_1N\epsilon _3p_2\right) {p_3}_{b} {\left(p_3\epsilon
   _2\right)}_{a}I_{2}+2 \left(p_2D\epsilon _3p_2\right) {p_3}_{b} {\left(p_3\epsilon
   _2\right)}_{a}I_{14}\right.\non\\
   & \qquad\left. +2 \left(p_2\epsilon _3Dp_3\right) {p_3}_{b} {\left(p_3\epsilon
   _2\right)}_{a}I'_{20}+2 \left(p_1N\epsilon _3Dp_2\right) {p_3}_{b}
   {\left(p_3\epsilon _2\right)}_{a}I'_{11}\right.\non\\
   & \qquad\left. +2 \left(p_2D\epsilon _3Dp_2\right)
   {p_3}_{b} {\left(p_3\epsilon _2\right)}_{a}I_{15}+2 \left(p_2D\epsilon _3Dp_3\right)
   {p_3}_{b} {\left(p_3\epsilon _2\right)}_{a}I_{22}\right.\non\\
   & \qquad\left. -2 \tr(D\epsilon _3)
   \left(p_2p_3\right) {p_3}_{b} {\left(p_3\epsilon _2\right)}_{a}I'_{20}-2
   \tr(D\epsilon_3) \left(p_1Np_2\right) {p_3}_{b} {\left(p_3\epsilon
   _2\right)}_{a}I'_{6}\right.\non\\
   & \qquad\left. -2 \left(p_1Np_2\right) {\left(p_2\epsilon _3\right)}_{b}
   {\left(p_3\epsilon _2\right)}_{a}I_{2}-2 \left(p_1Np_3\right) {\left(p_2\epsilon
   _3\right)}_{b} {\left(p_3\epsilon _2\right)}_{a}I_{2}\right.\non
\end{align}
\begin{align}
   & \qquad\left. -2 \left(p_1Np_2\right)
   \left(p_2p_3\right) {\left(\epsilon _2\epsilon _3\right)}_{ab}I_{2}+2
   \left(p_1Np_3\right) \left(p_2p_3\right) {\left(\epsilon _2\epsilon
   _3\right)}_{ab}I_{2}\right.\non\\
   & \qquad\left. +4 \left(p_1Np_2\right) \left(p_1Np_3\right) {\left(\epsilon
   _2\epsilon _3\right)}_{ab}I_{9}-2 \left(p_1Np_2\right) \left(p_2Dp_3\right)
   {\left(\epsilon _2\epsilon _3\right)}_{ab}I_{11}\right.\non\\
   & \qquad\left. +2 \left(p_1Np_3\right)
   \left(p_2Dp_3\right) {\left(\epsilon _2\epsilon _3\right)}_{ab}I'_{11}+2
   \left(p_2\epsilon _3p_2\right) {p_2}_{b} {\left(p_1N\epsilon _2\right)}_{a}I_{2}\right.\non\\
   & \qquad\left. -8
   \left(p_1N\epsilon _3p_2\right) {p_2}_{b} {\left(p_1N\epsilon _2\right)}_{a}I_{9}+4
   \left(p_2D\epsilon _3p_2\right) {p_2}_{b} {\left(p_1N\epsilon _2\right)}_{a}I_{11}\right.\non\\
   & \qquad\left. +4
   \left(p_2\epsilon _3Dp_3\right) {p_2}_{b} {\left(p_1N\epsilon _2\right)}_{a}I'_{6}-8
   \left(p_1N\epsilon _3Dp_2\right) {p_2}_{b} {\left(p_1N\epsilon _2\right)}_{a}I_{5}\right.\non\\
   & \qquad\left. -4
   \left(p_1N\epsilon _3Dp_3\right) {p_2}_{b} {\left(p_1N\epsilon
   _2\right)}_{a}I'_{4}+8 \left(p_1N\epsilon _3Np_1\right) {p_2}_{b}
   {\left(p_1N\epsilon _2\right)}_{a}I_{10}\right.\non\\
   & \qquad\left. +2 \left(p_2D\epsilon _3Dp_2\right)
   {p_2}_{b} {\left(p_1N\epsilon _2\right)}_{a}I_{1}-4 \left(p_2D\epsilon _3Dp_3\right)
   {p_2}_{b} {\left(p_1N\epsilon _2\right)}_{a}I'_{7}\right.\non\\
   & \qquad\left. -2 \tr(D\epsilon _3)
   \left(p_2p_3\right) {p_2}_{b} {\left(p_1N\epsilon _2\right)}_{a}I'_{6}+2
   \tr(D\epsilon _3) \left(p_1Np_3\right) {p_2}_{b} {\left(p_1N\epsilon
   _2\right)}_{a}I'_{4}\right.\non\\
   & \qquad\left. +2 \tr(D\epsilon _3) \left(p_2Dp_3\right) {p_2}_{b}
   {\left(p_1N\epsilon _2\right)}_{a}I'_{7}-4 \left(p_1N\epsilon _3p_2\right) {p_3}_{b}
   {\left(p_1N\epsilon _2\right)}_{a}I_{9}\right.\non\\
   & \qquad\left. +4 \left(p_2D\epsilon _3p_2\right) {p_3}_{b}
   {\left(p_1N\epsilon _2\right)}_{a}I_{11}+4 \left(p_2\epsilon _3Dp_3\right) {p_3}_{b}
   {\left(p_1N\epsilon _2\right)}_{a}I'_{6}\right.\non\\
   & \qquad\left. -4 \left(p_1N\epsilon _3Dp_2\right)
   {p_3}_{b} {\left(p_1N\epsilon _2\right)}_{a}I_{5}-4 \left(p_2D\epsilon _3Dp_3\right)
   {p_3}_{b} {\left(p_1N\epsilon _2\right)}_{a}I'_{7}\right.\non\\
   & \qquad\left. -2 \tr(D\epsilon _3)
   \left(p_2p_3\right) {p_3}_{b} {\left(p_1N\epsilon _2\right)}_{a}I'_{6}+2
   \tr(D\epsilon _3) \left(p_2Dp_3\right) {p_3}_{b} {\left(p_1N\epsilon
   _2\right)}_{a}I'_{7}\right.\non\\
   & \qquad\left. -4 \left(p_1Np_3\right) {\left(p_1N\epsilon _2\right)}_{a}
   {\left(p_2\epsilon _3\right)}_{b}I_{9}-2 \left(p_2\epsilon _3p_2\right) {p_3}_{a}
   {\left(p_1N\epsilon _2\right)}_{b}I_{2}\right.\non\\
   & \qquad\left. -2 \left(p_2D\epsilon _3Dp_2\right) {p_3}_{a}
   {\left(p_1N\epsilon _2\right)}_{b}I_{1}+4 \left(p_1N\epsilon _2p_3\right) {p_2}_{b}
   {\left(p_1N\epsilon _3\right)}_{a}I_{9}\right.\non\\
   & \qquad\left. +4 \left(p_2D\epsilon _2p_3\right) {p_2}_{b}
   {\left(p_1N\epsilon _3\right)}_{a}I_{6}+4 \left(p_3D\epsilon _2p_3\right) {p_2}_{b}
   {\left(p_1N\epsilon _3\right)}_{a}I'_{11}\right.\non\\
   & \qquad\left. +4 \left(p_1N\epsilon _2Dp_3\right)
   {p_2}_{b} {\left(p_1N\epsilon _3\right)}_{a}I_{5}+4 \left(p_2D\epsilon _2Dp_3\right)
   {p_2}_{b} {\left(p_1N\epsilon _3\right)}_{a}I_{7}\right.\non\\
   & \qquad\left. -4 \left(p_2p_3\right)
   {\left(p_1N\epsilon _3\right)}_{a} {\left(p_3\epsilon _2\right)}_{b}I_{2}-4
   \left(p_1Np_2\right) {\left(p_1N\epsilon _3\right)}_{a} {\left(p_3\epsilon
   _2\right)}_{b}I_{9}\right.\non\\
   & \qquad\left. +4 \left(p_2p_3\right) {\left(p_1N\epsilon _2\right)}_{a}
   {\left(p_1N\epsilon _3\right)}_{b}I_{9}+4 \left(p_2Dp_3\right) {\left(p_1N\epsilon
   _2\right)}_{a} {\left(p_1N\epsilon _3\right)}_{b}I_{5}\right.\non\\
   & \qquad\left. -2 \left(p_2\epsilon
   _3p_2\right) {p_2}_{b} {\left(p_2D\epsilon _2\right)}_{a}I_{20}-8 \left(p_1N\epsilon
   _3p_2\right) {p_2}_{b} {\left(p_2D\epsilon _2\right)}_{a}I_{6}\right.\non\\
   & \qquad\left. +4 \left(p_2D\epsilon
   _3p_2\right) {p_2}_{b} {\left(p_2D\epsilon _2\right)}_{a}I'_{22}-4 \left(p_2\epsilon
   _3Dp_3\right) {p_2}_{b} {\left(p_2D\epsilon _2\right)}_{a}I_{18}\right.\non\\
   & \qquad\left. -8\left(p_1N\epsilon _3Dp_2\right) {p_2}_{b} {\left(p_2D\epsilon _2\right)}_{a}I_{7}+8
   \left(p_1N\epsilon _3Dp_3\right) {p_2}_{b} {\left(p_2D\epsilon _2\right)}_{a}I_{3}\right.\non\\
   & \qquad\left. -4
   \left(p_1N\epsilon _3Np_1\right) {p_2}_{b} {\left(p_2D\epsilon _2\right)}_{a}I_{4}-2
   \left(p_2D\epsilon _3Dp_2\right) {p_2}_{b} {\left(p_2D\epsilon
   _2\right)}_{a}I_{21}\right.\non\\
   & \qquad\left. -4 \left(p_2D\epsilon _3Dp_3\right) {p_2}_{b}
   {\left(p_2D\epsilon _2\right)}_{a}I_{19}+2 \tr(D\epsilon _3)
   \left(p_2p_3\right) {p_2}_{b} {\left(p_2D\epsilon _2\right)}_{a}I_{18}\right.\non\\
   & \qquad\left. -4
   \tr(D\epsilon _3) \left(p_1Np_3\right) {p_2}_{b} {\left(p_2D\epsilon
   _2\right)}_{a}I_{3}+2 \tr(D\epsilon _3) \left(p_2Dp_3\right) {p_2}_{b}
   {\left(p_2D\epsilon _2\right)}_{a}I_{19}\right.\non\\
   & \qquad\left. -2 \left(p_2\epsilon _3p_2\right) {p_3}_{b}
   {\left(p_2D\epsilon _2\right)}_{a}I_{20}-4 \left(p_1N\epsilon _3p_2\right) {p_3}_{b}
   {\left(p_2D\epsilon _2\right)}_{a}I_{6}\right.\non\\
   & \qquad\left. +4 \left(p_2D\epsilon _3p_2\right) {p_3}_{b}
   {\left(p_2D\epsilon _2\right)}_{a}I'_{22}-4 \left(p_2\epsilon _3Dp_3\right)
   {p_3}_{b} {\left(p_2D\epsilon _2\right)}_{a}I_{18}\right.\non\\
   & \qquad\left. -4 \left(p_1N\epsilon
   _3Dp_2\right) {p_3}_{b} {\left(p_2D\epsilon _2\right)}_{a}I_{7}-2 \left(p_2D\epsilon
   _3Dp_2\right) {p_3}_{b} {\left(p_2D\epsilon _2\right)}_{a}I_{21}\right.\non\\
   & \qquad\left. -4
   \left(p_2D\epsilon _3Dp_3\right) {p_3}_{b} {\left(p_2D\epsilon
   _2\right)}_{a}I_{19}+2 \tr(D\epsilon _3) \left(p_2p_3\right) {p_3}_{b}
   {\left(p_2D\epsilon _2\right)}_{a}I_{18}\right.\non\\
   & \qquad\left. +2 \tr(D\epsilon _3)
   \left(p_2Dp_3\right) {p_3}_{b} {\left(p_2D\epsilon _2\right)}_{a}I_{19}-4
   \left(p_1Np_3\right) {\left(p_2D\epsilon _2\right)}_{a} {\left(p_2\epsilon
   _3\right)}_{b}I_{6}\right.\non\\
   & \qquad\left. +4 \left(p_2p_3\right) {\left(p_1N\epsilon _3\right)}_{b}
   {\left(p_2D\epsilon _2\right)}_{a}I_{6}+4 \left(p_2Dp_3\right) {\left(p_1N\epsilon
   _3\right)}_{b} {\left(p_2D\epsilon _2\right)}_{a}I_{7}\right.\non\\
   & \qquad\left. -2 \left(p_1N\epsilon
   _2p_3\right) {p_2}_{b} {\left(p_2D\epsilon _3\right)}_{a}I_{11}-2 \left(p_2D\epsilon
   _2p_3\right) {p_2}_{b} {\left(p_2D\epsilon _3\right)}_{a}I'_{22}\right.\non\\
   & \qquad\left. +2
   \left(p_3D\epsilon _2p_3\right) {p_2}_{b} {\left(p_2D\epsilon _3\right)}_{a}I_{15}-2
   \left(p_1N\epsilon _2Dp_3\right) {p_2}_{b} {\left(p_2D\epsilon _3\right)}_{a}I_{1}\right.\non\\
   & \qquad\left. +2
   \left(p_2D\epsilon _2Dp_3\right) {p_2}_{b} {\left(p_2D\epsilon
   _3\right)}_{a}I_{21}-2 \left(p_2D\epsilon _2p_3\right) {p_3}_{b} {\left(p_2D\epsilon
   _3\right)}_{a}I'_{22}\right.\non\\
   & \qquad\left. +2 \left(p_3D\epsilon _2p_3\right) {p_3}_{b}
   {\left(p_2D\epsilon _3\right)}_{a}I_{15}-2 \left(p_1N\epsilon _2Dp_3\right)
   {p_3}_{b} {\left(p_2D\epsilon _3\right)}_{a}I_{1}\right.\non
\end{align}
\begin{align}
   & \qquad\left. +2 \left(p_2D\epsilon _2Dp_3\right)
   {p_3}_{b} {\left(p_2D\epsilon _3\right)}_{a}I_{21}+4 \left(p_1Np_3\right)
   {\left(p_2D\epsilon _2\right)}_{b} {\left(p_2D\epsilon _3\right)}_{a}I_{7}\right.\non\\
   & \qquad\left. +2
   \left(p_1N\epsilon _2p_3\right) {p_3}_{a} {\left(p_2D\epsilon _3\right)}_{b}I_{12}-2
   \left(p_1Np_2\right) {\left(p_2D\epsilon _3\right)}_{b} {\left(p_3\epsilon
   _2\right)}_{a}I_{11}\right.\non\\
   & \qquad\left. +2 \left(p_1Np_3\right) {\left(p_2D\epsilon _3\right)}_{b}
   {\left(p_3\epsilon _2\right)}_{a}I'_{11}-4 \left(p_1Np_3\right) {\left(p_1N\epsilon
   _2\right)}_{a} {\left(p_2D\epsilon _3\right)}_{b}I_{5}\right.\non\\
   & \qquad\left. -4 {p_2}_{b}
   {\left(p_2\epsilon _3\epsilon _2\right)}_{a}I_{23}+2 \left(p_2p_3\right) {p_2}_{b}
   {\left(p_2\epsilon _3\epsilon _2\right)}_{a}I_{16}-2 \left(p_1Np_3\right) {p_2}_{b}
   {\left(p_2\epsilon _3\epsilon _2\right)}_{a}I_{2}\right.\non\\
   & \qquad\left. +4 {p_3}_{a} {\left(p_2\epsilon
   _3\epsilon _2\right)}_{b}I_{23}-2 \left(p_2p_3\right) {p_3}_{a} {\left(p_2\epsilon
   _3\epsilon _2\right)}_{b}I_{16}-2 \left(p_1Np_2\right) {p_3}_{a} {\left(p_2\epsilon
   _3\epsilon _2\right)}_{b}I_{2}\right.\non\\
   & \qquad\left. -2 \left(p_2\epsilon _3p_2\right) {p_2}_{b}
   {\left(p_3D\epsilon _2\right)}_{a}I_{14}-2 \left(p_1N\epsilon _3p_2\right) {p_2}_{b}
   {\left(p_3D\epsilon _2\right)}_{a}I'_{13}\right.\non\\
   & \qquad\left. -2 \left(p_2D\epsilon _3p_2\right)
   {p_2}_{b} {\left(p_3D\epsilon _2\right)}_{a}I_{15}-2 \left(p_2\epsilon _3Dp_3\right)
   {p_2}_{b} {\left(p_3D\epsilon _2\right)}_{a}I_{22}\right.\non\\
   & \qquad\left. +2 \left(p_1N\epsilon
   _3Dp_2\right) {p_2}_{b} {\left(p_3D\epsilon _2\right)}_{a}I_{1}-4 \left(p_1N\epsilon
   _3Dp_3\right) {p_2}_{b} {\left(p_3D\epsilon _2\right)}_{a}I'_{7}\right.\non\\
   & \qquad\left. -4
   \left(p_1N\epsilon _3Np_1\right) {p_2}_{b} {\left(p_3D\epsilon _2\right)}_{a}I_{5}-2
   \left(p_2D\epsilon _3Dp_3\right) {p_2}_{b} {\left(p_3D\epsilon
   _2\right)}_{a}I'_{21}\right.\non\\
   & \qquad\left. +2 \tr(D\epsilon _3) \left(p_1Np_3\right) {p_2}_{b}
   {\left(p_3D\epsilon _2\right)}_{a}I'_{7}+2 \tr(D\epsilon _3)
   \left(p_2Dp_3\right) {p_2}_{b} {\left(p_3D\epsilon _2\right)}_{a}I'_{21}\right.\non\\
   & \qquad\left. -2
   \left(p_2\epsilon _3p_2\right) {p_3}_{b} {\left(p_3D\epsilon _2\right)}_{a}I_{14}-2
   \left(p_1N\epsilon _3p_2\right) {p_3}_{b} {\left(p_3D\epsilon
   _2\right)}_{a}I'_{11}\right.\non\\
   & \qquad\left. -2 \left(p_2D\epsilon _3p_2\right) {p_3}_{b}
   {\left(p_3D\epsilon _2\right)}_{a}I_{15}-2 \left(p_2\epsilon _3Dp_3\right) {p_3}_{b}
   {\left(p_3D\epsilon _2\right)}_{a}I_{22}\right.\non\\
   & \qquad\left. +2 \left(p_1N\epsilon _3Dp_2\right)
   {p_3}_{b} {\left(p_3D\epsilon _2\right)}_{a}I_{1}-2 \left(p_2D\epsilon _3Dp_3\right)
   {p_3}_{b} {\left(p_3D\epsilon _2\right)}_{a}I'_{21}\right.\non\\
   & \qquad\left. +2 \tr(D\epsilon _3)
   \left(p_1Np_2\right) {p_3}_{b} {\left(p_3D\epsilon _2\right)}_{a}I'_{7}+2
   \tr(D\epsilon _3) \left(p_2Dp_3\right) {p_3}_{b} {\left(p_3D\epsilon
   _2\right)}_{a}I'_{21}\right.\non\\
   & \qquad\left. -2 \left(p_1Np_2\right) {\left(p_2\epsilon _3\right)}_{b}
   {\left(p_3D\epsilon _2\right)}_{a}I_{11}-2 \left(p_1Np_3\right) {\left(p_2\epsilon
   _3\right)}_{b} {\left(p_3D\epsilon _2\right)}_{a}I'_{11}\right.\non\\
   & \qquad\left. -2 \left(p_1Np_2\right)
   {\left(p_2D\epsilon _3\right)}_{b} {\left(p_3D\epsilon _2\right)}_{a}I_{1}+2
   \left(p_1Np_3\right) {\left(p_2D\epsilon _3\right)}_{b} {\left(p_3D\epsilon
   _2\right)}_{a}I_{1}\right.\non\\
   & \qquad\left. -4 \left(p_1Np_2\right) {\left(p_1N\epsilon _3\right)}_{a}
   {\left(p_3D\epsilon _2\right)}_{b}I_{5}+4 \left(p_2Dp_3\right) {\left(p_1N\epsilon
   _3\right)}_{a} {\left(p_3D\epsilon _2\right)}_{b}I_{1}\right.\non\\
   & \qquad\left. -4 {p_2}_{b}
   {\left(p_3\epsilon _2\epsilon _3\right)}_{a}I_{23}+2 \left(p_2p_3\right) {p_2}_{b}
   {\left(p_3\epsilon _2\epsilon _3\right)}_{a}I_{16}-2 \left(p_1Np_3\right) {p_2}_{b}
   {\left(p_3\epsilon _2\epsilon _3\right)}_{a}I_{2}\right.\non\\
   & \qquad\left. +4 {p_3}_{a} {\left(p_3\epsilon
   _2\epsilon _3\right)}_{b}I_{23}-2 \left(p_2p_3\right) {p_3}_{a} {\left(p_3\epsilon
   _2\epsilon _3\right)}_{b}I_{16}-2 \left(p_1Np_2\right) {p_3}_{a} {\left(p_3\epsilon
   _2\epsilon _3\right)}_{b}I_{2}\right.\non\\
   & \qquad\left. +2 \left(p_1Np_2\right) \left(p_2p_3\right)
   {\left(\epsilon _2D\epsilon _3\right)}_{ba}I_{11}+2 \left(p_1Np_3\right)
   \left(p_2p_3\right) {\left(\epsilon _2D\epsilon _3\right)}_{ba}I'_{11}\right.\non\\
   & \qquad\left. -4
   \left(p_1Np_2\right) \left(p_1Np_3\right) {\left(\epsilon _2D\epsilon
   _3\right)}_{ba}I_{5}+2 \left(p_1Np_2\right) \left(p_2Dp_3\right) {\left(\epsilon
   _2D\epsilon _3\right)}_{ba}I_{1}\right.\non\\
   & \qquad\left. +2 \left(p_1Np_3\right) \left(p_2Dp_3\right)
   {\left(\epsilon _2D\epsilon _3\right)}_{ba}I_{1}+2 \left(p_2p_3\right) {p_2}_{b}
   {\left(p_1N\epsilon _2\epsilon _3\right)}_{a}I_{2}\right.\non\\
   & \qquad\left. -4 \left(p_1Np_3\right) {p_2}_{b}
   {\left(p_1N\epsilon _2\epsilon _3\right)}_{a}I_{9}+2 \left(p_2Dp_3\right) {p_2}_{b}
   {\left(p_1N\epsilon _2\epsilon _3\right)}_{a}I_{11}\right.\non\\
   & \qquad\left. -2 \left(p_2p_3\right) {p_3}_{a}
   {\left(p_1N\epsilon _2\epsilon _3\right)}_{b}I_{2}-2 \left(p_2Dp_3\right) {p_3}_{a}
   {\left(p_1N\epsilon _2\epsilon _3\right)}_{b}I_{12}\right.\non\\
   & \qquad\left. -2 \left(p_2p_3\right) {p_2}_{b}
   {\left(p_1N\epsilon _3\epsilon _2\right)}_{a}I_{2}-2 \left(p_2Dp_3\right) {p_2}_{b}
   {\left(p_1N\epsilon _3\epsilon _2\right)}_{a}I'_{12}\right.\non\\
   & \qquad\left. +2 \left(p_2p_3\right) {p_3}_{a}
   {\left(p_1N\epsilon _3\epsilon _2\right)}_{b}I_{2}+4 \left(p_1Np_2\right) {p_3}_{a}
   {\left(p_1N\epsilon _3\epsilon _2\right)}_{b}I_{9}\right.\non\\
   & \qquad\left. +2 \left(p_2Dp_3\right) {p_3}_{a}
   {\left(p_1N\epsilon _3\epsilon _2\right)}_{b}I'_{11}-2 \left(p_2p_3\right) {p_2}_{b}
   {\left(p_2D\epsilon _2\epsilon _3\right)}_{a}I_{20}\right.\non\\
   & \qquad\left. -4 \left(p_1Np_3\right) {p_2}_{b}
   {\left(p_2D\epsilon _2\epsilon _3\right)}_{a}I_{6}+2 \left(p_2Dp_3\right) {p_2}_{b}
   {\left(p_2D\epsilon _2\epsilon _3\right)}_{a}I'_{22}\right.\non\\
   & \qquad\left. +2 \left(p_2p_3\right) {p_3}_{a}
   {\left(p_2D\epsilon _2\epsilon _3\right)}_{b}I_{20}-2 \left(p_2Dp_3\right) {p_3}_{a}
   {\left(p_2D\epsilon _2\epsilon _3\right)}_{b}I'_{22}\right.\non\\
   & \qquad\left. +2 \left(p_1Np_3\right)
   {p_2}_{b} {\left(p_2D\epsilon _3\epsilon _2\right)}_{a}I_{11}-2 \left(p_2Dp_3\right)
   {p_2}_{b} {\left(p_2D\epsilon _3\epsilon _2\right)}_{a}I_{15}\right.\non\\
   & \qquad\left. +2 \left(p_1Np_2\right)
   {p_3}_{b} {\left(p_2D\epsilon _3\epsilon _2\right)}_{a}I_{11}-2 \left(p_2Dp_3\right)
   {p_3}_{b} {\left(p_2D\epsilon _3\epsilon _2\right)}_{a}I_{15}\right.\non\\
   & \qquad\left. +2 \left(p_2p_3\right)
   {p_2}_{b} {\left(p_2\epsilon _3D\epsilon _2\right)}_{a}I_{14}-2 \left(p_1Np_3\right)
   {p_2}_{b} {\left(p_2\epsilon _3D\epsilon _2\right)}_{a}I_{11}\right.\non\\
   & \qquad\left. +2 \left(p_2p_3\right)
   {p_3}_{b} {\left(p_2\epsilon _3D\epsilon _2\right)}_{a}I_{14}+2 \left(p_1Np_2\right)
   {p_3}_{b} {\left(p_2\epsilon _3D\epsilon _2\right)}_{a}I_{11}\right.\non\\
   & \qquad\left. -2 \left(p_1Np_3\right)
   {p_2}_{b} {\left(p_3D\epsilon _2\epsilon _3\right)}_{a}I'_{11}-2
   \left(p_2Dp_3\right) {p_2}_{b} {\left(p_3D\epsilon _2\epsilon _3\right)}_{a}I_{15}\right.
\end{align}
\begin{align}
   & \qquad\left. -2
   \left(p_1Np_2\right) {p_3}_{b} {\left(p_3D\epsilon _2\epsilon
   _3\right)}_{a}I'_{11}-2 \left(p_2Dp_3\right) {p_3}_{b} {\left(p_3D\epsilon
   _2\epsilon _3\right)}_{a}I_{15}\right.\non\\
   & \qquad\left. +2 \left(p_2p_3\right) {p_2}_{b} {\left(p_3D\epsilon
   _3\epsilon _2\right)}_{a}I'_{20}-2 \left(p_2Dp_3\right) {p_2}_{b}
   {\left(p_3D\epsilon _3\epsilon _2\right)}_{a}I_{22}\right.\non\\
   & \qquad\left. +2 \left(p_2p_3\right) {p_3}_{b}
   {\left(p_3D\epsilon _3\epsilon _2\right)}_{a}I'_{20}+4 \left(p_1Np_2\right)
   {p_3}_{b} {\left(p_3D\epsilon _3\epsilon _2\right)}_{a}I'_{6}\right.\non\\
   & \qquad\left. -2 \left(p_2Dp_3\right)
   {p_3}_{b} {\left(p_3D\epsilon _3\epsilon _2\right)}_{a}I_{22}+2 \left(p_2p_3\right)
   {p_2}_{b} {\left(p_3\epsilon _2D\epsilon _3\right)}_{a}I_{14}\right.\non\\
   & \qquad\left. +2 \left(p_1Np_3\right)
   {p_2}_{b} {\left(p_3\epsilon _2D\epsilon _3\right)}_{a}I'_{11}+2 \left(p_2p_3\right)
   {p_3}_{b} {\left(p_3\epsilon _2D\epsilon _3\right)}_{a}I_{14}\right.\non\\
   & \qquad\left. -2 \left(p_1Np_2\right)
   {p_3}_{b} {\left(p_3\epsilon _2D\epsilon _3\right)}_{a}I'_{11}+2 \left(p_2p_3\right)
   {p_2}_{b} {\left(p_1N\epsilon _2D\epsilon _3\right)}_{a}I_{11}\right.\non\\
   & \qquad\left. -4
   \left(p_1Np_3\right) {p_2}_{b} {\left(p_1N\epsilon _2D\epsilon _3\right)}_{a}I_{5}+2
   \left(p_2Dp_3\right) {p_2}_{b} {\left(p_1N\epsilon _2D\epsilon _3\right)}_{a}I_{1}\right.\non\\
   & \qquad\left. -2
   \left(p_2p_3\right) {p_3}_{a} {\left(p_1N\epsilon _2D\epsilon _3\right)}_{b}I_{12}-2
   \left(p_2Dp_3\right) {p_3}_{a} {\left(p_1N\epsilon _2D\epsilon _3\right)}_{b}I_{1}\right.\non\\
   & \qquad\left. +2
   \left(p_2p_3\right) {p_2}_{b} {\left(p_1N\epsilon _3D\epsilon
   _2\right)}_{a}I'_{12}+2 \left(p_2Dp_3\right) {p_2}_{b} {\left(p_1N\epsilon
   _3D\epsilon _2\right)}_{a}I_{1}\right.\non\\
   & \qquad\left. -2 \left(p_2p_3\right) {p_3}_{a} {\left(p_1N\epsilon
   _3D\epsilon _2\right)}_{b}I'_{11}+4 \left(p_1Np_2\right) {p_3}_{a}
   {\left(p_1N\epsilon _3D\epsilon _2\right)}_{b}I_{5}\right.\non\\
   & \qquad\left. -2 \left(p_2Dp_3\right) {p_3}_{a}
   {\left(p_1N\epsilon _3D\epsilon _2\right)}_{b}I_{1}+2 \left(p_2p_3\right) {p_2}_{b}
   {\left(p_2D\epsilon _2D\epsilon _3\right)}_{a}I'_{22}\right.\non\\
   & \qquad\left. -4 \left(p_1Np_3\right)
   {p_2}_{b} {\left(p_2D\epsilon _2D\epsilon _3\right)}_{a}I_{7}-2 \left(p_2Dp_3\right)
   {p_2}_{b} {\left(p_2D\epsilon _2D\epsilon _3\right)}_{a}I_{21}\right.\non\\
   & \qquad\left. -2 \left(p_2p_3\right)
   {p_3}_{a} {\left(p_2D\epsilon _2D\epsilon _3\right)}_{b}I'_{22}+2
   \left(p_2Dp_3\right) {p_3}_{a} {\left(p_2D\epsilon _2D\epsilon
   _3\right)}_{b}I_{21}\right.\non\\
   & \qquad\left. +4 {p_2}_{b} {\left(p_2D\epsilon _3D\epsilon
   _2\right)}_{a}I_{24}+2 \left(p_1Np_3\right) {p_2}_{b} {\left(p_2D\epsilon
   _3D\epsilon _2\right)}_{a}I_{1}\right.\non\\
   & \qquad\left. -2 \left(p_2Dp_3\right) {p_2}_{b} {\left(p_2D\epsilon
   _3D\epsilon _2\right)}_{a}I_{17}-4 {p_3}_{a} {\left(p_2D\epsilon _3D\epsilon
   _2\right)}_{b}I_{24}\right.\non\\
   & \qquad\left. -2 \left(p_1Np_2\right) {p_3}_{a} {\left(p_2D\epsilon
   _3D\epsilon _2\right)}_{b}I_{1}+2 \left(p_2Dp_3\right) {p_3}_{a} {\left(p_2D\epsilon
   _3D\epsilon _2\right)}_{b}I_{17}\right.\non\\
   & \qquad\left. +4 {p_2}_{b} {\left(p_3D\epsilon _2D\epsilon
   _3\right)}_{a}I_{24}+2 \left(p_1Np_3\right) {p_2}_{b} {\left(p_3D\epsilon
   _2D\epsilon _3\right)}_{a}I_{1}\right.\non\\
   & \qquad\left. -2 \left(p_2Dp_3\right) {p_2}_{b} {\left(p_3D\epsilon
   _2D\epsilon _3\right)}_{a}I_{17}-4 {p_3}_{a} {\left(p_3D\epsilon _2D\epsilon
   _3\right)}_{b}I_{24}\right.\non\\
   & \qquad\left. -2 \left(p_1Np_2\right) {p_3}_{a} {\left(p_3D\epsilon
   _2D\epsilon _3\right)}_{b}I_{1}+2 \left(p_2Dp_3\right) {p_3}_{a} {\left(p_3D\epsilon
   _2D\epsilon _3\right)}_{b}I_{17}\right.\non\\
   & \qquad\left. +2 \left(p_2p_3\right) {p_2}_{b} {\left(p_3D\epsilon
   _3D\epsilon _2\right)}_{a}I_{22}-2 \left(p_2Dp_3\right) {p_2}_{b}
   {\left(p_3D\epsilon _3D\epsilon _2\right)}_{a}I'_{21}\right.\non\\
   & \qquad\left. +2 \left(p_2p_3\right)
   {p_3}_{b} {\left(p_3D\epsilon _3D\epsilon _2\right)}_{a}I_{22}-4
   \left(p_1Np_2\right) {p_3}_{b} {\left(p_3D\epsilon _3D\epsilon
   _2\right)}_{a}I'_{7}\right.\non\\
   & \qquad\left. -2 \left(p_2Dp_3\right) {p_3}_{b} {\left(p_3D\epsilon
   _3D\epsilon _2\right)}_{a}I'_{21}-\left(p_1Np_2\right) \left(p_2\epsilon
   _3p_2\right) {\epsilon _2}_{ab}I_{2}\right.\non\\
   & \qquad\left. -\left(p_1Np_3\right) \left(p_2\epsilon
   _3p_2\right) {\epsilon _2}_{ab}I_{2}+2 \left(p_1N\epsilon _3p_2\right)
   \left(p_2p_3\right) {\epsilon _2}_{ab}I_{2}\right.\non\\
   & \qquad\left. +4 \left(p_1N\epsilon _3p_2\right)
   \left(p_1Np_2\right) {\epsilon _2}_{ab}I_{9}-2 \left(p_1Np_2\right)
   \left(p_2D\epsilon _3p_2\right) {\epsilon _2}_{ab}I_{11}\right.\non\\
   & \qquad\left. -2 \left(p_1Np_2\right)
   \left(p_2\epsilon _3Dp_3\right) {\epsilon _2}_{ab}I'_{6}+4 \left(p_1N\epsilon
   _3Dp_2\right) \left(p_1Np_2\right) {\epsilon _2}_{ab}I_{5}\right.\non\\
   & \qquad\left. -2 \left(p_1N\epsilon
   _3Dp_2\right) \left(p_2Dp_3\right) {\epsilon _2}_{ab}I_{1}+2 \left(p_1N\epsilon
   _3Dp_3\right) \left(p_2p_3\right) {\epsilon _2}_{ab}I'_{6}\right.\non\\
   & \qquad\left. +2 \left(p_1N\epsilon
   _3Dp_3\right) \left(p_1Np_2\right) {\epsilon _2}_{ab}I'_{4}+2 \left(p_1N\epsilon
   _3Dp_3\right) \left(p_2Dp_3\right) {\epsilon _2}_{ab}I'_{7}\right.\non\\
   & \qquad\left. -2 \left(p_1N\epsilon
   _3Np_1\right) \left(p_2p_3\right) {\epsilon _2}_{ab}I_{9}-4 \left(p_1N\epsilon
   _3Np_1\right) \left(p_1Np_2\right) {\epsilon _2}_{ab}I_{10}\right.\non\\
   & \qquad\left. +2 \left(p_1N\epsilon
   _3Np_1\right) \left(p_2Dp_3\right) {\epsilon _2}_{ab}I_{5}-\left(p_1Np_2\right)
   \left(p_2D\epsilon _3Dp_2\right) {\epsilon _2}_{ab}I_{1}\right.\non\\
   & \qquad\left. +\left(p_1Np_3\right)
   \left(p_2D\epsilon _3Dp_2\right) {\epsilon _2}_{ab}I_{1}+2 \left(p_1Np_2\right)
   \left(p_2D\epsilon _3Dp_3\right) {\epsilon _2}_{ab}I'_{7}\right.\non\\
   & \qquad\left. +\tr(D\epsilon
   _3) \left(p_1Np_2\right) \left(p_2p_3\right) {\epsilon
   _2}_{ab}I'_{6}-\tr(D\epsilon _3) \left(p_1Np_3\right) \left(p_2p_3\right)
   {\epsilon _2}_{ab}I'_{6}\right.\non\\
   & \qquad\left. -\tr(D\epsilon _3) \left(p_1Np_2\right)
   \left(p_1Np_3\right) {\epsilon _2}_{ab}I'_{4}-\tr(D\epsilon _3)
   \left(p_1Np_2\right) \left(p_2Dp_3\right) {\epsilon _2}_{ab}I'_{7}\right.\non\\
   & \qquad\left. -\tr(D\epsilon
   _3) \left(p_1Np_3\right) \left(p_2Dp_3\right) {\epsilon _2}_{ab}I'_{7}\rp,
\end{align}


\begin{align}
\mathcal{A}_{C^{(p-1)}Bh}^{(1)}=& \frac{2 i^{p (p+1)} \sqrt{2}}{(p-2)!}{{C}^{i}}_{b_1...b_{p-2}}
   {\varepsilon}^{abcb_1...b_{p-2}}\lp -2 \tr(D\epsilon _3) {p_2}_{c} {p_3}_{b}
   {p_3}_{i} {\left(p_3\epsilon _2\right)}_{a}I'_{6}\right.\non\\
   & \qquad\left. -2 {p_2}_{c} {p_2}_{i}
   {\left(p_2\epsilon _3\right)}_{b} {\left(p_3\epsilon _2\right)}_{a}I_{2}-2 {p_2}_{i}
   {p_3}_{c} {\left(p_2\epsilon _3\right)}_{b} {\left(p_3\epsilon _2\right)}_{a}I_{2}\right.\non\\
   & \qquad\left. -2
   {p_2}_{c} {p_3}_{i} {\left(p_2\epsilon _3\right)}_{b} {\left(p_3\epsilon
   _2\right)}_{a}I_{2}-2 {p_3}_{c} {p_3}_{i} {\left(p_2\epsilon _3\right)}_{b}
   {\left(p_3\epsilon _2\right)}_{a}I_{2}\right.\non\\
   & \qquad\left. +2 \tr(D\epsilon _3) {p_2}_{c}
   {p_2}_{i} {p_3}_{a} {\left(p_3\epsilon _2\right)}_{b}I'_{6}-2 \left(p_2p_3\right)
   {p_2}_{c} {p_2}_{i} {\left(\epsilon _2\epsilon _3\right)}_{ab}I_{2}\right.\non\\
   & \qquad\left. +4
   \left(p_1Np_3\right) {p_2}_{c} {p_2}_{i} {\left(\epsilon _2\epsilon
   _3\right)}_{ab}I_{9}-2 \left(p_2Dp_3\right) {p_2}_{c} {p_2}_{i} {\left(\epsilon
   _2\epsilon _3\right)}_{ab}I_{11}\right.\non\\
   & \qquad\left. +2 \left(p_2p_3\right) {p_2}_{c} {p_3}_{i}
   {\left(\epsilon _2\epsilon _3\right)}_{ab}I_{2}+2 \left(p_2Dp_3\right) {p_2}_{c}
   {p_3}_{i} {\left(\epsilon _2\epsilon _3\right)}_{ab}I'_{11}\right.\non\\
   & \qquad\left. +2 \left(p_2p_3\right)
   {p_3}_{c} {p_3}_{i} {\left(\epsilon _2\epsilon _3\right)}_{ab}I_{2}+4
   \left(p_1Np_2\right) {p_3}_{c} {p_3}_{i} {\left(\epsilon _2\epsilon
   _3\right)}_{ab}I_{9}\right.\non\\
   & \qquad\left. +2 \left(p_2Dp_3\right) {p_3}_{c} {p_3}_{i} {\left(\epsilon
   _2\epsilon _3\right)}_{ab}I'_{11}-2 \left(p_2p_3\right) {p_2}_{i} {p_3}_{a}
   {\left(\epsilon _2\epsilon _3\right)}_{bc}I_{2}\right.\non\\
   & \qquad\left. -2 \left(p_2Dp_3\right) {p_2}_{i}
   {p_3}_{a} {\left(\epsilon _2\epsilon _3\right)}_{bc}I_{11}-8 \left(p_2Dp_3\right)
   {p_2}_{c} {p_3}_{a} {\left(\epsilon _2\epsilon _3\right)}_{bi}I_{0}\right.\non\\
   & \qquad\left. +8
   \left(p_2Dp_3\right) {p_2}_{c} {p_3}_{a} {\left(\epsilon _2\epsilon
   _3\right)}_{ib}I_{0}-2 \tr(D\epsilon _3) {p_2}_{c} {p_3}_{b} {p_3}_{i}
   {\left(p_1N\epsilon _2\right)}_{a}I'_{4}\right.\non\\
   & \qquad\left. -4 {p_2}_{c} {p_3}_{i} {\left(p_1N\epsilon
   _2\right)}_{a} {\left(p_2\epsilon _3\right)}_{b}I_{9}-4 {p_3}_{c} {p_3}_{i}
   {\left(p_1N\epsilon _2\right)}_{a} {\left(p_2\epsilon _3\right)}_{b}I_{9}\right.\non\\
   & \qquad\left. +4
   {p_2}_{c} {p_3}_{b} {\left(p_1N\epsilon _2\right)}_{a} {\left(p_2\epsilon
   _3\right)}_{i}I_{9}-4 {p_2}_{c} {p_2}_{i} {\left(p_1N\epsilon _3\right)}_{a}
   {\left(p_3\epsilon _2\right)}_{b}I_{9}\right.\non\\
   & \qquad\left. +8 {p_2}_{b} {p_3}_{i} {\left(p_1N\epsilon
   _2\right)}_{c} {\left(p_1N\epsilon _3\right)}_{a}I_{10}+4 {p_2}_{c} {p_3}_{i}
   {\left(p_1N\epsilon _3\right)}_{b} {\left(p_3\epsilon _2\right)}_{a}I_{9}\right.\non\\
   & \qquad\left. +4
   {p_2}_{c} {p_3}_{a} {\left(p_1N\epsilon _3\right)}_{i} {\left(p_3\epsilon
   _2\right)}_{b}I_{9}-8 {p_2}_{c} {p_3}_{b} {\left(p_1N\epsilon _2\right)}_{a}
   {\left(p_1N\epsilon _3\right)}_{i}I_{10}\right.\non\\
   & \qquad\left. +4 \tr(D\epsilon _3) {p_2}_{c}
   {p_3}_{b} {p_3}_{i} {\left(p_2D\epsilon _2\right)}_{a}I_{3}-4 {p_2}_{c} {p_3}_{i}
   {\left(p_2D\epsilon _2\right)}_{a} {\left(p_2\epsilon _3\right)}_{b}I_{6}\right.\non\\
   & \qquad\left. -4
   {p_3}_{c} {p_3}_{i} {\left(p_2D\epsilon _2\right)}_{a} {\left(p_2\epsilon
   _3\right)}_{b}I_{6}+4 {p_2}_{c} {p_3}_{b} {\left(p_2D\epsilon _2\right)}_{a}
   {\left(p_2\epsilon _3\right)}_{i}I_{6}\right.\non\\
   & \qquad\left. -4 {p_2}_{c} {p_3}_{i} {\left(p_1N\epsilon
   _3\right)}_{b} {\left(p_2D\epsilon _2\right)}_{a}I_{4}+4 {p_2}_{c} {p_3}_{b}
   {\left(p_1N\epsilon _3\right)}_{i} {\left(p_2D\epsilon _2\right)}_{a}I_{4}\right.\non\\
   & \qquad\left. +2
   {p_2}_{c} {p_2}_{i} {\left(p_2D\epsilon _3\right)}_{a} {\left(p_3\epsilon
   _2\right)}_{b}I_{11}+2 {p_2}_{i} {p_3}_{c} {\left(p_2D\epsilon _3\right)}_{a}
   {\left(p_3\epsilon _2\right)}_{b}I_{11}\right.\non\\
   & \qquad\left. -4 {p_2}_{b} {p_3}_{i} {\left(p_1N\epsilon
   _2\right)}_{c} {\left(p_2D\epsilon _3\right)}_{a}I_{5}+4 {p_2}_{c} {p_3}_{i}
   {\left(p_2D\epsilon _2\right)}_{b} {\left(p_2D\epsilon _3\right)}_{a}I_{7}\right.\non\\
   & \qquad\left. +4
   {p_3}_{c} {p_3}_{i} {\left(p_2D\epsilon _2\right)}_{b} {\left(p_2D\epsilon
   _3\right)}_{a}I_{7}+2 {p_2}_{c} {p_3}_{i} {\left(p_2D\epsilon _3\right)}_{b}
   {\left(p_3\epsilon _2\right)}_{a}I'_{11}\right.\non\\
   & \qquad\left. +2 {p_3}_{c} {p_3}_{i} {\left(p_2D\epsilon
   _3\right)}_{b} {\left(p_3\epsilon _2\right)}_{a}I'_{11}+8 {p_2}_{c} {p_3}_{a}
   {\left(p_2D\epsilon _3\right)}_{b} {\left(p_3\epsilon _2\right)}_{i}I_{0}\right.\non\\
   & \qquad\left. -4
   {p_3}_{c} {p_3}_{i} {\left(p_1N\epsilon _2\right)}_{a} {\left(p_2D\epsilon
   _3\right)}_{b}I_{5}-4 {p_2}_{c} {p_3}_{a} {\left(p_2D\epsilon _3\right)}_{i}
   {\left(p_3\epsilon _2\right)}_{b}I_{8}\right.\non\\
   & \qquad\left. +4 {p_2}_{c} {p_3}_{b} {\left(p_1N\epsilon
   _2\right)}_{a} {\left(p_2D\epsilon _3\right)}_{i}I_{5}+4 {p_2}_{c} {p_3}_{b}
   {\left(p_2D\epsilon _2\right)}_{a} {\left(p_2D\epsilon _3\right)}_{i}I_{7}\right.\non\\
   & \qquad\left. -2
   {p_2}_{c} {p_2}_{i} {p_3}_{a} {\left(p_2\epsilon _3\epsilon _2\right)}_{b}I_{2}-2
   {p_2}_{c} {p_3}_{a} {p_3}_{i} {\left(p_2\epsilon _3\epsilon _2\right)}_{b}I_{2}\right.\non\\
   & \qquad\left. -2
   \tr(D\epsilon _3) {p_2}_{c} {p_3}_{b} {p_3}_{i} {\left(p_3D\epsilon
   _2\right)}_{a}I'_{7}-2 {p_2}_{c} {p_3}_{i} {\left(p_2\epsilon _3\right)}_{b}
   {\left(p_3D\epsilon _2\right)}_{a}I'_{11}\right.\non\\
   & \qquad\left. -2 {p_3}_{c} {p_3}_{i} {\left(p_2\epsilon
   _3\right)}_{b} {\left(p_3D\epsilon _2\right)}_{a}I'_{11}-4 {p_2}_{c} {p_3}_{b}
   {\left(p_2\epsilon _3\right)}_{i} {\left(p_3D\epsilon _2\right)}_{a}I_{8}\right.\non\\
   & \qquad\left. -4
   {p_2}_{c} {p_3}_{i} {\left(p_1N\epsilon _3\right)}_{b} {\left(p_3D\epsilon
   _2\right)}_{a}I_{5}+4 {p_2}_{c} {p_3}_{b} {\left(p_1N\epsilon _3\right)}_{i}
   {\left(p_3D\epsilon _2\right)}_{a}I_{5}\right.\non\\
   & \qquad\left. +2 {p_2}_{c} {p_3}_{i} {\left(p_2D\epsilon
   _3\right)}_{b} {\left(p_3D\epsilon _2\right)}_{a}I_{1}+2 {p_3}_{c} {p_3}_{i}
   {\left(p_2D\epsilon _3\right)}_{b} {\left(p_3D\epsilon _2\right)}_{a}I_{1}\right.\non\\
   & \qquad\left. -2
   \tr(D\epsilon _3) {p_2}_{c} {p_2}_{i} {p_3}_{a} {\left(p_3D\epsilon
   _2\right)}_{b}I'_{7}+2 {p_2}_{c} {p_2}_{i} {\left(p_2\epsilon _3\right)}_{a}
   {\left(p_3D\epsilon _2\right)}_{b}I_{11}\right.\non\\
   & \qquad\left. +2 {p_2}_{i} {p_3}_{c} {\left(p_2\epsilon
   _3\right)}_{a} {\left(p_3D\epsilon _2\right)}_{b}I_{11}-4 {p_2}_{c} {p_2}_{i}
   {\left(p_1N\epsilon _3\right)}_{a} {\left(p_3D\epsilon _2\right)}_{b}I_{5}\right.\non\\
   & \qquad\left. +2
   {p_2}_{c} {p_2}_{i} {\left(p_2D\epsilon _3\right)}_{a} {\left(p_3D\epsilon
   _2\right)}_{b}I_{1}+2 {p_2}_{i} {p_3}_{c} {\left(p_2D\epsilon _3\right)}_{a}
   {\left(p_3D\epsilon _2\right)}_{b}I_{1}\right.\non\\
   & \qquad\left. +8 {p_2}_{c} {p_3}_{a} {\left(p_2\epsilon
   _3\right)}_{b} {\left(p_3D\epsilon _2\right)}_{i}I_{0}+4 {p_2}_{c} {p_3}_{b}
   {\left(p_3D\epsilon _3\right)}_{i} {\left(p_3\epsilon _2\right)}_{a}I'_{6}\right.\non
\end{align}
\begin{align}
   & \qquad\left. +4
   {p_2}_{c} {p_3}_{b} {\left(p_1N\epsilon _2\right)}_{a} {\left(p_3D\epsilon
   _3\right)}_{i}I'_{4}-8 {p_2}_{c} {p_3}_{b} {\left(p_2D\epsilon _2\right)}_{a}
   {\left(p_3D\epsilon _3\right)}_{i}I_{3}\right.\non\\
   & \qquad\left. +4 {p_2}_{c} {p_3}_{b} {\left(p_3D\epsilon
   _2\right)}_{a} {\left(p_3D\epsilon _3\right)}_{i}I'_{7}-2 {p_2}_{c} {p_2}_{i}
   {p_3}_{a} {\left(p_3\epsilon _2\epsilon _3\right)}_{b}I_{2}\right.\non\\
   & \qquad\left. -2 {p_2}_{c} {p_3}_{a}
   {p_3}_{i} {\left(p_3\epsilon _2\epsilon _3\right)}_{b}I_{2}-2 \left(p_2p_3\right)
   {p_2}_{c} {p_2}_{i} {\left(\epsilon _2D\epsilon _3\right)}_{ab}I_{11}\right.\non\\
   & \qquad\left. +4
   \left(p_1Np_3\right) {p_2}_{c} {p_2}_{i} {\left(\epsilon _2D\epsilon
   _3\right)}_{ab}I_{5}-2 \left(p_2Dp_3\right) {p_2}_{c} {p_2}_{i} {\left(\epsilon
   _2D\epsilon _3\right)}_{ab}I_{1}\right.\non\\
   & \qquad\left. +2 \left(p_2p_3\right) {p_2}_{c} {p_3}_{i}
   {\left(\epsilon _2D\epsilon _3\right)}_{ba}I'_{11}+2 \left(p_2Dp_3\right) {p_2}_{c}
   {p_3}_{i} {\left(\epsilon _2D\epsilon _3\right)}_{ba}I_{1}\right.\non\\
   & \qquad\left. +2 \left(p_2p_3\right)
   {p_3}_{c} {p_3}_{i} {\left(\epsilon _2D\epsilon _3\right)}_{ba}I'_{11}-4
   \left(p_1Np_2\right) {p_3}_{c} {p_3}_{i} {\left(\epsilon _2D\epsilon
   _3\right)}_{ba}I_{5}\right.\non\\
   & \qquad\left. +2 \left(p_2Dp_3\right) {p_3}_{c} {p_3}_{i} {\left(\epsilon
   _2D\epsilon _3\right)}_{ba}I_{1}-2 \left(p_2p_3\right) {p_2}_{i} {p_3}_{a}
   {\left(\epsilon _2D\epsilon _3\right)}_{bc}I_{11}\right.\non\\
   & \qquad\left. -2 \left(p_2Dp_3\right) {p_2}_{i}
   {p_3}_{a} {\left(\epsilon _2D\epsilon _3\right)}_{bc}I_{1}+8 \left(p_2p_3\right)
   {p_2}_{c} {p_3}_{a} {\left(\epsilon _2D\epsilon _3\right)}_{bi}I_{0}\right.\non\\
   & \qquad\left. +8
   \left(p_2p_3\right) {p_2}_{c} {p_3}_{a} {\left(\epsilon _2D\epsilon
   _3\right)}_{ib}I_{0}-4 {p_2}_{c} {p_3}_{a} {p_3}_{i} {\left(p_1N\epsilon _2\epsilon
   _3\right)}_{b}I_{9}\right.\non\\
   & \qquad\left. +4 {p_2}_{c} {p_2}_{i} {p_3}_{a} {\left(p_1N\epsilon _3\epsilon
   _2\right)}_{b}I_{9}-4 {p_2}_{c} {p_3}_{a} {p_3}_{i} {\left(p_2D\epsilon _2\epsilon
   _3\right)}_{b}I_{6}-2 {p_2}_{c} {p_2}_{i} {p_3}_{a} {\left(p_2D\epsilon _3\epsilon
   _2\right)}_{b}I_{11}\right.\non\\
   & \qquad\left. +2 {p_2}_{c} {p_3}_{a} {p_3}_{i} {\left(p_2D\epsilon _3\epsilon
   _2\right)}_{b}I_{11}-2 {p_2}_{c} {p_2}_{i} {p_3}_{a} {\left(p_2\epsilon _3D\epsilon
   _2\right)}_{b}I_{11}-2 {p_2}_{c} {p_3}_{a} {p_3}_{i} {\left(p_2\epsilon _3D\epsilon
   _2\right)}_{b}I_{11}\right.\non\\
   & \qquad\left. +2 {p_2}_{c} {p_2}_{i} {p_3}_{a} {\left(p_3D\epsilon _2\epsilon
   _3\right)}_{b}I'_{11}-2 {p_2}_{c} {p_3}_{a} {p_3}_{i} {\left(p_3D\epsilon _2\epsilon
   _3\right)}_{b}I'_{11}-4 {p_2}_{c} {p_2}_{i} {p_3}_{a} {\left(p_3D\epsilon _3\epsilon
   _2\right)}_{b}I'_{6}\right.\non\\
   & \qquad\left. +2 {p_2}_{c} {p_2}_{i} {p_3}_{a} {\left(p_3\epsilon _2D\epsilon
   _3\right)}_{b}I'_{11}+2 {p_2}_{c} {p_3}_{a} {p_3}_{i} {\left(p_3\epsilon _2D\epsilon
   _3\right)}_{b}I'_{11}\right.\non\\
   & \qquad\left. -4 {p_2}_{c} {p_3}_{a} {p_3}_{i} {\left(p_1N\epsilon
   _2D\epsilon _3\right)}_{b}I_{5}+4 {p_2}_{c} {p_2}_{i} {p_3}_{a} {\left(p_1N\epsilon
   _3D\epsilon _2\right)}_{b}I_{5}\right.\non\\
   & \qquad\left. -4 {p_2}_{c} {p_3}_{a} {p_3}_{i} {\left(p_2D\epsilon
   _2D\epsilon _3\right)}_{b}I_{7}-2 {p_2}_{c} {p_2}_{i} {p_3}_{a} {\left(p_2D\epsilon
   _3D\epsilon _2\right)}_{b}I_{1}\right.\non\\
   & \qquad\left. +2 {p_2}_{c} {p_3}_{a} {p_3}_{i} {\left(p_2D\epsilon
   _3D\epsilon _2\right)}_{b}I_{1}-2 {p_2}_{c} {p_2}_{i} {p_3}_{a} {\left(p_3D\epsilon
   _2D\epsilon _3\right)}_{b}I_{1}\right.\non\\
   & \qquad\left. +2 {p_2}_{c} {p_3}_{a} {p_3}_{i} {\left(p_3D\epsilon
   _2D\epsilon _3\right)}_{b}I_{1}+4 {p_2}_{c} {p_2}_{i} {p_3}_{a} {\left(p_3D\epsilon
   _3D\epsilon _2\right)}_{b}I'_{7}-\left(p_2\epsilon _3p_2\right) {p_2}_{c} {p_2}_{i}
   {\epsilon _2}_{ab}I_{2}\right.\non\\
   & \qquad\left. +4 \left(p_1N\epsilon _3p_2\right) {p_2}_{c} {p_2}_{i}
   {\epsilon _2}_{ab}I_{9}-2 \left(p_2D\epsilon _3p_2\right) {p_2}_{c} {p_2}_{i}
   {\epsilon _2}_{ab}I_{11}\right.\non\\
   & \qquad\left. -2 \left(p_2\epsilon _3Dp_3\right) {p_2}_{c} {p_2}_{i}
   {\epsilon _2}_{ab}I'_{6}+4 \left(p_1N\epsilon _3Dp_2\right) {p_2}_{c} {p_2}_{i}
   {\epsilon _2}_{ab}I_{5}\right.\non\\
   & \qquad\left. +2 \left(p_1N\epsilon _3Dp_3\right) {p_2}_{c} {p_2}_{i}
   {\epsilon _2}_{ab}I'_{4}-4 \left(p_1N\epsilon _3Np_1\right) {p_2}_{c} {p_2}_{i}
   {\epsilon _2}_{ab}I_{10}\right.\non\\
   & \qquad\left. -\left(p_2D\epsilon _3Dp_2\right) {p_2}_{c} {p_2}_{i}
   {\epsilon _2}_{ab}I_{1}+2 \left(p_2D\epsilon _3Dp_3\right) {p_2}_{c} {p_2}_{i}
   {\epsilon _2}_{ab}I'_{7}\right.\non\\
   & \qquad\left. +\tr(D\epsilon _3) \left(p_2p_3\right) {p_2}_{c}
   {p_2}_{i} {\epsilon _2}_{ab}I'_{6}-\tr(D\epsilon _3) \left(p_1Np_3\right)
   {p_2}_{c} {p_2}_{i} {\epsilon _2}_{ab}I'_{4}\right.\non\\
   & \qquad\left. -\tr(D\epsilon _3)
   \left(p_2Dp_3\right) {p_2}_{c} {p_2}_{i} {\epsilon _2}_{ab}I'_{7}-\left(p_2\epsilon
   _3p_2\right) {p_2}_{i} {p_3}_{c} {\epsilon _2}_{ab}I_{2}\right.\non\\
   & \qquad\left. +2 \left(p_1N\epsilon
   _3p_2\right) {p_2}_{i} {p_3}_{c} {\epsilon _2}_{ab}I_{9}-2 \left(p_2D\epsilon
   _3p_2\right) {p_2}_{i} {p_3}_{c} {\epsilon _2}_{ab}I_{11}\right.\non\\
   & \qquad\left. -2 \left(p_2\epsilon
   _3Dp_3\right) {p_2}_{i} {p_3}_{c} {\epsilon _2}_{ab}I'_{6}+2 \left(p_1N\epsilon
   _3Dp_2\right) {p_2}_{i} {p_3}_{c} {\epsilon _2}_{ab}I_{5}\right.\non\\
   & \qquad\left. -\left(p_2D\epsilon
   _3Dp_2\right) {p_2}_{i} {p_3}_{c} {\epsilon _2}_{ab}I_{1}+2 \left(p_2D\epsilon
   _3Dp_3\right) {p_2}_{i} {p_3}_{c} {\epsilon _2}_{ab}I'_{7}\right.\non\\
   & \qquad\left. +\tr(D\epsilon
   _3) \left(p_2p_3\right) {p_2}_{i} {p_3}_{c} {\epsilon
   _2}_{ab}I'_{6}-\tr(D\epsilon _3) \left(p_2Dp_3\right) {p_2}_{i} {p_3}_{c}
   {\epsilon _2}_{ab}I'_{7}\right.\non\\
   & \qquad\left. -\left(p_2\epsilon _3p_2\right) {p_2}_{c} {p_3}_{i}
   {\epsilon _2}_{ab}I_{2}+2 \left(p_1N\epsilon _3p_2\right) {p_2}_{c} {p_3}_{i}
   {\epsilon _2}_{ab}I_{9}\right.\non\\
   & \qquad\left. -2 \left(p_1N\epsilon _3Dp_2\right) {p_2}_{c} {p_3}_{i}
   {\epsilon _2}_{ab}I_{5}+\left(p_2D\epsilon _3Dp_2\right) {p_2}_{c} {p_3}_{i}
   {\epsilon _2}_{ab}I_{1}\right.\non\\
   & \qquad\left. -\tr(D\epsilon _3) \left(p_2p_3\right) {p_2}_{c}
   {p_3}_{i} {\epsilon _2}_{ab}I'_{6}-\tr(D\epsilon _3) \left(p_2Dp_3\right)
   {p_2}_{c} {p_3}_{i} {\epsilon _2}_{ab}I'_{7}\right.\non\\
   & \qquad\left. -\left(p_2\epsilon _3p_2\right)
   {p_3}_{c} {p_3}_{i} {\epsilon _2}_{ab}I_{2}+\left(p_2D\epsilon _3Dp_2\right)
   {p_3}_{c} {p_3}_{i} {\epsilon _2}_{ab}I_{1}\right.\non\\
   & \qquad\left. +2 \left(p_1Np_3\right) {p_2}_{i}
   {\left(p_2\epsilon _3\right)}_{c} {\epsilon _2}_{ab}I_{9}-2 \left(p_1Np_2\right)
   {p_3}_{i} {\left(p_2\epsilon _3\right)}_{c} {\epsilon _2}_{ab}I_{9}\right.\non\\
   & \qquad\left. +2
   \left(p_2p_3\right) {p_2}_{c} {\left(p_2\epsilon _3\right)}_{i} {\epsilon
   _2}_{ab}I_{2}-2 \left(p_1Np_3\right) {p_2}_{c} {\left(p_2\epsilon _3\right)}_{i}
   {\epsilon _2}_{ab}I_{9}\right.\non\\
   & \qquad\left. +2 \left(p_2p_3\right) {p_3}_{c} {\left(p_2\epsilon
   _3\right)}_{i} {\epsilon _2}_{ab}I_{2}+2 \left(p_1Np_2\right) {p_3}_{c}
   {\left(p_2\epsilon _3\right)}_{i} {\epsilon _2}_{ab}I_{9}\right.\non
\end{align}
\begin{align}
   & \qquad\left. -2 \left(p_2p_3\right)
   {p_2}_{i} {\left(p_1N\epsilon _3\right)}_{c} {\epsilon _2}_{ab}I_{9}-2
   \left(p_2Dp_3\right) {p_2}_{i} {\left(p_1N\epsilon _3\right)}_{c} {\epsilon
   _2}_{ab}I_{5}\right.\non\\
   & \qquad\left. +2 \left(p_2p_3\right) {p_3}_{i} {\left(p_1N\epsilon _3\right)}_{c}
   {\epsilon _2}_{ab}I_{9}+4 \left(p_1Np_2\right) {p_3}_{i} {\left(p_1N\epsilon
   _3\right)}_{c} {\epsilon _2}_{ab}I_{10}\right.\non\\
   & \qquad\left. -2 \left(p_2Dp_3\right) {p_3}_{i}
   {\left(p_1N\epsilon _3\right)}_{c} {\epsilon _2}_{ab}I_{5}-2 \left(p_2p_3\right)
   {p_2}_{c} {\left(p_1N\epsilon _3\right)}_{i} {\epsilon _2}_{ab}I_{9}\right.\non\\
   & \qquad\left. +2
   \left(p_2Dp_3\right) {p_2}_{c} {\left(p_1N\epsilon _3\right)}_{i} {\epsilon
   _2}_{ab}I_{5}+2 \left(p_1Np_3\right) {p_2}_{i} {\left(p_2D\epsilon _3\right)}_{c}
   {\epsilon _2}_{ab}I_{5}\right.\non\\
   & \qquad\left. -2 \left(p_1Np_2\right) {p_3}_{i} {\left(p_2D\epsilon
   _3\right)}_{c} {\epsilon _2}_{ab}I_{5}+2 \left(p_1Np_3\right) {p_2}_{c}
   {\left(p_2D\epsilon _3\right)}_{i} {\epsilon _2}_{ab}I_{5}\right.\non\\
   & \qquad\left. -2 \left(p_2Dp_3\right)
   {p_2}_{c} {\left(p_2D\epsilon _3\right)}_{i} {\epsilon _2}_{ab}I_{1}+2
   \left(p_1Np_2\right) {p_3}_{c} {\left(p_2D\epsilon _3\right)}_{i} {\epsilon
   _2}_{ab}I_{5}\right.\non\\
   & \qquad\left. -2 \left(p_2Dp_3\right) {p_3}_{c} {\left(p_2D\epsilon _3\right)}_{i}
   {\epsilon _2}_{ab}I_{1}+2 \left(p_2p_3\right) {p_2}_{c} {\left(p_3D\epsilon
   _3\right)}_{i} {\epsilon _2}_{ab}I'_{6}\right.\non\\
   & \qquad\left. +2 \left(p_2Dp_3\right) {p_2}_{c}
   {\left(p_3D\epsilon _3\right)}_{i} {\epsilon _2}_{ab}I'_{7}-\tr(D\epsilon
   _3) \left(p_2p_3\right) {p_3}_{a} {p_3}_{i} {\epsilon
   _2}_{bc}I'_{6}\right.\non\\
   & \qquad\left. -\tr(D\epsilon _3) \left(p_1Np_2\right) {p_3}_{a} {p_3}_{i}
   {\epsilon _2}_{bc}I'_{4}-\tr(D\epsilon _3) \left(p_2Dp_3\right) {p_3}_{a}
   {p_3}_{i} {\epsilon _2}_{bc}I'_{7}\right.\non\\
   & \qquad\left. -2 \left(p_2p_3\right) {p_3}_{a}
   {\left(p_1N\epsilon _3\right)}_{i} {\epsilon _2}_{bc}I_{9}-4 \left(p_1Np_2\right)
   {p_3}_{a} {\left(p_1N\epsilon _3\right)}_{i} {\epsilon _2}_{bc}I_{10}\right.\non\\
   & \qquad\left. +2
   \left(p_2Dp_3\right) {p_3}_{a} {\left(p_1N\epsilon _3\right)}_{i} {\epsilon
   _2}_{bc}I_{5}+2 \left(p_2p_3\right) {p_3}_{a} {\left(p_3D\epsilon _3\right)}_{i}
   {\epsilon _2}_{bc}I'_{6}\right.\non\\
   & \qquad\left. +2 \left(p_1Np_2\right) {p_3}_{a} {\left(p_3D\epsilon
   _3\right)}_{i} {\epsilon _2}_{bc}I'_{4}+2 \left(p_2Dp_3\right) {p_3}_{a}
   {\left(p_3D\epsilon _3\right)}_{i} {\epsilon _2}_{bc}I'_{7}\right.\non\\
   & \qquad\left. -4 \left(p_2p_3\right)
   {p_2}_{c} {\left(p_3\epsilon _2\right)}_{b} {\epsilon _3}_{ai}I_{2}+4
   \left(p_1Np_3\right) {p_2}_{c} {\left(p_3\epsilon _2\right)}_{b} {\epsilon
   _3}_{ai}I_{9}\right.\non\\
   & \qquad\left. -4 \left(p_2p_3\right) {p_3}_{c} {\left(p_3\epsilon _2\right)}_{b}
   {\epsilon _3}_{ai}I_{2}-4 \left(p_1Np_2\right) {p_3}_{c} {\left(p_3\epsilon
   _2\right)}_{b} {\epsilon _3}_{ai}I_{9}\right.\non\\
   & \qquad\left. -4 \left(p_2p_3\right) {p_2}_{c}
   {\left(p_2D\epsilon _2\right)}_{b} {\epsilon _3}_{ai}I_{6}-4 \left(p_1Np_3\right)
   {p_2}_{c} {\left(p_2D\epsilon _2\right)}_{b} {\epsilon _3}_{ai}I_{4}\right.\non\\
   & \qquad\left. -4
   \left(p_2Dp_3\right) {p_2}_{c} {\left(p_2D\epsilon _2\right)}_{b} {\epsilon
   _3}_{ai}I_{7}-4 \left(p_1Np_3\right) {p_2}_{c} {\left(p_3D\epsilon _2\right)}_{b}
   {\epsilon _3}_{ai}I_{5}\right.\non\\
   & \qquad\left. +4 \left(p_2Dp_3\right) {p_2}_{c} {\left(p_3D\epsilon
   _2\right)}_{b} {\epsilon _3}_{ai}I_{1}-4 \left(p_1Np_2\right) {p_3}_{c}
   {\left(p_3D\epsilon _2\right)}_{b} {\epsilon _3}_{ai}I_{5}\right.\non\\
   & \qquad\left. +4 \left(p_2Dp_3\right)
   {p_3}_{c} {\left(p_3D\epsilon _2\right)}_{b} {\epsilon _3}_{ai}I_{1}+4
   \left(p_1N\epsilon _2p_3\right) {p_2}_{c} {p_3}_{a} {\epsilon _3}_{bi}I_{9}\right.\non\\
   & \qquad\left. +4
   \left(p_2D\epsilon _2p_3\right) {p_2}_{c} {p_3}_{a} {\epsilon _3}_{bi}I_{6}+4
   \left(p_3D\epsilon _2p_3\right) {p_2}_{c} {p_3}_{a} {\epsilon _3}_{bi}I'_{11}\right.\non\\
   & \qquad\left. +4
   \left(p_1N\epsilon _2Dp_3\right) {p_2}_{c} {p_3}_{a} {\epsilon _3}_{bi}I_{5}+4
   \left(p_2D\epsilon _2Dp_3\right) {p_2}_{c} {p_3}_{a} {\epsilon _3}_{bi}I_{7}\right.\non\\
   & \qquad\left. +4
   \left(p_2p_3\right) {p_2}_{c} {\left(p_1N\epsilon _2\right)}_{a} {\epsilon
   _3}_{bi}I_{9}-8 \left(p_1Np_3\right) {p_2}_{c} {\left(p_1N\epsilon _2\right)}_{a}
   {\epsilon _3}_{bi}I_{10}\right.\non\\
   & \qquad\left. +4 \left(p_2Dp_3\right) {p_2}_{c} {\left(p_1N\epsilon
   _2\right)}_{a} {\epsilon _3}_{bi}I_{5}-4 \left(p_2p_3\right) {p_3}_{a}
   {\left(p_2D\epsilon _2\right)}_{c} {\epsilon _3}_{bi}I_{6}\right.\non\\
   & \qquad\left. -4 \left(p_2Dp_3\right)
   {p_3}_{a} {\left(p_2D\epsilon _2\right)}_{c} {\epsilon _3}_{bi}I_{7}-4
   \left(p_2p_3\right) {p_3}_{b} {\left(p_1N\epsilon _2\right)}_{a} {\epsilon
   _3}_{ci}I_{9}\right.\non\\
   & \qquad\left. -4 \left(p_2Dp_3\right) {p_3}_{b} {\left(p_1N\epsilon _2\right)}_{a}
   {\epsilon _3}_{ci}I_{5}+2 \left(p_1Np_2\right) \left(p_2p_3\right) {\epsilon
   _2}_{ab} {\epsilon _3}_{ci}I_{9}\right.\non\\
   & \qquad\left. -2 \left(p_1Np_3\right) \left(p_2p_3\right)
   {\epsilon _2}_{ab} {\epsilon _3}_{ci}I_{9}-4 \left(p_1Np_2\right)
   \left(p_1Np_3\right) {\epsilon _2}_{ab} {\epsilon _3}_{ci}I_{10}\right.\non\\
   & \qquad\left. +2
   \left(p_1Np_2\right) \left(p_2Dp_3\right) {\epsilon _2}_{ab} {\epsilon
   _3}_{ci}I_{5}+2 \left(p_1Np_3\right) \left(p_2Dp_3\right) {\epsilon _2}_{ab}
   {\epsilon _3}_{ci}I_{5}\rp,
\end{align}


\begin{align}
\mathcal{A}_{C^{(p-1)}Bh}^{(2)}=& \frac{4 i^{p (p+1)} \sqrt{2}}{(p-3)!}{{C}^{ij}}_{b_1...b_{p-3}}
   {\varepsilon}^{abcdb_1...b_{p-3}}\left(2 {p_2}_{d} {p_2}_{j} {p_3}_{a} {p_3}_{i}
   {\left(\epsilon _2\epsilon _3\right)}_{bc}I_{9}\right.\non\\
   & \qquad\left. +2 {p_2}_{d} {p_2}_{j} {p_3}_{a}
   {p_3}_{i} {\left(\epsilon _2D\epsilon _3\right)}_{bc}I_{5}+{p_2}_{d} {p_2}_{i}
   {p_3}_{j} {\left(p_2\epsilon _3\right)}_{c} {\epsilon _2}_{ab}I_{9}\right.\non\\
   & \qquad\left. +{p_2}_{i}
   {p_3}_{d} {p_3}_{j} {\left(p_2\epsilon _3\right)}_{c} {\epsilon _2}_{ab}I_{9}-2
   {p_2}_{d} {p_2}_{i} {p_3}_{j} {\left(p_1N\epsilon _3\right)}_{c} {\epsilon
   _2}_{ab}I_{10}\right.\non\\
   & \qquad\left. +{p_2}_{d} {p_2}_{i} {p_3}_{j} {\left(p_2D\epsilon _3\right)}_{c}
   {\epsilon _2}_{ab}I_{5}+{p_2}_{i} {p_3}_{d} {p_3}_{j} {\left(p_2D\epsilon
   _3\right)}_{c} {\epsilon _2}_{ab}I_{5}\right.\non
\end{align}
\begin{align}
   & \qquad\left. -\frac{1}{2} \tr(D\epsilon _3)
   {p_2}_{d} {p_2}_{j} {p_3}_{a} {p_3}_{i} {\epsilon _2}_{bc}I'_{4}+{p_2}_{d} {p_2}_{j}
   {p_3}_{a} {\left(p_2\epsilon _3\right)}_{i} {\epsilon _2}_{bc}I_{9}+{p_2}_{d}
   {p_3}_{a} {p_3}_{j} {\left(p_2\epsilon _3\right)}_{i} {\epsilon _2}_{bc}I_{9}\right.\non\\
   & \qquad\left. -2
   {p_2}_{d} {p_2}_{j} {p_3}_{a} {\left(p_1N\epsilon _3\right)}_{i} {\epsilon
   _2}_{bc}I_{10}+{p_2}_{d} {p_2}_{j} {p_3}_{a} {\left(p_2D\epsilon _3\right)}_{i}
   {\epsilon _2}_{bc}I_{5}\right.\non\\
   & \qquad\left. +{p_2}_{d} {p_3}_{a} {p_3}_{i} {\left(p_2D\epsilon
   _3\right)}_{j} {\epsilon _2}_{bc}I_{5}+{p_2}_{d} {p_2}_{j} {p_3}_{a}
   {\left(p_3D\epsilon _3\right)}_{i} {\epsilon _2}_{bc}I'_{4}+2 {p_2}_{d} {p_2}_{i}
   {p_3}_{a} {\left(p_3\epsilon _2\right)}_{c} {\epsilon _3}_{bj}I_{9}\right.\non\\
   & \qquad\left. -2 {p_2}_{a}
   {p_3}_{d} {p_3}_{i} {\left(p_3\epsilon _2\right)}_{c} {\epsilon _3}_{bj}I_{9}-2
   {p_2}_{d} {p_3}_{a} {p_3}_{i} {\left(p_2D\epsilon _2\right)}_{c} {\epsilon
   _3}_{bj}I_{4}\right.\non\\
   & \qquad\left. +2 {p_2}_{d} {p_2}_{i} {p_3}_{a} {\left(p_3D\epsilon _2\right)}_{c}
   {\epsilon _3}_{bj}I_{5}+2 {p_2}_{a} {p_3}_{d} {p_3}_{i} {\left(p_3D\epsilon
   _2\right)}_{c} {\epsilon _3}_{bj}I_{5}\right.\non\\
   & \qquad\left. +4 {p_2}_{d} {p_3}_{b} {p_3}_{i}
   {\left(p_1N\epsilon _2\right)}_{a} {\epsilon _3}_{cj}I_{10}-\left(p_2p_3\right)
   {p_2}_{d} {p_2}_{i} {\epsilon _2}_{ab} {\epsilon _3}_{cj}I_{9}\right.\non\\
   & \qquad\left. +2
   \left(p_1Np_3\right) {p_2}_{d} {p_2}_{i} {\epsilon _2}_{ab} {\epsilon
   _3}_{cj}I_{10}-\left(p_2Dp_3\right) {p_2}_{d} {p_2}_{i} {\epsilon _2}_{ab} {\epsilon
   _3}_{cj}I_{5}\right.\non\\
   & \qquad\left. +\left(p_2p_3\right) {p_2}_{d} {p_3}_{i} {\epsilon _2}_{ab} {\epsilon
   _3}_{cj}I_{9}-\left(p_2Dp_3\right) {p_2}_{d} {p_3}_{i} {\epsilon _2}_{ab} {\epsilon
   _3}_{cj}I_{5}-\left(p_2p_3\right) {p_2}_{j} {p_3}_{a} {\epsilon _2}_{bc} {\epsilon
   _3}_{di}I_{9}\right.\non\\
   & \qquad\left. -\left(p_2Dp_3\right) {p_2}_{j} {p_3}_{a} {\epsilon _2}_{bc} {\epsilon
   _3}_{di}I_{5}+\left(p_2p_3\right) {p_3}_{a} {p_3}_{j} {\epsilon _2}_{bc} {\epsilon
   _3}_{di}I_{9}\right.\non\\
   & \qquad\left. +2 \left(p_1Np_2\right) {p_3}_{a} {p_3}_{j} {\epsilon _2}_{bc}
   {\epsilon _3}_{di}I_{10}-\left(p_2Dp_3\right) {p_3}_{a} {p_3}_{j} {\epsilon _2}_{bc}
   {\epsilon _3}_{di}I_{5}\right),
\end{align}


\be
\mathcal{A}_{C^{(p-1)}Bh}^{(3)}=\frac{8 i^{p (p+1)} \sqrt{2}}{(p-4)!}{{C}^{ijk}}_{b_1...b_{p-4}}
   {\varepsilon}^{abcdeb_1...b_{p-4}}{p_2}_{e} {p_2}_{j} {p_3}_{a} {p_3}_{i} {\epsilon
   _2}_{bc} {\epsilon _3}_{dk}I_{10}.
\ee

\subsection{$C^{(p-3)}$ amplitudes}


\be
\mathcal{A}_{C^{(p-3)}BB}=\mathcal{A}_{C^{(p-3)}BB}^{(0)}+\mathcal{A}_{C^{(p-3)}BB}^{(1)}+\mathcal
{A}_{C^{(p-3)}BB}^{(2)}.
\ee

\begin{align}
\mathcal{A}_{C^{(p-3)}BB}^{(0)}=& \frac{2 i^{p (p+1)} \sqrt{2}}{(p-3)!}{C}_{b_1...b_{p-3}}
   {\varepsilon}^{abcdb_1...b_{p-3}}\lp -4 \left(p_2Dp_3\right) {p_2}_{d} {p_3}_{a}
   {\left(\epsilon _2\epsilon _3\right)}_{bc}I_{0}\right.\non\\
   & \qquad\left. +4 {p_2}_{d} {p_3}_{b}
   {\left(p_1N\epsilon _2\right)}_{a} {\left(p_2\epsilon _3\right)}_{c}I_{9}+4
   {p_2}_{c} {p_3}_{a} {\left(p_1N\epsilon _2\right)}_{d} {\left(p_1N\epsilon
   _3\right)}_{b}I_{10}\right.\non\\
   & \qquad\left. +4 {p_2}_{d} {p_3}_{b} {\left(p_2D\epsilon _2\right)}_{a}
   {\left(p_2\epsilon _3\right)}_{c}I_{6}+4 {p_2}_{d} {p_3}_{b} {\left(p_1N\epsilon
   _3\right)}_{c} {\left(p_2D\epsilon _2\right)}_{a}I_{4}\right.\non\\
   & \qquad\left. -4 {p_2}_{c} {p_3}_{a}
   {\left(p_1N\epsilon _2\right)}_{d} {\left(p_2D\epsilon _3\right)}_{b}I_{5}+4
   {p_2}_{d} {p_3}_{a} {\left(p_2D\epsilon _2\right)}_{c} {\left(p_2D\epsilon
   _3\right)}_{b}I_{7}\right.\non\\
   & \qquad\left. -4 {p_2}_{d} {p_3}_{a} {\left(p_2D\epsilon _3\right)}_{c}
   {\left(p_3\epsilon _2\right)}_{b}I_{8}-4 {p_2}_{d} {p_3}_{a} {\left(p_2D\epsilon
   _2\right)}_{c} {\left(p_3D\epsilon _3\right)}_{b}I_{3}\right.\non\\
   & \qquad\left. -4 \left(p_2p_3\right)
   {p_2}_{d} {p_3}_{a} {\left(\epsilon _2D\epsilon _3\right)}_{cb}I_{0}+2
   \left(p_2p_3\right) {p_2}_{d} {\left(p_2\epsilon _3\right)}_{c} {\epsilon
   _2}_{ab}I_{2}\right.\non\\
   & \qquad\left. -2 \left(p_1Np_3\right) {p_2}_{d} {\left(p_2\epsilon _3\right)}_{c}
   {\epsilon _2}_{ab}I_{9}+2 \left(p_2p_3\right) {p_3}_{d} {\left(p_2\epsilon
   _3\right)}_{c} {\epsilon _2}_{ab}I_{2}\right.\non\\
   & \qquad\left. +2 \left(p_1Np_2\right) {p_3}_{d}
   {\left(p_2\epsilon _3\right)}_{c} {\epsilon _2}_{ab}I_{9}-2 \left(p_2p_3\right)
   {p_2}_{d} {\left(p_1N\epsilon _3\right)}_{c} {\epsilon _2}_{ab}I_{9}\right.\non\\
   & \qquad\left. +2
   \left(p_2Dp_3\right) {p_2}_{d} {\left(p_1N\epsilon _3\right)}_{c} {\epsilon
   _2}_{ab}I_{5}+2 \left(p_2p_3\right) {p_2}_{d} {\left(p_3D\epsilon _3\right)}_{c}
   {\epsilon _2}_{ab}I'_{6}\right.\non\\
   & \qquad\left. +2 \left(p_2Dp_3\right) {p_2}_{d} {\left(p_3D\epsilon
   _3\right)}_{c} {\epsilon _2}_{ab}I'_{7}-2 \left(p_1N\epsilon _3p_2\right) {p_2}_{a}
   {p_3}_{d} {\epsilon _2}_{bc}I_{9}\right.\non\\
   & \qquad\left. -2 \left(p_2\epsilon _3Dp_3\right) {p_2}_{a}
   {p_3}_{d} {\epsilon _2}_{bc}I'_{6}+2 \left(p_1N\epsilon _3Dp_2\right) {p_2}_{a}
   {p_3}_{d} {\epsilon _2}_{bc}I_{5}\right.\non\\
   & \qquad\left. -2 \left(p_2D\epsilon _3Dp_3\right) {p_2}_{a}
   {p_3}_{d} {\epsilon _2}_{bc}I'_{7}+2 \left(p_2p_3\right) {p_3}_{a}
   {\left(p_1N\epsilon _3\right)}_{d} {\epsilon _2}_{bc}I_{9}\right.\non\\
   & \qquad\left. +4 \left(p_1Np_2\right)
   {p_3}_{a} {\left(p_1N\epsilon _3\right)}_{d} {\epsilon _2}_{bc}I_{10}-2
   \left(p_2Dp_3\right) {p_3}_{a} {\left(p_1N\epsilon _3\right)}_{d} {\epsilon
   _2}_{bc}I_{5}\right.\non\\
   & \qquad\left. +2 \left(p_1Np_3\right) {p_2}_{d} {\left(p_2D\epsilon _3\right)}_{a}
   {\epsilon _2}_{bc}I_{5}-2 \left(p_2Dp_3\right) {p_2}_{d} {\left(p_2D\epsilon
   _3\right)}_{a} {\epsilon _2}_{bc}I_{1}\right.\non
\end{align}
\begin{align}
   & \qquad\left. +2 \left(p_1Np_2\right) {p_3}_{d}
   {\left(p_2D\epsilon _3\right)}_{a} {\epsilon _2}_{bc}I_{5}-2 \left(p_2Dp_3\right)
   {p_3}_{d} {\left(p_2D\epsilon _3\right)}_{a} {\epsilon _2}_{bc}I_{1}\right.\non\\
   & \qquad\left. -2
   \left(p_2p_3\right) {p_3}_{a} {\left(p_3D\epsilon _3\right)}_{d} {\epsilon
   _2}_{bc}I'_{6}-2 \left(p_1Np_2\right) {p_3}_{a} {\left(p_3D\epsilon _3\right)}_{d}
   {\epsilon _2}_{bc}I'_{4}\right.\non\\
   & \qquad\left. -2 \left(p_2Dp_3\right) {p_3}_{a} {\left(p_3D\epsilon
   _3\right)}_{d} {\epsilon _2}_{bc}I'_{7}+2 \left(p_3D\epsilon _2p_3\right) {p_2}_{d}
   {p_3}_{a} {\epsilon _3}_{bc}I'_{11}\right.\non\\
   & \qquad\left. -\left(p_1Np_2\right) \left(p_2p_3\right)
   {\epsilon _2}_{ab} {\epsilon _3}_{cd}I_{9}+\left(p_1Np_2\right) \left(p_1Np_3\right)
   {\epsilon _2}_{ab} {\epsilon _3}_{cd}I_{10}\right.\non\\
   & \qquad\left. -\left(p_1Np_2\right)
   \left(p_2Dp_3\right) {\epsilon _2}_{ab} {\epsilon _3}_{cd}I_{5}\rp+(2\leftrightarrow 3),
\end{align}

\begin{align}
\mathcal{A}_{C^{(p-3)}BB}^{(1)}=& \frac{2 i^{p (p+1)} \sqrt{2}}{(p-4)!}{{C}^{i}}_{b_1...b_{p-4}}
   {\varepsilon}^{abcdeb_1...b_{p-4}}\lp -2 {p_2}_{e} {p_2}_{i} {p_3}_{a}
   {\left(p_2\epsilon _3\right)}_{d} {\epsilon _2}_{bc}I_{9}\right.\non\\
   & \qquad\left. +2 {p_2}_{a} {p_3}_{e}
   {p_3}_{i} {\left(p_2\epsilon _3\right)}_{d} {\epsilon _2}_{bc}I_{9}+4 {p_2}_{e}
   {p_2}_{i} {p_3}_{a} {\left(p_1N\epsilon _3\right)}_{d} {\epsilon _2}_{bc}I_{10}\right.\non\\
   & \qquad\left. -2
   {p_2}_{e} {p_2}_{i} {p_3}_{a} {\left(p_2D\epsilon _3\right)}_{d} {\epsilon
   _2}_{bc}I_{5}-2 {p_2}_{a} {p_3}_{e} {p_3}_{i} {\left(p_2D\epsilon _3\right)}_{d}
   {\epsilon _2}_{bc}I_{5}\right.\non\\
   & \qquad\left. -2 {p_2}_{e} {p_2}_{i} {p_3}_{a} {\left(p_3D\epsilon
   _3\right)}_{d} {\epsilon _2}_{bc}I'_{4}-\left(p_2p_3\right) {p_2}_{e} {p_2}_{i}
   {\epsilon _2}_{ab} {\epsilon _3}_{cd}I_{9}\right.\non\\
   & \qquad\left. +2 \left(p_1Np_3\right) {p_2}_{e}
   {p_2}_{i} {\epsilon _2}_{ab} {\epsilon _3}_{cd}I_{10}-\left(p_2Dp_3\right) {p_2}_{e}
   {p_2}_{i} {\epsilon _2}_{ab} {\epsilon _3}_{cd}I_{5}\right.\non\\
   & \qquad\left. +\left(p_2p_3\right) {p_2}_{e}
   {p_3}_{i} {\epsilon _2}_{ab} {\epsilon _3}_{cd}I_{9}-\left(p_2Dp_3\right) {p_2}_{e}
   {p_3}_{i} {\epsilon _2}_{ab} {\epsilon _3}_{cd}I_{5}\rp+(2\leftrightarrow 3),
\end{align}

\be
\mathcal{A}_{C^{(p-3)}BB}^{(2)}=\frac{2 \sqrt{2} i^{p(p+1)}}{(p-5)!}{{C}^{ij}}_{b_1...b_{p-5}}
   {\varepsilon}^{abcdefb_1...b_{p-5}}{\epsilon _2}_{bc} {\epsilon _3}_{de} {p_2}_{f}
   {p_2}_{j} {p_3}_{a} {p_3}_{i}I_{10}+(2\leftrightarrow 3),
\ee

\be
\mathcal{A}_{C^{(p-3)}hh}=\mathcal{A}_{C^{(p-3)}hh}^{(0)}.
\ee

\begin{align}
\mathcal{A}_{C^{(p-3)}hh}^{(0)}=& \frac{8 \sqrt{2} i^{p (p+1)}}{(p-3)!}{C}_{b_1...b_{p-3}} {p_2}_{d} {p_3}_{a}
   {\varepsilon}^{abcdb_1...b_{p-3}} \lp\left(p_2Dp_3\right) {\left(\epsilon _2\epsilon
   _3\right)}_{bc}I_{0}\right.\non\\
   & \qquad\left. -2 {\left(p_2D\epsilon _3\right)}_{c} {\left(p_3\epsilon
   _2\right)}_{b}I_{0}-\left(p_2p_3\right) {\left(\epsilon _2D\epsilon
   _3\right)}_{cb}I_{0}\rp+(2\leftrightarrow 3).
\end{align}

\section{Conclusion}
In this paper we calculated all tree level string theory vacuum to Dp-brane disc amplitudes involving an arbitrary RR-state and two NS-NS vertex operators. This computation was performed in~\cite{Becker:2011ar} for the simplest case of a RR-state of type $C^{(p-3)}$. Here we used the aid of a computer to calculate all possible amplitudes involving a RR-vertex operator of type $C^{(p+1+2k)}$.
Our calculation was checked for consistency against previous results from the literature~\cite{Becker:2011ar}, as well as its symmetry under the exchange of the NS-NS vertex operators. The evaluation of the effective action that follows from our result is work in progress. Aspects of this work will appear in a forthcoming publication~\cite{paper2}.

\section*{Acknowledgements}
We thank S.Sethi for useful discussions. 
This work
was supported by the grant PHY-1214344, the George P. and Cynthia W. Mitchell Institute for Fundamental Physics and Astronomy
and the Department of Physics at Texas A{\&}M.

\appendix

\section{Evaluation of each sector}
\label{app:Sectors}

In this appendix we present the correlators that we need from each sector.  More details can be found in  \cite{Becker:2011bw}.  Note that we have already dealt with the ghosts by the equation (\ref{eq:GhostSector}).  This leaves only the matter fields $\psi$ and $X$.

\vspace{0.5cm}

{\bf{$\psi$ sector:}}

For each explicit $\widetilde{\psi}(\bar{z})$, we first bring it to the right and convert it to a $\psi(\bar{z}^{-1})$ using the boundary state $|B\rangle$, via
\be
\widetilde{\psi}^\m(\bar{z})\left|B\right\rangle=-i\bar{z}^{-1}D^\m_{\hph{\m}\n}\psi^\n(\bar{z}^{-1})\left|B\right\rangle.
\ee

This gives
\begin{multline}
f_{AB}\left\langle A,B\left|\psi^{\m_1}(z_1)\cdots\psi^{\m_n}(z_n)\right|B\right\rangle=\lp -1\rp^{n+1}2^{-\frac{n}{2}}\lp z_1\cdots z_n\rp^{-\hlf}\\
\times\left\{ T^{\m_1\cdots\m_n}+\frac{z_1+z_2}{z_1-z_2}\eta^{\m_1\m_2}T^{\m_3\cdots\m_n}+\cdots\right.\\
\left. +\frac{z_1+z_2}{z_1-z_2}\frac{z_3+z_4}{z_3-z_4}\eta^{\m_1\m_2}\eta^{\m_3\m_4}T^{\m_5\cdots\m_n}+\cdots\right\},
\end{multline}
where $\cdots$ represents all the possible contractions, keeping track of appropriate signs from anticommuting the $\psi$'s or $\G$'s, and where the objects $T^{\m_1\cdots\m_n}$ are given by
\be
T^{a_1\cdots a_ki_1\cdots i_\ell}=\lp -1\rp^{\hlf\lp p^2+p+k^2+k\rp+p\ell+1}\frac{32}{\lp p+1-k\rp!}\e^{a_1\cdots a_k}_{\hph{a_1\cdots a_k}b_1\cdots b_{p+1-k}}F^{b_1\cdots b_{p+1-k}i_1\cdots i_\ell}.
\ee
Recall that we use notation where $a$, $b$, etc.\ represent directions along the D-brane, while $i$, $j$, etc.\ are normal to the D-brane.


%
%

\vspace{0.5cm}

{\bf{$X$ sector:}}

For the bosons, we again use the boundary state to convert anti-holomorphic operators to holomorphic ones.  Indeed, if we split the exponential into left- and right-moving parts\footnote{This neglects the zero-mode, but the zero-mode piece is correctly accounted for in the full correlators (\ref{eq:XCor1}) and (\ref{eq:XCor2}).},
\be
:e^{ipX(z,\bar{z})}:=:e^{ipX_L(z)}e^{ipX_R(\bar{z})}:,
\ee
then we can use
\be
e^{ipX_R(\bar{z})}\left|B\right\rangle=e^{ipDX_L(\bar{z}^{-1})}\left|B\right\rangle.
\ee
Then for a correlator with only exponentials, we have
\begin{multline}
\label{eq:XCor1}
\left\langle 0\left|:e^{ip_1X(z_1,\barz_1)}:\cdots:e^{ip_nX(z_n,\barz_n)}:\right|B\right\rangle=\lp 2\pi\rp^{p+1}\d^{p+1}(\hlf\lp 1+D\rp\sum_{i=1}^np_i)\\
\times\prod_{k=1}^n\lp\left|z_k\right|^2-1\rp^{p_kDp_k}
\prod_{1\le\ell<m\le n}\left|z_\ell-z_m\right|^{2p_\ell p_m}\left|z_\ell\barz_m-1\right|^{2p_\ell Dp_m}.
\end{multline}

Similarly, if we have explicit factors of $\bar{\p}X(\bar{z})$, we use
\be
\overline{\p}X^\m(\barz)\left|B\right\rangle=-\barz^{-2}D^\m_{\hphantom{\m}\n}\p X^\n(\barz^{-1})\left|B\right\rangle,
\ee
to convert them to holomorphic operators.  Then for a correlator that involves these as well, we have for example
\begin{multline}
\label{eq:XCor2}
\left\langle 0\left|:e^{ip_1X(z_1,\barz_1)}:\cdots :e^{ip_{n-1}X(z_{n-1},\barz_{n-1})}::\p X^\m(z_n)e^{ip_nX(z_n,\barz_n)}:\right|B\right\rangle\\
=\left\langle 0\left|:e^{ip_1X(z_1,\barz_1)}:\cdots :e^{ip_nX(z_n,\barz_n)}:\right|B\right\rangle\\
\times\lp\frac{ip_1}{z_1-z_n}+\cdots +\frac{ip_{n-1}}{z_{n-1}-z_n}-\frac{i\barz_1Dp_1}{z_n\barz_1-1}-\cdots -\frac{i\barz_nDp_n}{\left|z_n\right|^2-1}\rp^\m.
\end{multline}
If there is more than one $\p X(z)$, then we must also include in the usual way terms where they contract with each other.

\section{Some integrals}
\label{app:Integrals}

The integrals appearing in the amplitudes section are defined as
\begin{equation}
I_n=\int _{\left|z_i\right|\leq 1}d^2z_2d^2z_3A_n
\end{equation}
where \(A_n\) are\\
\\
\(A_0=\frac{\left(-\bar{z}_3 z_2+\bar{z}_2 z_3\right){}^2}{2 \left|z_2\right|{}^2 \left|1-\bar{z}_3 z_2\right|{}^2 \left|z_2-z_3\right|{}^2 \left|z_3\right|{}^2}\\
\\
A_1=\frac{\left|1+\bar{z}_3 z_2\right|{}^2}{\left|z_2\right|{}^2 \left|1-\bar{z}_3 z_2\right|{}^2 \left|z_3\right|{}^2}\\
\\
A_2=\frac{\left|z_2+z_3\right|{}^2}{\left|z_2\right|{}^2 \left|z_2-z_3\right|{}^2 \left|z_3\right|{}^2}\\
\\
A_3=\frac{\left(1+\left|z_2\right|{}^2\right) \left(1+\left|z_3\right|{}^2\right)}{\left|z_2\right|{}^2 \left(1-\left|z_2\right|{}^2\right) \left|z_3\right|{}^2
\left(1-\left|z_3\right|{}^2\right)}\\
\\
A_4=\frac{2 \left(1+\left|z_2\right|{}^2\right)}{\left|z_2\right|{}^2 \left(1-\left|z_2\right|{}^2\right) \left|z_3\right|{}^2}\\
\\
A_5=\frac{1-\left|z_2\right|{}^2 \left|z_3\right|{}^2}{\left|z_2\right|{}^2 \left|1-\bar{z}_3 z_2\right|{}^2 \left|z_3\right|{}^2}\\
\\
A_6=\frac{\left(-1-\left|z_2\right|{}^2\right) \left(\left|z_2\right|{}^2-\left|z_3\right|{}^2\right)}{\left|z_2\right|{}^2 \left(1-\left|z_2\right|{}^2\right)
\left|z_2-z_3\right|{}^2 \left|z_3\right|{}^2}\\
\\
A_7=\frac{\left(-1-\left|z_2\right|{}^2\right) \left(1-\left|z_2\right|{}^2 \left|z_3\right|{}^2\right)}{\left|z_2\right|{}^2 \left(1-\left|z_2\right|{}^2\right)
\left|1-\bar{z}_3 z_2\right|{}^2 \left|z_3\right|{}^2}\\
\\
A_8=\frac{\left(\left|z_2\right|{}^2-\left|z_3\right|{}^2\right) \left(1-\left|z_2\right|{}^2 \left|z_3\right|{}^2\right)}{\left|z_2\right|{}^2 \left|1-\bar{z}_3
z_2\right|{}^2 \left|z_2-z_3\right|{}^2 \left|z_3\right|{}^2}\\
\\
A_9=\frac{\left|z_2\right|{}^2-\left|z_3\right|{}^2}{\left|z_2\right|{}^2 \left|z_2-z_3\right|{}^2 \left|z_3\right|{}^2}\\
\\
A_{10}=\frac{1}{\left|z_2\right|{}^2 \left|z_3\right|{}^2}\\
\\
A_{11}=\frac{\left(\left|z_2\right|{}^2-\left|z_3\right|{}^2\right) \left(1-\left|z_2\right|{}^2 \left|z_3\right|{}^2\right)}{\left|z_2\right|{}^2
\left|1-\bar{z}_3 z_2\right|{}^2 \left|z_2-z_3\right|{}^2 \left|z_3\right|{}^2}-\frac{\left(-\bar{z}_3 z_2+\bar{z}_2 z_3\right){}^2}{\left|z_2\right|{}^2
\left|1-\bar{z}_3 z_2\right|{}^2 \left|z_2-z_3\right|{}^2 \left|z_3\right|{}^2}=A_8-2A_0\\
\\
A_{12}=\frac{\left(\left|z_2\right|{}^2-\left|z_3\right|{}^2\right) \left(1-\left|z_2\right|{}^2 \left|z_3\right|{}^2\right)}{\left|z_2\right|{}^2
\left|1-\bar{z}_3 z_2\right|{}^2 \left|z_2-z_3\right|{}^2 \left|z_3\right|{}^2}-\frac{3 \left(-\bar{z}_3 z_2+\bar{z}_2 z_3\right){}^2}{\left|z_2\right|{}^2
\left|1-\bar{z}_3 z_2\right|{}^2 \left|z_2-z_3\right|{}^2 \left|z_3\right|{}^2}=A_8-6A_0\\
\\
A_{13}=\frac{3 \left(\left|z_2\right|{}^2-\left|z_3\right|{}^2\right) \left(1-\left|z_2\right|{}^2 \left|z_3\right|{}^2\right)}{\left|z_2\right|{}^2
\left|1-\bar{z}_3 z_2\right|{}^2 \left|z_2-z_3\right|{}^2 \left|z_3\right|{}^2}-\frac{\left(-\bar{z}_3 z_2+\bar{z}_2 z_3\right){}^2}{\left|z_2\right|{}^2
\left|1-\bar{z}_3 z_2\right|{}^2 \left|z_2-z_3\right|{}^2 \left|z_3\right|{}^2}=3A_8-2A_0\\
\\
A_{14}=\frac{\left(1-\left|z_2\right|{}^2 \left|z_3\right|{}^2\right) \left|z_2+z_3\right|{}^2}{\left|z_2\right|{}^2 \left|z_2-z_3\right|{}^2 \left|z_3\right|{}^2
\left|-1+\bar{z}_2 z_3\right|{}^2}\\
\\
A_{15}=\frac{\left(\left|z_2\right|{}^2-\left|z_3\right|{}^2\right) \left|1+\bar{z}_2 z_3\right|{}^2}{\left|z_2\right|{}^2 \left|z_2-z_3\right|{}^2
\left|z_3\right|{}^2 \left|-1+\bar{z}_2 z_3\right|{}^2}\\
\\
A_{16}=\frac{\left(\left|z_2\right|{}^2-\left|z_3\right|{}^2\right) \left|z_2+z_3\right|{}^2}{\left|z_2\right|{}^2 \left|z_2-z_3\right|{}^4 \left|z_3\right|{}^2}\\
\\
A_{17}=\frac{\left(1-\left|z_2\right|{}^2 \left|z_3\right|{}^2\right) \left|1+\bar{z}_2 z_3\right|{}^2}{\left|z_2\right|{}^2 \left|z_3\right|{}^2
\left|-1+\bar{z}_2 z_3\right|{}^4}\\
\\
A_{18}=\frac{\left(1+\left|z_2\right|{}^2\right) \left(\left|z_2\right|{}^2-\left|z_3\right|{}^2\right) \left(1+\left|z_3\right|{}^2\right)}{\left|z_2\right|{}^2
\left(-1+\left|z_2\right|{}^2\right) \left|z_2-z_3\right|{}^2 \left|z_3\right|{}^2 \left(-1+\left|z_3\right|{}^2\right)}\\
\\
A_{19}=\frac{\left(1+\left|z_2\right|{}^2\right) \left(1+\left|z_3\right|{}^2\right) \left(1-\left|z_2\right|{}^2 \left|z_3\right|{}^2\right)}{\left|z_2\right|{}^2
\left(-1+\left|z_2\right|{}^2\right) \left|z_3\right|{}^2 \left(-1+\left|z_3\right|{}^2\right) \left|-1+\bar{z}_2 z_3\right|{}^2}\\
\\
A_{20}=\frac{\left(1+\left|z_2\right|{}^2\right) \left|z_2+z_3\right|{}^2}{\left|z_2\right|{}^2 \left(1-\left|z_2\right|{}^2\right) \left|z_2-z_3\right|{}^2
\left|z_3\right|{}^2}\\
\\
A_{21}=\frac{\left(1+\left|z_2\right|{}^2\right) \left|1+\bar{z}_2 z_3\right|{}^2}{\left|z_2\right|{}^2 \left(1-\left|z_2\right|{}^2\right) \left|z_3\right|{}^2
\left|-1+\bar{z}_2 z_3\right|{}^2}\\
\\
A_{22}=\frac{\left(1+\left|z_3\right|{}^2\right) \left(\left(\left|z_2\right|{}^2-\left|z_3\right|{}^2\right) \left(1-\left|z_2\right|{}^2 \left|z_3\right|{}^2\right)+\left(\bar{z}_3
z_2-\bar{z}_2 z_3\right){}^2\right)}{\left|z_2\right|{}^2 \left|z_2-z_3\right|{}^2 \left|z_3\right|{}^2 \left(1-\left|z_3\right|{}^2\right) \left|-1+\bar{z}_2
z_3\right|{}^2}\\
\\
A_{23}=\frac{\left(\left|z_2\right|{}^2-\left|z_3\right|{}^2\right) \left(\bar{z}_3 z_2+\bar{z}_2 z_3\right)}{\left|z_2\right|{}^2 \left|z_2-z_3\right|{}^4
\left|z_3\right|{}^2}\\
\\
A_{24}=\frac{\left(1-\left|z_2\right|{}^2 \left|z_3\right|{}^2\right) \left(\bar{z}_3 z_2+\bar{z}_2 z_3\right)}{\left|z_2\right|{}^2 \left|z_3\right|{}^2
\left|-1+\bar{z}_2 z_3\right|{}^4}\)\\

Each of these could be expanded out in terms of the $I_{a,b,c,d,e,f}$ defined in section \ref{subsec:Integrands}, but it is these combinations which appear naturally from the contractions.

Finally, when we put a prime on an integral, $I_n'$, we obtain it from $I_n$ by exchanging $z_2$ with $z_3$ in $A_n$ before performing the integration, so for instance
\be
I_4'=\int _{\left|z_i\right|\leq 1}d^2z_2d^2z_3\frac{2 \left(1+\left|z_3\right|{}^2\right)}{\left|z_2\right|{}^2 \left(1-\left|z_3\right|{}^2\right) \left|z_3\right|{}^2}.
\ee


\providecommand{\href}[2]{#2}\begingroup\raggedright
\endgroup
\end{document}